\documentclass[a4paper,11pt]{article}
\pdfoutput=1 

\usepackage{jheppub} 

\usepackage[T1]{fontenc} 
\usepackage{tensor} 
\usepackage{soul}
\usepackage{color}

\usepackage{empheq}
\usepackage{mathrsfs} 
\usepackage{url}
\usepackage[utf8]{inputenc}
\usepackage[toc,page]{appendix}
\usepackage[autostyle]{csquotes}
\usepackage{slashed}
\usepackage{amssymb}
\usepackage{graphicx}
\usepackage{dirtytalk}
\usepackage{longtable}
\usepackage{array}
\usepackage{bm} 
\usepackage{amsmath}
\usepackage{nccmath}
\usepackage{cancel}
\usepackage{mathtools}
\usepackage{amsfonts}
\usepackage{amssymb}
\usepackage[labelfont=bf]{caption}
\usepackage{paralist} 
\usepackage{hyperref}
\usepackage{cleveref}
\usepackage{float}

\title{\boldmath Generalized Einstein-Maxwell theory: Seeley-DeWitt coefficients and logarithmic corrections to the entropy of extremal and non-extremal black holes}

\author{Sudip Karan}
\author{and Binata Panda}
\affiliation{Department of Physics,\\ Indian Institute of Technology (Indian School of Mines),\\ Dhanbad, Jharkhand-826004, India}

\emailAdd{sudip.karan@ap.ism.ac.in}
\emailAdd{binata@iitism.ac.in}

\abstract{We present a consolidated manual of Euclidean gravity approaches for finding the logarithmic corrections to the entropy of the full Kerr-Newman family of black holes in both extremal and non-extremal limits. Seeley-DeWitt coefficients for the quadratic fluctuations of a concern gravity theory appear to be the key ingredients in this manual. Following the manual, we calculate the first three Seeley-DeWitt coefficients and logarithmic corrections to the entropy of extremal and non-extremal black holes in a generalized Einstein-Maxwell theory minimally-coupled to additional massless scalar, vector, spin-1/2 Dirac and spin-3/2 Rarita-Schwinger fields. We finally employ the Seeley-DeWitt data to reproduce the logarithmic entropy corrections for extremal black holes in all $\mathcal{N} \geq 2$ Einstein-Maxwell supergravity via an alternative local supersymmetrization method.}

\begin{document} 
\maketitle
\flushbottom

\section{Introduction}\label{A}

In any quantum gravity model, including string theory, it has been found that the leading quantum correction to the Bekenstein-Hawking entropy formula of black holes carrying large charges\footnote{In the large-charge limit, the charge, angular momentum, mass and other black hole parameters are scaled so that the black hole becomes large (i.e., the horizon area $\mathcal{A}_{H}\gg l_P^2$, $l_P$ is the Planck length), keeping different dimensionless ratios unchanged.} is proportional to the logarithm of horizon area, called the logarithmic correction \cite{Solodukhin:1995na,Solodukhin:1995nb,Fursaev:1995df,Mavromatos:1996kc,Mann:1996bi,Mann:1998hm,Kaul:2000rk,Carlip:2000nv,Govindarajan:2001ee,Gupta:2002bg,Medved:2004eh,Page:2005xp,Banerjee:2008cf,Banerjee:2009fz,Majhi:2009gi,Cai:2010ua,Aros:2010jb,Solodukhin:2010pk,Mukherji:2002de}. These quantum corrections are special features of black hole entropy that can be entirely computable macroscopically only using IR or low-energy gravity data (i.e., the massless fields and their couplings to the black hole background), without any prior knowledge about the UV completion of the gravity theory \cite{Banerjee:2011oo,Banerjee:2011pp,Sen:2012rr,Sen:2012qq,Chowdhury:2014np,Gupta:2014ns,Bhattacharyya:2012ss,Karan:2019sk,Keeler:2014nn,Charles:2015nn,Larsen:2015nx,Castro:2018tg,Karan:2020sk,Sen:2013ns,Banerjee:2020wbr}. Also, the logarithmic entropy corrections are quite robust for being unaffected by the massive fields as well as the classical higher-derivative corrections \cite{Banerjee:2011oo}. Logarithmic corrections to the Bekenstein-Hawking formula must match with the entropy calculated from the UV complete microscopic side (via logarithm of microstate degeneracy) and hence serve as a strong \say{infrared window} into the microphysics of black holes.

On the macroscopic side, quantum corrections to the Bekenstein-Hawking entropy of black holes are generally realized by different loop contributions in the saddle point expansion \cite{Banerjee:2011oo,Banerjee:2011pp,Sen:2012rr,Sen:2012qq,Chowdhury:2014np,Gupta:2014ns,Bhattacharyya:2012ss,Karan:2019sk,Banerjee:2020wbr,Karan:2020sk} of the partition function describing the black hole geometry. Logarithmic corrections are filtered by tracking only massless states in the one-loop, requiring evaluation of the one-loop quantum effective action. Euclidean gravity approaches have been an excellent success in computing logarithmic entropy corrections for the extremal \cite{Banerjee:2011oo,Banerjee:2011pp,Sen:2012rr,Sen:2012qq,Chowdhury:2014np,Gupta:2014ns,Bhattacharyya:2012ss,Karan:2019sk,Karan:2020sk,Banerjee:2020wbr,Keeler:2014nn,Larsen:2015nx} as well as non-extremal \cite{Sen:2013ns,Charles:2015nn,Castro:2018tg} black holes. For extremal black holes, the most popular and efficient approach is the quantum entropy function formalism \cite{Sen:2008wa,Sen:2009wb,Sen:2009wc} that only demands data from the finite part of Euclideanized extremal near-horizon geometry.\footnote{Readers are also encouraged to review \cite{Majhi:2015pra}, where the formalism of entropy function for extremal near-horizon black holes is identical to Sen's original work \cite{Sen:2008wa,Sen:2009wb,Sen:2009wc} but uses only the surface term (Gibbons-Hawking-York type) of the gravitational action.} For non-extremal black holes, we cast the strategy developed in \cite{Sen:2013ns} that analyzes the whole Euclideanized black hole geometry, including the near-horizon. The main technical aspects of these Euclidean gravity approaches are --
\begin{inparaenum}[(i)]
	\item easy to track down the appropriate logarithmic terms,
	\item not limited to any particular type of space-time, and
	\item a special treatment to deal with the zero-mode contributions to the one-loop effective action.	
\end{inparaenum}

A conventional way to estimate the necessary one-loop effective action is the heat kernel treatment. In this treatment, one-loop effective action is represented as the proper time integral \cite{Schwinger:1951sp,DeWitt:1975ps} of heat kernel of the kinetic operator controlling fluctuations in the one-loop. One can further expand the heat kernel perturbatively in terms of the Seeley-DeWitt expansion coefficients \cite{Seeley:1966tt,Seeley:1969uu,DeWitt:1965ff,DeWitt:1967gg,DeWitt:1967hh,DeWitt:1967ii,Vassilevich:2003ll} for a short proper time. Therefore, the computation of logarithmic corrections now involves finding these coefficients (especially the third Seeley-DeWitt coefficient $a_4(x)$ introduced in \eqref{B5}) for massless fluctuations around the black hole background. In this paper, we will follow a standard but indirect approach \cite{Vassilevich:2003ll} to compute the Seeley-DeWitt coefficients. The major highlight of this computation approach is that the Seeley-DeWitt coefficients are expressed only into different invariants induced from the background fields and background geometry. Henceforth one can utilize the results for any arbitrary black hole background of the concerned theory without any limitations.

Einstein-Maxwell theory (EMT), also known as electrovacuum, is a typical source-free gravitational field theory. The most general, static/stationary solutions to the EMT field equations describe the Kerr-Newman family of black holes.\footnote{See \cite{Adamo:2014lk} for a review.} In this family, the Kerr-Newman background is rotating and electrically charged, while it's non-rotating-uncharged, non-rotating-charged and rotating-uncharged limits respectively describe the Schwarzschild, Reissner-Nordstr\"om and Kerr black holes. Quantum corrections to the Bekenstein-Hawking entropy of Kerr-Newman family of black holes play a vital role in macroscopic (low-energy) analysis of any quantum theory of gravity. For example, in the works \cite{Banerjee:2011pp,Sen:2012qq,Karan:2019sk,Charles:2015nn,Karan:2020sk,Gupta:2014ns,Keeler:2014nn,Larsen:2015nx,Banerjee:2020wbr}, Kerr-Newman black holes are interpreted as solutions to the supergravity-embedded Einstein-Maxwell theories and corresponding logarithmic entropy corrections are obtained. These supergravity theories are low-energy string theory models (generally type-II string theory compactified on a Calabi–Yau three-fold \cite{Grana:2006mg,Freedman:2012xp}), where the EMT serves as a basic building block.\footnote{EMT describes the bosonic-sector of supergravity multiplet of the Einstein-Maxwell supergravity theories (e.g., see \cite{Sen:2012qq,Karan:2020sk,Charles:2015nn}).} So it will always be fundamental to investigate quantum corrections to the entropy of black holes in a simple EMT.

Logarithmic corrections to the entropy of black holes in a simple four-dimensional ($d=4$) EMT are not new; results for the extremal Kerr-Newman black holes are already achieved in \cite{Bhattacharyya:2012ss} via the quantum entropy function formalism. The current paper aims to extend the work \cite{Bhattacharyya:2012ss} for both extremal and non-extremal black holes in a more generalized EMT. We investigate a \say{minimally-coupled} EMT where the simple four-dimensional Einstein-Maxwell system is coupled minimally\footnote{The minimally-coupled fields are only coupled to gravity via the metric, without any other interactions.} to additional massless scalar, vector (\textit{a.k.a.} the Maxwell field or U(1) gauge field), spin-1/2 Dirac, spin-3/2 Rarita-Schwinger fields. If the couplings were set as \say{non-minimal}, the generalized theory would have new black hole solutions beyond the Kerr-Newman family. For example, in the non-minimally coupled Einstein-Maxwell-scalar models \cite{Astefanesei:2019pk}, various dilatonic and scalarised black holes are possible. But, the \say{minimally-coupled} EMT is structured so that the fields are minimal and fluctuate around the pure Einstein-Maxwell backgrounds. Thus, there will not be any new type of black hole solutions except the Kerr-Newman family of black holes. As a result, the minimally-coupled massless fields will give rise to additional contributions to the pure EMT results (both Seeley-DeWitt coefficients and logarithmic entropy corrections), and our primary goal in this paper is to evaluate all these contributions.

The technical aim of this paper is three-fold. First, we design a consolidated and compact logarithmic correction manual based on the Euclidean gravity approaches \cite{Sen:2008wa,Sen:2009wb,Sen:2009wc,Sen:2013ns}, followed by the standard Seeley-DeWitt computation approach \cite{Vassilevich:2003ll}. The manual is global for the full Kerr-Newman family of black holes, which does not even depend on supersymmetry. In the second part, we calculate the first three Seeley-DeWitt coefficients for the fluctuations of the $d=4$ \say{minimally-coupled} EMT and employ them in obtaining logarithmic entropy corrections for both the extremal and non-extremal Kerr-Newman family of black holes. Finally, we generalize the \say{minimally-coupled} EMT for arbitrary numbers of fields. The generalized Seeley-DeWitt coefficients and logarithmic correction results are recorded in \cref{gem1,gem5,gem8,gem11,gem12}. 
The non-extremal results exhibit a perfect match with that of \cite{Sen:2013ns}, where the $a_4(x)$ coefficients for individual fields are not evaluated directly but arranged from secondary data provided in some earlier research works and schemes \cite{Duff:1977ay,Christensen:1979md,Christensen:1980iy,Duff:1980qv,Christensen:1980ee}. In contrast, our work follows a generic path: we set up the action of concerned theory, analyze the quadratic fluctuations, calculate all the first three Seeley-DeWitt coefficients and finally find the logarithmic corrections. The contributions due to vector and Rarita-Schwinger fields in the Schwarzschild formula \eqref{gem12} perfectly match the results obtained in \cite{Majhi:2009uk} via the tunneling approach. All the corrections to extremal black hole entropy are found to be new reports. The calculated \say{minimally-coupled} EMT results have crucial utility in derivations of various minimally-coupled sectors of supergravity-embedded Einstein-Maxwell theories. As an indirect application, we locally supersymmetrize the generalized $a_4(x)$ formula \eqref{gem1} and derive logarithmic entropy corrections for the extremal Kerr-Newman, Kerr and Reissner-Nordstr\"om black holes in $\mathcal{N} \geq 2,d=4$ Einstein-Maxwell supergravity theories (see \cref{loc6,loc8}).

The plan of this paper is as follows. \Cref{B} serves as an effective manual of computing logarithmic correction to the entropy of extremal and non-extremal Kerr-Newman family of black holes. In \cref{em}, we calculate necessary Seeley-DeWitt coefficients and logarithmic corrections to the entropy of extremal and non-extremal Kerr-Newman family of black holes in the $d=4$ \say{minimally-coupled} Einstein-Maxwell theory. We end in \cref{gem} by  summarizing and discussing the generalized results. \Cref{App1} discusses the treatment of separating the local and zero-mode contributions of Euclidean one-loop quantum effective action that eventually leads to the working formula for logarithmic corrections.  \Cref{App2} derives the Einstein equation as well as includes some particular identities for the four-dimensional Einstein-Maxwell background. In \cref{App3}, we briefly present the  cumbersome trace calculations for the Einstein-Maxwell sector.


\section{An effective manual for logarithmic correction to black hole entropy}\label{B}
In this section, we present a consolidated manual of how Euclidean gravity approaches provide a standard path to evaluate logarithmic correction to the entropy of extremal and non-extremal Kerr-Newman family of black holes using the Seeley-DeWitt coefficients.
\subsection{The working formula}\label{Ba}
Let us consider a four-dimensional Euclidean gravitational theory with the matter fields $\xi$ and the metric $g$ describing corresponding space-time geometry over a compact manifold. If we fluctuate $g$ and $\xi$ around an arbitrary classical background solution ($\bar{g},\bar{\xi}$) for small quantum fluctuations $\tilde{\xi}_m = \lbrace \tilde{{g}}, \tilde{\xi} \rbrace $,
\begin{align}\label{B1}
	{g} = \bar{g}+ \tilde{{g}},\enspace \xi = \bar\xi + \tilde{\xi},
\end{align}
then the action $\mathcal{S}[g,\xi]$ describing the theory is expanded perturbatively as
\begin{equation}
	\mathcal{S}[{g},\xi] = \mathcal{S}[\bar{g}, \bar{\xi}]+ \delta^2\mathcal{S}[\tilde{\xi}_m]+ \text{higher order terms}.
\end{equation}
The quadratic-fluctuated action $\delta^2\mathcal{S}[\tilde{\xi}_m]$ can take the schematic form,\footnote{All the tensor indices of the fluctuations have been suppressed here for simplicity. We will consider these labelings from \cref{nonzeromode} onward.}
\begin{align}\label{B2}
	\delta^2\mathcal{S}[\tilde{\xi}_m] = \int \mathrm{d}^4x \sqrt{\text{det}\thinspace \bar{g}}\thinspace \tilde{\xi}_m \Lambda\tilde{\xi}_n,
\end{align}
where $\Lambda$ is the kinetic differential operator that characterizes the quadratic fluctuations. We can now introduce the heat kernel $K(x,y;s)$ that encodes all the data about the spectrum of the operator $\Lambda$ \cite{Banerjee:2011oo,Banerjee:2011pp,Sen:2012qq,Sen:2012rr,Vassilevich:2003ll},
\begin{equation}\label{B13}
	K(x,y;s) = \sum_i e^{-\lambda_is}f_i(x)f_i(y),
\end{equation}
followed by the trace of the heat kernel, called the heat trace $D(s)$,
\begin{align}\label{B4}
	D(s)= \text{tr}\thinspace (e^{-s\Lambda}) = \int \mathrm{d}^4x \sqrt{\text{det}\thinspace \bar{g}}\thinspace K(x,x;s).
\end{align}
Here $\lbrace f_i\rbrace$ are the eigenfunctions of the operator $\Lambda$ with eigenvalues $\lbrace \lambda_i \rbrace$ and $s$ is a proper time with units of (length)$^2$, called the heat kernel time. 
By the proper time representation \cite{Schwinger:1951sp,DeWitt:1975ps}, the quantum corrected one-loop effective action $\mathcal{W}$ is then expressed in terms of the heat trace $D(s)$ as \cite{Karan:2019sk} 
\begin{align}\label{B3}
	\mathcal{W}= -\frac{\chi}{2}\int_\epsilon^\infty \frac{\mathrm{d}s}{s} D(s),
\end{align}
where $\chi = \pm 1$ for bosonic and fermionic fluctuations, respectively; $\epsilon$ is a UV cutoff, restricted by \ $\epsilon\sim {l_p}^2\sim G_N$.\footnote{This work considers the units $\hbar = c = k_B = 1$.} In order to evaluate $K(x,x;s)$ and $D(s)$, one can cast the Seeley-DeWitt expansion as $s\to 0$,
\begin{align}\label{B5}
	K(x,x;s) \cong \sum_{n=0}^\infty s^{n-2}a_{2n}(x),
\end{align}
where the coefficients $a_{2n}(x)$ of the perturbative expansion are known as the {Seeley-DeWitt coefficients} \cite{Seeley:1966tt,Seeley:1969uu,DeWitt:1965ff,DeWitt:1967gg,DeWitt:1967hh,DeWitt:1967ii}. The $s$ independent part of the expansion \eqref{B5} allows us to find a logarithmic term from the integration \eqref{B3} in the range $\epsilon\ll s\ll \mathcal{A}_H$ ($\mathcal{A}_H$ is the black hole horizon area),
\begin{align}\label{B6}
	\int_\epsilon^\infty \frac{\mathrm{d}s}{s} \chi D(s) = \cdots+ \int \mathrm{d}^4x \sqrt{\text{det}\thinspace \bar{g}}\thinspace a_4(x)\thinspace\text{ln}\thinspace \left(\frac{\mathcal{A}_{H}}{G_N}\right)+ \cdots .
\end{align}
In any Euclidean gravity approach, this logarithmic term corrects the black hole entropy if one integrates out only massless modes in the one-loop effective action $\mathcal{W}$ \cite{Banerjee:2011oo,Banerjee:2011pp,Sen:2012rr,Sen:2012qq,Chowdhury:2014np,Gupta:2014ns,Bhattacharyya:2012ss,Karan:2019sk,Karan:2020sk,Keeler:2014nn,Charles:2015nn,Larsen:2015nx,Castro:2018tg,Sen:2013ns}. Therefore, the logarithmic correction to black hole entropy is calculated by the general formula,\footnote{See \cref{App1} of how a revised form of $\mathcal{W}$ leads to the $\Delta S_{\text{BH}}$ formula \eqref{B7t}.}
\begin{subequations}\label{B7t}
	\begin{align}\label{B7}
		\Delta S_{\text{BH}} &= \frac{1}{2}(\mathcal{C}_{\text{local}}+\mathcal{C}_{\text{zm}})\thinspace\text{ln}\thinspace \left(\frac{\mathcal{A}_{H}}{G_N}\right),
	\end{align}
	with the following local ($\mathcal{C}_{\text{local}}$) and zero-mode ($\mathcal{C}_{\text{zm}}$) contributions
	\begin{align}
		\mathcal{C}_{\text{local}} &= \int \mathrm{d}^4x \sqrt{\text{det}\thinspace \bar{g}}\thinspace a_4(x),\label{B8}\\
		\mathcal{C}_{\text{zm}}&= \sum_{\tilde{\xi}_m}\chi (\beta_{\tilde{\xi}_m}-1)n^0_{\tilde{\xi}_m},\label{B9}
	\end{align}
\end{subequations}
where $n^0_{\tilde{\xi}_m}$ serves the zero-mode counts for a particular fluctuation $\tilde{\xi}_m$ having scaling dimension $\beta_{\tilde{\xi}_m}$. 

It is also reasonable to investigate whether the universal nature of black hole entropy sustains after incorporating the logarithmic corrections. In the describe Euclidean gravity framework, the logarithmic corrections $\Delta S_{\text{BH}}$ are obtained as $\mathcal{C}\,\text{ln}\thinspace \big(\frac{\mathcal{A}_{H}}{G_N}\big)$ where the prefactor $\mathcal{C}=\frac{1}{2}(\mathcal{C}_{\text{local}}+\mathcal{C}_{\text{zm}})$ is termed as the coefficient of logarithmic correction. The $\text{ln}\thinspace \big(\frac{\mathcal{A}_{H}}{G_N}\big)$ part is global in any generic gravity theory, while the coefficient of logarithmic correction $\mathcal{C}$ is generally \say{geometric} (i.e., depends on the black hole geometric characteristics like mass, charge, angular momentum, etc.). But $\mathcal{C}$ may also become \say{non-geometric}, then the particular logarithmic correction result is fully universal.

\subsection{Computation of the zero-mode part of logarithmic entropy correction}\label{zeromode}
The contributions of various fields to the zero-mode correction $\mathcal{C}_{\text{zm}}$ are not new; they have been computed and analyzed in many works \cite{Banerjee:2011pp,Sen:2012rr,Sen:2012qq,Sen:2013ns}. A generalized and concise review can also be found in \cite{Charles:2015nn}. For both the scalar and spin-1/2 fields $n^0_{0}=n^0_{1/2}=0$, hence they have no contribution in the $\mathcal{C}_{\text{zm}}$ formula \eqref{B9}. The four-dimensional vector fields have $\beta_1=1$, so they also contribute nothing to $\mathcal{C}_{\text{zm}}$ via the formula \eqref{B9}. For a spin-3/2 field $\beta_{3/2}=3$ in $d=4$; $n^0_{3/2}=-4$ for BPS solutions in $\mathcal{N} \geq 2,d=4$ supergravities and 0 for all kinds of non-supersymmetric black holes. The four-dimensional metric has $\beta_2=2$ with $n^0_{2}=-3-\mathbb{K}$ for the extremal case and $n^0_{2}=-\mathbb{K}$ for the non-extremal case. Here $\mathbb{K}$ (number of rotational isometries) is 3 for the non-rotating black holes and 1 for the rotating black holes. In summary, $\mathcal{C}_{\text{zm}}$ receives a contribution from only the metric for non-supersymmetric black holes, while both the metric and gravitino field contribute for BPS black holes. And $\chi$ in the formula \eqref{B9} takes care of the bosonic and fermionic nature of the fields. Make sure to set $\chi=-1$ for fermions and $\chi=+1$ for bosons.
\subsection{Computation of the local part of logarithmic entropy correction}\label{nonzeromode}
We will now outline computation strategies of the local correction $\mathcal{C}_{\text{local}}$ for both the extremal and non-extremal Kerr-Newman family of black holes. This analysis is twofold -- a standard computation approach of Seeley-DeWitt coefficients for any arbitrary gravity theory and the appropriate limits of necessary background invariants over the extremal and non-extremal Kerr-Newman black hole geometries.
\subsubsection{A standard approach for computing Seeley-DeWitt coefficients}\label{sdc}
In order to compute $a_4(x)$, we will pursue a standard and efficient approach reviewed in \cite{Vassilevich:2003ll}. The basic technical set-up of this approach is briefly depicted as follows. First, one needs to adjust (up to a total derivative\footnote{In our choice of $d=4$ compact manifold (without boundary), all the total derivative terms emerge as boundary terms in the integration \eqref{B4} via the relation \eqref{B5} and hence contribute nothing. This particular setting is useful in dealing with quantum fluctuations around the boundary of asymptotically flat black holes (e.g., Kerr-Newman family of black holes).}) the quadratic-fluctuated action \eqref{B2} so that the kinetic operator $\Lambda$ becomes Hermitian, Laplace-type and minimal of the following form 
\begin{align}\label{Bx}
	\tilde{\xi}_m \Lambda^{\tilde{\xi}_m \tilde{\xi}_n}\tilde{\xi}_n = \pm\tilde{\xi}_m\Big((D_\rho D^\rho)I^{\tilde{\xi}_m \tilde{\xi}_n} +(N^\rho D_\rho)^{\tilde{\xi}_m \tilde{\xi}_n}+P^{\tilde{\xi}_m \tilde{\xi}_n}\Big)\tilde{\xi}_n.
\end{align}
Here $D_\rho$ is a Christoffel-spin-connected covariant derivative. $I$ is an arbitrary matrix induced from combinations of the background metric $\bar{g}$ and the identity operator in spin space. $I$ acts as an effective metric that simultaneously contracts all the indices of any particular fluctuation. $N^\rho, P$ are also arbitrary matrices in terms of background fields $\bar{\xi}$ and the background metric $\bar{g}$. The standard form \eqref{Bx} can be  generalized further so that it incorporates interactions between the fluctuations. The generalized operator form is prescribed as
\begin{align}\label{B10}
	\tilde{\xi}_m \Lambda^{\tilde{\xi}_m \tilde{\xi}_n}\tilde{\xi}_n = \pm\tilde{\xi}_m\Big((\mathcal{D}_\rho\mathcal{D}^\rho)I^{\tilde{\xi}_m \tilde{\xi}_n}+E^{\tilde{\xi}_m \tilde{\xi}_n}\Big)\tilde{\xi}_n,
\end{align}
where the redefined covariant derivative $\mathcal{D}_\rho$, the gauge connection $\omega_\rho$, the commutator curvature $\Omega_{\rho\sigma}\equiv [\mathcal{D}_\rho,\mathcal{D}_\sigma]$ and the matrix-valued potential $E$ are defined as
\begin{subequations}\label{B11}
	\begin{align}
		\mathcal{D}_\rho \tilde{\xi}_m &= D_\rho \tilde{\xi}_m + {(\omega_\rho)_{\tilde{\xi}_m}}^{\tilde{\xi}_n} \tilde{\xi}_n,\thinspace (\omega_\rho)^{\tilde{\xi}_m \tilde{\xi}_n} = \frac{1}{2}(N_\rho)^{\tilde{\xi}_m \tilde{\xi}_n} \quad\forall m\neq n,\label{B11a}\\
		(\Omega_{\rho\sigma})^{\tilde{\xi}_m \tilde{\xi}_n}&= [D_\rho,D_\sigma]^{\tilde{\xi}_m \tilde{\xi}_n}+{D_{[\rho}\omega_{\sigma]}} ^{\tilde{\xi}_m \tilde{\xi}_n}+[\omega_\rho,\omega_\sigma]^{\tilde{\xi}_m \tilde{\xi}_n},\label{B11b}\\
		E^{\tilde{\xi}_m \tilde{\xi}_n} &= P^{\tilde{\xi}_m \tilde{\xi}_n}-(D^\rho\omega_\rho)^{\tilde{\xi}_m \tilde{\xi}_n}-(\omega^\rho)^{\tilde{\xi}_m \tilde{\xi}_p}{(\omega_\rho)_{\tilde{\xi}_p}}^{\tilde{\xi}_n}. \label{B11c}
	\end{align}
\end{subequations}
Note that all the matrices are labeled by \say{$\tilde{\xi}_m\tilde{\xi}_n$}, describing with which pair of fluctuations the matrices are contracted where $\tilde{\xi}_m$ includes particular fluctuations-types along with their tensor indices. Any fluctuation is considered as the \say{minimally-coupled} if it evaluates $\omega_\rho=0$ (i.e., no other interactions except the one with background gravity via the $\sqrt{\text{det}\thinspace \bar{g}}$ term in its quadratic-fluctuated action). The commutations of covariant derivatives acting on the scalar $\phi$, vector $a_\mu$, spin-1/2 Dirac $\lambda$, spin-3/2 Rarita-Schwinger $\psi_\mu$ and metric $h_{\mu\nu}$ fluctuations have the following standard definitions
{
	\allowdisplaybreaks
	\begin{subequations}\label{BB}
		\begin{align}
			[D_\rho,D_\sigma]\phi &= 0, \label{cs}\\
			[D_\rho,D_\sigma]a_\mu &= R\indices{_\mu^\nu_{\rho\sigma}} a_\nu,\label{cv}\\
			[D_\rho,D_\sigma]\lambda &= \frac{1}{4}\gamma^\alpha\gamma^\beta R_{\alpha\beta\rho\sigma}\lambda,\label{cd}\\
			[D_\rho,D_\sigma]\psi_\mu &=R\indices{_\mu^\nu_{\rho\sigma}}\psi_\nu +\frac{1}{4}\gamma^\alpha\gamma^\beta R_{\alpha\beta\rho\sigma}\psi_\mu,\label{crs}\\
			[D_\rho,D_\sigma]h_{\mu\nu} &= R\indices{_\mu^\alpha_{\rho\sigma}}h_{\alpha\nu} + R\indices{_\nu^\alpha_{\rho\sigma}}h_{\mu\alpha}.\label{ct}
		\end{align}
\end{subequations}}
With all these data, the formulae for the first three Seeley-DeWitt coefficients are listed as \cite{Vassilevich:2003ll}
\begin{equation}\label{B12}
	\begin{split}
		a_0(x) &= \frac{\chi}{16\pi^2}\thinspace\text{tr}(I),\\
		a_2(x) &=\frac{\chi}{16\pi^2 \times 6}\thinspace \text{tr}(6E+RI),\\
		a_4(x) &= \frac{\chi}{16\pi^2 \times 360}\thinspace \text{tr}\Big( 60RE+ 180E^2+ 30\Omega_{\rho\sigma}\Omega^{\rho\sigma}\\
		&\qquad\qquad\qquad +( 2R_{\mu\nu\rho\sigma}R^{\mu\nu\rho\sigma}-2R_{\mu\nu}R^{\mu\nu}+5 R^2)I\Big),
	\end{split}
\end{equation}
where $\chi=+1,-1$ and -1/2 for the fluctuation of bosons, Dirac spinors and Majorana spinors, respectively. {The above described approach naturally recognizes any fermionic fluctuation as a Dirac spinor \cite{Karan:2020sk,Sen:2012qq,Charles:2015nn}. Majorana spinors have half the degrees of freedom of Dirac fermions, and hence for casting them via the current approach, one needs to employ an additional 1/2 factor in the formulae \eqref{B12}. Weyl spinors are prohibited in this approach due to having both the right and left chiral states and must be redefined into Dirac or Majorana forms.} The crucial benefit of the present Seeley-DeWitt computation approach is that after taking quadratic fluctuations around a classical background, we have direct formulae to calculate the Seeley-DeWitt coefficients in terms of background invariants like $R_{\mu\nu\rho\sigma}R^{\mu\nu\rho\sigma}, R_{\mu\nu}R^{\mu\nu}, R^2, R_{\mu\nu\rho\sigma}\bar{F}^{\mu\nu}\bar{F}^{\rho\sigma},\bar{F}_{\mu\nu}\bar{F}^{\mu\nu}, (\bar{F}_{\mu\nu}\bar{F}^{\mu\nu})^2$, etc. Thus the results are global and not limited to any particular background of the theory. Apart from several useful applications, the Seeley-DeWitt coefficients have essential utility in one-loop quantum corrections. Since the logarithmic corrections are evaluated by $a_4(x)$, we find it sufficient to calculate the Seeley-DeWitt coefficients only up to this order.

\subsubsection{Strategies for extremal and non-extremal Kerr-Newman black holes}\label{Bc}
Kerr-Newman black holes are the most general stationary  solutions to the equations of motion of Einstein-Maxwell theory \cite{Adamo:2014lk}. In standard spherical coordinates $(t,r,\psi,\phi)$, a Kerr-Newman black hole with charge $Q$, mass $M$ and angular momentum $J$ is characterized by the metric \cite{Bhattacharyya:2012ss},
{
	\allowdisplaybreaks
	\begin{equation}\label{lc8}
		\begin{split}
			ds^2 &= \bar{g}_{\mu\nu}dx^\mu dx^\nu\\
			&= -\frac{r^2+b^2{\cos}^2 \psi-2Mr+Q^2}{(r^2+b^2{\cos}^2 \psi)}dt^2+\frac{r^2+b^2{\cos}^2 \psi}{(r^2+b^2-2Mr+Q^2)}dr^2 \\
			&\quad +\frac{(r^2+b^2{\cos}^2 \psi)(r^2+b^2)+(2Mr-Q^2)b^2{\sin}^2 \psi}{(r^2+b^2{\cos}^2 \psi)}{\sin}^2 \psi d\phi^2\\
			&\quad +(r^2+b^2{\cos}^2 \psi)d\psi^2+\frac{2(Q^2-2Mr)b}{(r^2+b^2{\cos}^2 \psi)}{\sin}^2 \psi\thinspace dt\thinspace d\phi,
		\end{split}
\end{equation}}
with the following geometric invariants \cite{Henry:2000wd,Cherubini:2002we}
{
	\allowdisplaybreaks
	\begin{align}\label{B14}
		\begin{split}
			R_{\mu\nu}R^{\mu\nu}&= \frac{4Q^4}{(r^2+b^2{\cos}^2 \psi)^4},\\
			R_{\mu\nu\rho\sigma}R^{\mu\nu\rho\sigma} &= \frac{8}{(r^2+b^2{\cos}^2 \psi)^6}\Big(Q^4(7r^4-34r^2b^2{\cos}^2 \psi+7b^4 {\cos}^4 \psi)\\
			&\qquad -12MQ^2r(r^4-10r^2b^2{\cos}^2 \psi+ 5b^4{\cos}^4 \psi)\\
			&\qquad +6M^2(r^6-15b^2r^4{\cos}^2 \psi+ 15b^4r^2{\cos}^4 \psi-b^6{\cos}^6 \psi)\Big),
		\end{split}
\end{align}}
where $b=J/M$. One can achieve the metric forms of Schwarzschild, Kerr and Reissner-Nordstr\"om black holes in appropriate limits of the Kerr-Newman metric \eqref{lc8}. We are now going to discuss two separate strategies for calculating $\mathcal{C}_{\text{local}}$ corrections to the entropy of extremal and non-extremal Kerr-Newman family of black holes. For both the strategies, all the charges of the Kerr-Newman black hole need to be scaled by a common large scale, say $L$, so that the angular momentum, charge and horizon area of the black hole are scaled as $J \sim L^2$, $Q \sim L$ and $\mathcal{A}_{H} \sim L^2$ (the large-charge limits) \cite{Bhattacharyya:2012ss,Sen:2013ns}.
\subsubsection*{Strategy A (for extremal black holes):}\label{ST1}
For the extremal Kerr-Newman family of black holes, we follow a strategy that casts the quantum entropy function formalism \cite{Sen:2008wa,Sen:2009wb,Sen:2009wc}. This Euclidean gravity approach is quite efficient in calculating the one-loop quantum corrections of extremal black holes by only using the near-horizon geometry data \cite{Banerjee:2011oo,Banerjee:2011pp,Sen:2012rr,Sen:2012qq,Chowdhury:2014np,Gupta:2014ns,Bhattacharyya:2012ss,Karan:2019sk,Karan:2020sk,Banerjee:2020wbr}. The near-horizon geometry of an extremal black hole is structured as $AdS_2 \times \mathcal{K}$ ($\mathcal{K}$ is a compact space that includes the angular coordinates) and can be described by a Euclidean path integral partition function $\mathcal{Z}_{AdS_2}$ of various fields asymptotically approaching the classical near-horizon background. $\mathcal{Z}_{AdS_2}$ could be expressed in the form $e^{-\alpha \ell}\times \mathcal{Z}^{\text{finite}}_{AdS_2}$ for some constant $\alpha$ and boundary length $\ell$ of the regulated $AdS_2$. Then the principles of $AdS_2/CFT_1$ correspondence allow the quantum entropy function formalism to identify the finite part $\mathcal{Z}^{\text{finite}}_{AdS_2}$ as an alternative definition of the quantum degeneracy for extremal black holes from the macroscopic side. Therefore, the $\mathcal{C}_{\text{local}}$ formula for extremal black holes becomes
\begin{align}\label{ext1}
	\mathcal{C}^{\thinspace\text{extremal}}_{\text{local}} &= \int_{\text{near-horizon}} \mathrm{d}^4x \sqrt{\text{det}\thinspace \bar{g}}\thinspace a_4(x),
\end{align}
where the $a_4(x)$ coefficient needs to be integrated only over the extremal near-horizon geometry (of the form $AdS_2 \times \mathcal{K}$) by dropping all the terms proportional to the boundary of IR regulated $AdS_2$ \cite{Banerjee:2011pp,Sen:2012rr,Sen:2012qq,Bhattacharyya:2012ss,Karan:2020sk,Banerjee:2020wbr}. Now for the fluctuations of Einstein-Maxwell theories, $a_4(x)$ is always electro-magnetic dual invariant (i.e., the $R_{\mu\nu\rho\sigma}\bar{F}^{\mu\nu}\bar{F}^{\rho\sigma}$ and $(\bar{F}_{\mu\nu}\bar{F}^{\mu\nu})^2$ terms are absent in the final expression\footnote{For example, see the $a_4(x)$ results in \cref{pem} of this paper and also in \cite{Karan:2020sk,Banerjee:2020wbr}.}) and the theories also satisfy the $R=0$ condition\footnote{See \cref{App2}.} for any background solution. Therefore, one needs only the \say{finite} near-horizon extremal limits of Ricci and Riemann tensor squares for calculating logarithmic corrections to the entropy of extremal Kerr-Newman black holes \cite{Bhattacharyya:2012ss}, 
{
	\allowdisplaybreaks
	\begin{subequations}\label{ext2}
		\begin{align}
			\begin{split}
				\int_{\text{near-horizon}}  \mathrm{d}^4x \sqrt{\text{det}\thinspace \bar{g}}R_{\mu\nu}R^{\mu\nu}&= -4\pi^2\Big( 3\mathcal{B} +(8{b^\prime}^2+5)\mathcal{B}^\prime\Big),\\
				\int_{\text{near-horizon}}  \mathrm{d}^4x \sqrt{\text{det}\thinspace \bar{g}}\thinspace R_{\mu\nu\rho\sigma}R^{\mu\nu\rho\sigma} &= -16\pi^2\Big( 3\mathcal{B}-(8{b^\prime}^6+20{b^\prime}^4+8{b^\prime}^2-1)\mathcal{B}^\prime\Big),
			\end{split}
		\end{align}
		where
		\begin{align}
			\begin{split}
				{b^\prime}&=b/Q= J/MQ,\\
				\mathcal{B} &= \frac{2{b^\prime}^2+1}{{b^\prime}({b^\prime}^2+1)^{5/2}}{\tan}^{-1}\Bigg(\frac{{b^\prime}}{\sqrt{{b^\prime}^2+1}} \Bigg),\\
				\mathcal{B}^\prime &=\frac{1}{({b^\prime}^2+1)^2(2{b^\prime}^2+1)}.
			\end{split}
		\end{align}
\end{subequations}}
The strategy of finding $\mathcal{C}_{\text{local}}$ for the extremal Kerr-Newman family of black holes involves the following algorithm --
\begin{enumerate}
	\item Calculate $a_4(x)$ pursuing the standard technique of \cref{sdc} for the quadratic fluctuations of massless fields in the theory embedded with Einstein-Maxwell backgrounds.
	
	\item Simplify the final form of $a_4(x)$ only into the curvature invariants $R_{\mu\nu\rho\sigma}R^{\mu\nu\rho\sigma}$ and $R_{\mu\nu}R^{\mu\nu}$ using appropriate equations of motion of the Einstein-Maxwell backgrounds (refer to \cref{App2}).
	
	\item Compute $\mathcal{C}_{\text{local}}$ for an extremal Kerr-Newman black hole by employing the $a_4(x)$ result into the formula \eqref{ext1}, along with the limits \eqref{ext2}.
	
	\item For extremal Kerr ($Q=0, J\neq 0$) and Reissner-Nordstr\"om ($Q\neq 0,J=0$) black holes, set $b^\prime \to \infty$ and $b^\prime \to 0$ respectively in the Kerr-Newman $\mathcal{C}_{\text{local}}$ result. The undetermined value of  $b^\prime$ for Schwarzschild black holes ($Q= 0,J=0$) justifies that the extremal-Schwarzschild limit is not possible.	
\end{enumerate}
\subsubsection*{Strategy B (for non-extremal black holes):}\label{ST2}
In \hyperref[ST1]{Strategy A}, the quantum entropy function formalism exactly predicts the degeneracy of extremal black holes by alternatively defining a near-horizon partition function. But for the generic non-extremal Kerr-Newman family of black holes, we cast the Euclidean gravity approach developed in \cite{Sen:2013ns} where a special treatment\footnote{All technical details are elaborately depicted in section 2.3 of \cite{Sen:2013ns}. Also, see the note mentioned in \cref{App1} of this paper.} is used to extract out the particular black hole partition function by eliminating the thermal gas contribution of all {particles} (massless and massive) present in the theory. This treatment effectively leads to the logarithmic corrections for a particular choice of integration range of the heat kernel time $s$ and writes the following $\mathcal{C}_{\text{local}}$ formula for non-extremal black holes 
\begin{align}\label{nonextclocal}
	\mathcal{C}^{\thinspace\text{non-extremal}}_{\text{local}} &= \int_{\text{full geometry}} \mathrm{d}^4x \sqrt{\text{det}\thinspace \bar{g}}\thinspace a_4(x),
\end{align} 
where $a_4(x)$ needs to be integrated over the full black hole geometry. Now the Seeley-DeWitt coefficient  $a_4(x)$ encodes all the trace anomaly data related to the logarithmic corrections via \eqref{nonextclocal} and can be written in the following form 
\begin{align}\label{next1}
	a_4(x) = \frac{c}{16\pi^2}W_{\mu\nu\rho\sigma}W^{\mu\nu\rho\sigma}-\frac{a}{16\pi^2}E_4, 
\end{align}
where the anomalies $W_{\mu\nu\rho\sigma}W^{\mu\nu\rho\sigma}$ and $E_4$ are recognized as Weyl tensor square and 4D Euler-Gauss-Bonnet density, respectively, for the constant coefficients $c$ and $a$ (a.k.a. the central charges of corresponding conformal anomalies). For any arbitrary backgrounds, one can use the standard forms --
\begin{align}\label{next2}
	\begin{gathered}
		W_{\mu\nu\rho\sigma}W^{\mu\nu\rho\sigma} = R_{\mu\nu\rho\sigma}R^{\mu\nu\rho\sigma}-2R_{\mu\nu}R^{\mu\nu} + \frac{1}{3}R^2,\\
		E_4 = R_{\mu\nu\rho\sigma}R^{\mu\nu\rho\sigma}-4R_{\mu\nu}R^{\mu\nu} + R^2.
	\end{gathered}	
\end{align} 
The standard definition of four-dimensional Euler characteristic suggests the integral of $E_4$ is a pure number for non-extremal black holes, i.e.,
\begin{align}\label{Euler}
	\frac{1}{32\pi^2}\int_{\text{full geometry}} \mathrm{d}^4x \sqrt{\text{det}\thinspace \bar{g}}\thinspace E_4 &= 2.
\end{align}
On the other hand, the integral of  $W_{\mu\nu\rho\sigma}W^{\mu\nu\rho\sigma}$ over the full non-extremal Kerr-Newman geometry \eqref{lc8} can be evaluated in terms of different dimensionless ratios of black hole parameters as \cite{Sen:2013ns}
\begin{subequations}\label{next3}
	\begin{align}
		\int_{\text{full geometry}} \mathrm{d}^4x \sqrt{\text{det}\thinspace \bar{g}}\thinspace W_{\mu\nu\rho\sigma}W^{\mu\nu\rho\sigma} &= 64\pi^2 + \frac{\pi \beta Q^4 \mathcal{B}^{\prime\prime}}{b^5 r_H^4(b^2+r_H^2)},
	\end{align}
	where
	\begin{align}
		\begin{split}
			r_H &= M + \sqrt{M^2-Q^2-b^2},\\
			\beta &= \frac{32\pi^2}{\sqrt{M^2-Q^2-b^2}}\left(2M^2-Q^2+2M\sqrt{M^2-Q^2-b^2} \right),\\
			\mathcal{B}^{\prime\prime} &= 3 b^5r_H + 2b^3 r_H^3+ 3 b r_H^5 +3(b^2-r_H^2)(b^2+r_H^2)^2\thinspace {\tan}^{-1}\left(\frac{b}{r_H}\right).
		\end{split}
	\end{align}
\end{subequations}  
Thus, we arrive at a modified working formula of $\mathcal{C}_{\text{local}}$ for the non-extremal black holes,\footnote{At any point, one can also solve \cref{Euler,next3} for the integrals of $R_{\mu\nu\rho\sigma}R^{\mu\nu\rho\sigma}$ and $R_{\mu\nu}R^{\mu\nu}$ over the Kerr-Newman geometry and proceed with the primary formula \eqref{nonextclocal}.}
\begin{align}\label{next5}
	\mathcal{C}^{\thinspace\text{non-extremal}}_{\text{local}} &= \frac{1}{16\pi^2}\bigg( c\int_{\text{full geometry}} \mathrm{d}^4x \sqrt{\text{det}\thinspace \bar{g}}\thinspace W_{\mu\nu\rho\sigma}W^{\mu\nu\rho\sigma} \nonumber\\
	&\qquad\qquad\quad -a\int_{\text{full geometry}} \mathrm{d}^4x \sqrt{\text{det}\thinspace \bar{g}}\thinspace E_4 \bigg).
\end{align}  
The strategy of finding $\mathcal{C}_{\text{local}}$ for the non-extremal Kerr-Newman family of black holes involves the following algorithm --
\begin{enumerate}
	\item Calculate $a_4(x)$ following the method of \cref{sdc}, and express them only into $R_{\mu\nu\rho\sigma}R^{\mu\nu\rho\sigma}$ and $R_{\mu\nu}R^{\mu\nu}$ invariants of the Einstein-Maxwell backgrounds.
	
	\item Compare the obtained $a_4(x)$ result with the standard form \eqref{next1} and extract out the coefficients of trace anomalies or the central charges $(c, a)$.
	
	\item Evaluate $\mathcal{C}_{\text{local}}$ for a non-extremal Kerr-Newman black hole entropy by employing the $(c,a)$ data into the formula \eqref{next5}, along with the limits \eqref{Euler} and \eqref{next3}.
	
	\item For non-extremal Reissner-Nordstr\"om ($Q\neq 0,J=0$) black holes, set $b \to 0$ in the Kerr-Newman  $\mathcal{C}_{\text{local}}$ result, while put $Q=0$ for both non-extremal Kerr ($Q=0, J\neq 0$) and Schwarzschild ($Q= 0,J=0$) black holes.
\end{enumerate}

We, therefore, have all the necessary ingredients for evaluating logarithmic entropy corrections for all the Kerr-Newman family of black holes. The \say{\hyperref[ST1]{Strategy A}} and \say{\hyperref[ST2]{Strategy B}} provide the local corrections ($\mathcal{C}_{\text{local}}$) for the extremal and non-extremal cases, respectively.\footnote{In principle, one can use \say{\hyperref[ST2]{Strategy B}} for the extremal Kerr-Newman black holes. But it will be challenging to achieve the finite integration values of $W_{\mu\nu\rho\sigma}W^{\mu\nu\rho\sigma}$ and $E_4$ (or $R_{\mu\nu\rho\sigma}R^{\mu\nu\rho\sigma}$ and $R_{\mu\nu}R^{\mu\nu}$) in the extremal limit $M = \sqrt{Q^2+b^2}$. In contrast, the quantum entropy function formalism \cite{Sen:2008wa,Sen:2009wb,Sen:2009wc} used in \say{\hyperref[ST1]{Strategy A}} is a popular trick that easily determines the logarithmic entropy corrections for extremal black holes using finite near-horizon limits of the relevant background-geometric invariants.} The zero-mode corrections ($\mathcal{C}_{\text{zm}}$) can be extracted using the inputs of \cref{zeromode} in the general formula \eqref{B9}. Finally, the central formula \eqref{B7} evaluates the necessary logarithmic correction results for the Kerr-Newman family of black holes. It is important to highlight that the whole framework \ref{nonzeromode} of calculating logarithmic corrections does not rely on supersymmetry and hence entirely appropriate for all extremal and non-extremal black holes in supergravity embedded Einstein-Maxwell theories.

\section{Seeley-DeWitt coefficients and logarithmic entropy corrections in the ``minimally-coupled" Einstein-Maxwell theory}\label{em}

A pure or simple EMT casts a vector field $A_\mu$ coupled minimally to metric $g_{\mu\nu}$ in four dimensions via the action ($G_N=1/16\pi$),
\begin{align}\label{E1}
	\mathcal{S}_{\text{EM}}= \int \mathrm{d}^4x \sqrt{\text{det}\thinspace {g}} \thinspace \mathcal{L}_{\text{EM}} , \enspace \mathcal{L}_{\text{EM}} = \left(\mathcal{R}-F_{\mu\nu}F^{\mu\nu} \right),
\end{align}
where $\mathcal{R}$ is the Ricci scalar constructed from $g^{\mu\nu}$ and $F_{\mu\nu}= \partial_\mu A_\nu-\partial_\nu A_\mu$ is the field strength tensor of $A_\mu$.  The Einstein equation for any general classical background solution ($\bar{g}_{\mu\nu},\bar{A}_\mu$) to \eqref{E1} is\footnote{The derivation of the Einstein equation \eqref{pem2} is provided in \cref{App2}.}
\begin{equation}\label{pem2}
	R_{\mu\nu}=2\bar{F}_{\mu\rho}\bar{F}\indices{_\nu^\rho}-\frac{1}{2}\bar{g}_{\mu\nu}\bar{F}_{\rho\sigma}\bar{F}^{\rho\sigma},\thinspace R=0,
\end{equation}
where $\bar{F}_{\mu\nu} = \partial_\mu \bar{A}_\nu-\partial_\nu \bar{A}_\mu$ is the background field strength, $R_{\mu\nu}$ and $R$ are background Ricci parameters induced from $\bar{g}_{\mu\nu}$. We now turn to a generalization of the simple EMT: a massless scalar field $\phi$, an additional massless vector field $a^{\prime}_\mu$, a massless spin-1/2 Dirac field $\lambda$ and a massless spin-3/2 Rarita-Schwinger field $\psi_\mu$ (Majorana form) are minimally coupled to the pure Einstein-Maxwell system \eqref{E1}. The generalized theory is structured such that all the additionally-coupled fields must fluctuate around the background of the pure Einstein-Maxwell system for the requirement of sharing the common Kerr-Newman family of solutions. The action describing the resultant $d=4$ \say{minimally-coupled} Einstein-Maxwell theory, denoted as $\mathcal{S}_{\text{EM(mc)}}$, can be structured by coupling the free actions of the massless fields minimally to the pure Einstein-Maxwell action \eqref{E1},
\begin{align}\label{E5}
	\begin{gathered}
		\mathcal{S}_{\text{EM(mc)}}= \int \mathrm{d}^4x \sqrt{\text{det}\thinspace {g}}\thinspace \mathcal{L}_{\text{EM(mc)}},\\
		\mathcal{L}_{\text{EM(mc)}} =  \mathcal{L}_{\text{EM}}+ D_\rho \phi D^\rho \phi-\frac{1}{4}f^{\prime}_{\mu\nu}{f^\prime}^{\mu\nu}+i\bar{\lambda}\gamma^\rho D_\rho\lambda-\bar{\psi}_{\mu }\gamma^{\mu\rho\nu}D_\rho \psi_{\nu },
	\end{gathered}
\end{align}
where ${f}^{\prime}_{\mu\nu} = \partial_\mu a^{\prime}_\nu-\partial_\nu a^{\prime}_\mu$, $\bar{\lambda} = \lambda^\dagger$, $\bar{\psi}_{\mu}=\psi_{\mu}^\dagger$, and $\gamma^{\mu\rho\nu}$  is an antisymmetrized product\footnote{In our convention, the antisymmetrized products of gamma matrices are defined as $\gamma^{\alpha_1 \alpha_2\ldots \alpha_n}=\frac{1}{n!}\sum_{\mathcal{P}}(-1)^\mathcal{P}\gamma^{\alpha_1}\gamma^{\alpha_2}\cdots\gamma^{\alpha_n}$, where $\mathcal{P}$ stands for the type (i.e., even or odd) of permutations.} of Euclidean gamma matrices $\gamma^\mu$ which follow the 4D Clifford algebra (with the identity matrix $\mathbb{I}_4$),
\begin{align}\label{E6}
	\gamma^\mu\gamma^\nu+\gamma^\nu\gamma^\mu = 2g^{\mu\nu}\mathbb{I}_4.
\end{align}
For investigating quadratic fluctuation of the content in the \say{minimally-coupled} EMT \eqref{E5}, we consider the following fluctuations --
\begin{enumerate}[(a)]
	\item the metric $g_{\mu\nu}$ and vector field $A_\mu$  fluctuate around the classical background ($\bar{g}_{\mu\nu},\bar{A}_\mu$) of the pure Einstein-Maxwell system \eqref{E1} for small fluctuation $(\tilde{g}_{\mu\nu},\tilde{A}_\mu)$,
	\begin{align}\label{pem3}
		\tilde{g}_{\mu\nu} = \sqrt{2}h_{\mu\nu},\enspace  \tilde{A}_\mu = \frac{1}{2} a_\mu,
	\end{align}
	which yields (up to quadratic order),
	\begin{align}\label{pem4}
		\begin{split}
			\sqrt{\text{det}\thinspace {g}} &=\sqrt{\text{det}\thinspace \bar{g}}\left(1+ \frac{1}{\sqrt{2}}h\indices{^\mu_\mu}-\frac{1}{2}h\indices{^\mu_\nu}h\indices{_\nu^\mu}+\frac{1}{4}(h\indices{^\mu_\mu})^2 \right),\\
			g^{\mu\nu} &= \bar{g}^{\mu\nu}-\sqrt{2}h^{\mu\nu}+2h^{\mu\rho}h\indices{_\rho^\nu},\\
			F_{\mu\nu} &= \bar{F}_{\mu\nu}+\frac{1}{2}f_{\mu\nu},
		\end{split}
	\end{align}
	where ${f}_{\mu\nu} = \partial_\mu a_\nu-\partial_\nu a_\mu$,
	
	\item all other minimally-coupled fields ($\phi,a^{\prime}_\mu,\lambda,\psi_{\mu }$) are supposed to have no background values and must fluctuate around the Einstein-Maxwell background ($\bar{g}_{\mu\nu},\bar{A}_\mu$).	
\end{enumerate} 
As a result, the \say{minimally-coupled} EMT \eqref{E5} satisfies the same Einstein equation \eqref{pem2} and the other equations of motion as the simple EMT \eqref{E1} for the common background solution ($\bar{g}_{\mu\nu},\bar{A}_\mu$). 
We then execute the quadratic fluctuation of the action \eqref{E5}. The particular kind of couplings allows us to distribute the fluctuations $\tilde{\xi}_m = \lbrace h_{\mu\nu}, a_\mu, \phi, a^\prime_\mu, \lambda, \psi_{\mu } \rbrace $ into various sectors,
\begin{align}
	\delta^2\mathcal{S}_{\text{EM(mc)}}[\tilde{\xi}_m] = \delta^2\mathcal{S}_{\text{EM}}[h_{\mu\nu},a_\mu] + \delta^2\mathcal{S}_{\text{scalar}}[\phi] + \delta^2\mathcal{S}_{\text{vector}}[a^{\prime}_\mu]+ \delta^2\mathcal{S}_{\text{Dirac}}[\lambda]+ \delta^2\mathcal{S}_{\text{RS}}[\psi_{\mu }],
\end{align}
where the quadratic-fluctuated Einstein-Maxwell sector $\delta^2\mathcal{S}_{\text{EM}}$ and the additionally-coupled field sectors $\delta^2\mathcal{S}_{\text{scalar}}, \delta^2\mathcal{S}_{\text{vector}}, \delta^2\mathcal{S}_{\text{Dirac}}, \delta^2\mathcal{S}_{\text{RS}}$ are expressed as well as analyzed in \cref{pem,scalar,vector,spinor,RS}. Furthermore, the \say{minimally-coupled} EMT does not give rise to new black holes beyond the Kerr-Newman family of solutions, and hence the minimally-coupled fields provide additional contributions to the logarithmic correction results of the pure Einstein-Maxwell system. Our purpose is to compute all these logarithmic correction contributions for the Kerr-Newman family of black holes in both extremal and non-extremal limits. For that, we need to analyze the quadratic fluctuated action components and evaluate the Seeley-DeWitt coefficients. Following the prescription of \cref{B}, we now pursue this direction further. Note that all the significant terms and data relevant to the pure Einstein-Maxwell sector and the additionally-coupled scalar field, vector field, spin-1/2 Dirac field, spin-3/2 Rarita-Schwinger field sectors are respectively labeled by \say{EM}, \say{scalar}, \say{vector}, \say{Dirac} and \say{RS}.

\subsection{Contributions of the Einstein-Maxwell sector}\label{pem}
Investigating the Einstein-Maxwell part of the action \eqref{E5} via the Seeley-DeWitt approach is not an easy task; it involves a lengthy but systematic process. The initial challenge is preparing the quadratic order fluctuated action for the fluctuations \eqref{pem3} and expressing it into the prescribed Laplace-type form \eqref{Bx}. Then one needs to encounter a mountain of tedious trace calculations. For the quadratic fluctuations $\tilde{\xi}_m = \lbrace h_{\mu\nu}, a_\mu \rbrace $, the Einstein-Maxwell sector in the action \eqref{E5} can be decoupled into two separate subparts,
\begin{subequations}
	\begin{align}
		\delta^2\mathcal{S}_{\text{EM}}[h_{\mu\nu},a_\mu]&= \delta^2\mathcal{S}_{\text{Ricci}}[h_{\mu\nu}] + \delta^2\mathcal{S}_{\text{Maxwell}}[h_{\mu\nu},a_\mu],
	\end{align}
	where $\delta^2\mathcal{S}_{\text{Ricci}}$ and $\delta^2\mathcal{S}_{\text{Maxwell}}$ respectively denote the quadratic fluctuated Ricci scalar part and Maxwell part:
	\begin{align}
		\delta^2\mathcal{S}_{\text{Ricci}}[h_{\mu\nu}] &= \int \mathrm{d}^4x\thinspace \delta^2(\sqrt{\text{det}\thinspace {g}} \thinspace \mathcal{R}) ,\label{pem5}\\
		\delta^2\mathcal{S}_{\text{Maxwell}}[h_{\mu\nu},a_\mu] &= -\int \mathrm{d}^4x\thinspace \delta^2(\sqrt{\text{det}\thinspace {g}} \thinspace F_{\mu\nu}F^{\mu\nu}).\label{pem17}
	\end{align}
\end{subequations}

\subsubsection*{The Ricci scalar part}
We begin by expressing the standard form of Christoffel symbol $\Gamma\indices{^\rho_{\mu\nu}}$ in terms of the fluctuated $g_{\mu\nu}$ and $g^{\mu\nu}$,
\begin{align}\label{pem7}
	\Gamma\indices{^\rho_{\mu\nu}} &= (\bar{g}^{\rho\sigma}-\sqrt{2}h^{\rho\sigma}+2 h^{\rho\alpha}h\indices{_\alpha^\sigma}) \Big(\bar{\Gamma}_{\sigma\mu\nu}+ \frac{\sqrt{2}}{2}(D_\mu h_{\nu\sigma}+ D_\nu h_{\mu\sigma}-D_\sigma h_{\mu\nu}+2 \bar{\Gamma}\indices{^\alpha_{\mu\nu}}h_{\alpha\sigma})\Big),
\end{align}
where the background Christoffel symbol $\bar{\Gamma}_{\sigma\mu\nu}$ and the covariant derivative $D_\rho$ operating on the metric fluctuation $h_{\mu\nu}$ are defined as
\begin{align}\label{pem8}
	\begin{split}
		\bar{\Gamma}_{\sigma\mu\nu} &= \frac{1}{2} (\partial_\mu \bar{g}_{\nu\sigma}+ \partial_\nu \bar{g}_{\mu\sigma} - \partial_\sigma \bar{g}_{\mu\nu}),\\
		D_\sigma h_{\mu\nu} &= \partial_\sigma h_{\mu\nu}- \bar{\Gamma}\indices{^\alpha_{\mu\sigma}} h_{\alpha\nu}- \bar{\Gamma}\indices{^\alpha_{\nu\sigma}} h_{\mu\alpha}.
	\end{split}
\end{align}
One can now execute the product \eqref{pem7} by eliminating all the terms higher than second order of the fluctuation $h_{\mu\nu}$ and express,
\begin{align}\label{pem9}
	\begin{split}
		\Gamma^{\rho}_{\;\mu\nu} &= \bar{\Gamma}\indices{^\rho_{\mu\nu}} + \delta \Gamma\indices{^\rho_{\mu\nu}},\\
		\delta \Gamma\indices{^\rho_{\mu\nu}} &= \frac{\sqrt{2}}{2}(D_\mu h\indices{_\nu^\rho}+ D_\nu h\indices{_\mu^\rho} - D^\rho h_{\mu\nu})-h^{\rho\sigma}(D_\mu h_{\nu\sigma} + D_\nu h_{\mu\sigma} - D_\sigma h_{\mu\nu}),
	\end{split}
\end{align}
which adjusts the expression of Ricci tensor $\mathcal{R}_{\mu\nu} = g^{\rho\sigma}\mathcal{R}\indices{_{\sigma\mu\rho\nu}}$ up to quadratic order by avoiding all the total derivative terms as
\begin{align}\label{pem10}
	\mathcal{R}_{\mu\nu} &= R_{\mu\nu}+ D_\rho (\delta \Gamma\indices{^\rho_{\mu\nu}})- D_\nu (\delta \Gamma\indices{^\rho_{\mu\rho}})+ \delta \Gamma\indices{^\alpha_{\mu\nu}}\delta \Gamma\indices{^\rho_{\alpha\rho}}- \delta \Gamma\indices{^\alpha_{\mu\rho}}\delta \Gamma\indices{^\rho_{\alpha\nu}},\nonumber\\
	&= R_{\mu\nu}+ \frac{\sqrt{2}}{2}(D_\rho D_\mu h\indices{_\nu^\rho}+ D_\rho D_\nu h\indices{_\mu^\rho} - D_\rho D^\rho h_{\mu\nu}-D_\nu D_\mu h\indices{^\rho_\rho})\nonumber\\
	&\quad + \frac{1}{2}\big(h_{\rho\sigma} D_\mu D_\nu h^{\rho\sigma} + 2 h\indices{_\mu^\sigma}D_\rho D_\sigma h\indices{_\nu^\rho}-2 h\indices{_\mu^\sigma}D_\rho D^\rho h_{\sigma\nu}\nonumber\\
	&\quad -h\indices{_\nu^\rho} D_\mu D_\rho h\indices{^\sigma_\sigma}- h\indices{_\mu^\rho}D_\nu D_\rho h\indices{^\sigma_\sigma}+ h_{\mu\nu} D_\rho D^\rho h\indices{^\sigma_\sigma}\big).
\end{align}
Here $R_{\mu\nu}$ is the background Ricci tensor induced from $\bar{\Gamma}\indices{^\rho_{\mu\nu}}$. 
Further, contracting the $\mathcal{R}_{\mu\nu}$ expression \eqref{pem10} by the fluctuated $g^{\mu\nu}$ form \eqref{pem4}, one can achieve a simplified quadratic fluctuated Ricci scalar $\mathcal{R}$ form as
\begin{align}\label{pem12}
	\mathcal{R} 
	&= R + \sqrt{2}\big(D_\mu D_\nu h^{\mu\nu}-D_\rho D^\rho h^{\mu}_{\;\mu}-{R}_{\mu\nu}h^{\mu\nu}\big)\nonumber\\
	&\quad + \frac{1}{2}\left(h_{\mu\nu}D_\rho D^\rho h^{\mu\nu}+ h\indices{^\mu_\mu}D_\rho D^\rho h\indices{^\nu_\nu}-2 h^{\nu\rho}D_\mu D_\nu h\indices{^\mu_\rho}+4 {R}_{\mu\nu}h^{\mu\rho}h\indices{^\nu_\rho}\right),
\end{align} 
where $R=\bar{g}^{\mu\nu}R_{\mu\nu}$ is the background Ricci scalar. We, therefore, obtain,
{
	\allowdisplaybreaks
	\begin{align}\label{pem13}
		\delta^2(\sqrt{\text{det}\thinspace {g}} \thinspace \mathcal{R}) &= \sqrt{\text{det}\thinspace \bar{g}}\left(1+ \frac{1}{\sqrt{2}}h\indices{^\alpha_\alpha}-\frac{1}{2}h\indices{^\alpha_\beta}h\indices{^\beta_\alpha}+\frac{1}{4}(h\indices{^\alpha_\alpha})^2 \right)\mathcal{R},\nonumber\\
		& =\frac{1}{2}\sqrt{\text{det}\thinspace \bar{g}} \bigg(h_{\mu\nu}D_\rho D^\rho h^{\mu\nu}- h\indices{^\mu_\mu}D_\rho D^\rho h\indices{^\nu_\nu}-2 h^{\nu\rho}D_\mu D_\nu h\indices{^\mu_\rho}+2h^{\mu\nu}D_\mu D_\nu h\indices{^\alpha_\alpha}\nonumber\\
		&\qquad  +2 {R}_{\mu\nu}\big(2h^{\mu\rho}h\indices{^\nu_\rho}-h\indices{^\alpha_\alpha} h_{\mu\nu}\big)  -R \Big(h_{\mu\nu}h^{\mu\nu}-\frac{1}{2}(h\indices{^\alpha_\alpha})^2\Big) \bigg),
\end{align}}
where only second-order product terms of the metric fluctuation $h_{\mu\nu}$ are considered. At this stage, we can choose a harmonic gauge $D_\mu h^{\mu\rho}-\frac{1}{2}D^\rho h\indices{^\alpha_\alpha}=0$ in the form of the gauge-fixing term,
\begin{align}\label{pem14}
	-\sqrt{\text{det}\thinspace \bar{g}}\Big(D_\mu h^{\mu\rho}-\frac{1}{2}D^\rho h\indices{^\alpha_\alpha}\Big)\Big(D^\nu h_{\nu\rho}-\frac{1}{2}D_\rho h\indices{^\beta_\beta}\Big),
\end{align} 
which provides the following simplified, gauge-fixed and quadratic fluctuated form of the Ricci scalar part \eqref{pem5}
\begin{align}\label{pem15}
	{\delta^2\mathcal{S}}_{\text{Ricci}}[h_{\mu\nu}]&= \frac{1}{2}\int \mathrm{d}^4x \sqrt{\text{det}\thinspace \bar{g}}\Big(h_{\mu\nu}D_\rho D^\rho h^{\mu\nu}- \frac{1}{2}h\indices{^\mu_\mu}D_\rho D^\rho h\indices{^\nu_\nu} \nonumber\\
	&\qquad+2R_{\mu\rho\nu\sigma}h^{\mu\nu}h^{\rho\sigma} +2R_{\mu\nu}h^{\mu\rho}h\indices{^\nu_\rho}-2R_{\mu\nu}h^{\mu\nu}h\indices{^\rho_\rho}\Big).
\end{align}
Derivation of the above quadratic fluctuated form involves the elimination of all total derivative terms, the use of the commutation relation \eqref{ct}, and the condition $R=0$ for the \say{minimally-coupled} EMT background. 
\subsubsection*{The Maxwell part}
With the help of fluctuations \eqref{pem4} and considering up to the second-order fluctuated terms for the fluctuations $\tilde{\xi}_m = \lbrace h_{\mu\nu}, a_\mu\rbrace $, we obtain 
\begin{align}\label{pem20}
	\delta^2(\sqrt{\text{det}\thinspace {g}} \thinspace F_{\mu\nu}F^{\mu\nu}) &= \sqrt{\text{det}\thinspace \bar{g}}\bigg(1+ \frac{1}{\sqrt{2}}h\indices{^\rho_\rho}-\frac{1}{2}h\indices{^\rho_\sigma}h\indices{^\sigma_\rho}+\frac{1}{4}(h\indices{^\rho_\rho})^2 \bigg)g^{\mu\alpha}g^{\nu\beta} F_{\mu\nu}F_{\alpha\beta},\nonumber\\
	&= \frac{1}{2}\sqrt{\text{det}\thinspace \bar{g}}\bigg(\frac{1}{2}f_{\mu\nu}f^{\mu\nu} + 4 \bar{F}_{\mu\nu}\bar{F}_{\alpha\beta}h^{\mu\alpha}h^{\nu\beta} +8 \bar{F}_{\mu\nu}\bar{F}^{\mu\alpha}h^{\nu\beta}h_{\alpha\beta}\nonumber\\
	&\quad -4\sqrt{2}\bar{F}_{\mu\nu}h^{\mu\alpha}f\indices{_\alpha^\nu}+\sqrt{2}\bar{F}_{\mu\nu}h\indices{^\rho_\rho} f^{\mu\nu}- 4\bar{F}_{\mu\nu}\bar{F}\indices{_\alpha^\nu}h\indices{^\rho_\rho}h^{\mu\alpha}\nonumber\\
	&\quad -\bar{F}_{\mu\nu}\bar{F}^{\mu\nu}\Big(h_{\alpha\beta}h^{\alpha\beta}-\frac{1}{2}(h\indices{^\rho_\rho})^2\Big)\bigg),
\end{align}
where only second-order product terms of the metric and gauge field fluctuations are considered. Furthermore, up to a total derivative and with the help of $a_\mu$ commutation relation \eqref{cv}, one can also express,
\begin{align}\label{pem21}
	-\frac{1}{2}f_{\mu\nu}f^{\mu\nu} = \bar{g}^{\mu\nu}a_\mu D_\rho D^\rho a_\nu - a_\mu R^{\mu\nu} a_\nu+ (D_\mu a^\mu)^2. 
\end{align}
After substituting the forms \eqref{pem20} and \eqref{pem21}, we gauge fix the action \eqref{pem17} by choosing a Lorenz gauge $D_\mu a^\mu=0$ in the form of the following gauge-fixing term\footnote{The ghost contributions compensating the effects of the gauge-fixing terms \eqref{pem14} and \eqref{pem22} are provided later in the form of action \eqref{pemghost}.}
\begin{align}\label{pem22}
	-\frac{1}{2}\sqrt{\text{det}\thinspace \bar{g}}(D_\mu a^\mu)^2,
\end{align}
and obtain the simplified, gauge-fixed and quadratic fluctuated form of the Maxwell part \eqref{pem17} as
\begin{align}\label{pem23}
	{\delta^2\mathcal{S}}_{\text{Maxwell}}[h_{\mu\nu},a_\mu]&= \frac{1}{2}\int \mathrm{d}^4x \sqrt{\text{det}\thinspace \bar{g}}\bigg(a_\mu D_\rho D^\rho a^\mu-a_\mu R^{\mu\nu} a_\nu +4\sqrt{2}\bar{F}_{\mu\nu}h^{\mu\alpha}f\indices{_\alpha^\nu}\nonumber\\
	&\quad -\sqrt{2}\bar{F}_{\mu\nu}h\indices{^\rho_\rho} f^{\mu\nu}- 4 \bar{F}_{\mu\nu}\bar{F}_{\alpha\beta}h^{\mu\alpha}h^{\nu\beta} -8 \bar{F}_{\mu\nu}\bar{F}^{\mu\alpha}h^{\nu\beta}h_{\alpha\beta}\nonumber\\
	&\quad + 4\bar{F}_{\mu\nu}\bar{F}\indices{_\alpha^\nu}h\indices{^\rho_\rho}h^{\mu\alpha}+ \bar{F}_{\mu\nu}\bar{F}^{\mu\nu}\Big(h_{\alpha\beta}h^{\alpha\beta}-\frac{1}{2}(h\indices{^\rho_\rho})^2\Big) \bigg).
\end{align}

We now need to recombine ${\delta^2\mathcal{S}}_{\text{Ricci}}$ \eqref{pem15} and ${\delta^2\mathcal{S}}_{\text{Maxwell}}$ \eqref{pem23} parts in order to present the Einstein-Maxwell sector's contribution in the quadratic fluctuation of the action \eqref{E5},
\begin{align}\label{pem24}
	\delta^2\mathcal{S}_{\text{EM}}[h_{\mu\nu},a_\mu] &= \frac{1}{2}\int \mathrm{d}^4x \sqrt{\text{det}\thinspace \bar{g}}\thinspace \bigg(h_{\mu\nu}D_\rho D^\rho h^{\mu\nu}- \frac{1}{2}h\indices{^\mu_\mu}D_\rho D^\rho h\indices{^\nu_\nu}+a_\alpha D_\rho D^\rho a^\alpha \nonumber\\
	&\quad  -a_\alpha R^{\alpha\beta} a_\beta+ h_{\mu\nu}\Big(2R^{\mu\alpha\nu\beta}+\bar{g}^{\mu\alpha} R^{\nu\beta}-3\bar{g}^{\nu\beta} R^{\mu\alpha}-4\bar{F}^{\mu\alpha} \bar{F}^{\nu\beta}\nonumber\\
	&\quad + \frac{1}{2}\bar{F}_{\rho\sigma}\bar{F}^{\rho\sigma}(\bar{g}^{\mu\nu}\bar{g}^{\alpha\beta}-2\bar{g}^{\mu\alpha} \bar{g}^{\nu\beta})\Big)h_{\alpha\beta}\nonumber\\
	&\quad +2\sqrt{2}h_{\mu\nu}\big(2\bar{g}^{\nu\rho} \bar{F}^{\mu\alpha}- 2\bar{g}^{\alpha\nu} \bar{F}^{\mu\rho}-\bar{g}^{\mu\nu}\bar{F}^{\rho\alpha}\big) (D_\rho a_\alpha) \bigg).
\end{align}
We further perform multiple customizations over the above form to extract the necessary Laplace-type operator $\Lambda$. This includes bypassing the kinetic term $h\indices{^\mu_\mu}D_\rho D^\rho h\indices{^\nu_\nu}$ via casting an effective metric, neglecting all total derivative terms, and considering all symmetric properties of the fluctuations and their all possible pairs. All these lead to the following Laplace-type structure for the Einstein-Maxwell fluctuations $\tilde{\xi}_m = \lbrace h_{\mu\nu},a_\mu\rbrace $,
\begin{align}\label{pem25}
	\begin{split}
		\delta^2\mathcal{S}_{\text{EM}}[\tilde{\xi}_m] &= \frac{1}{2}\int \mathrm{d}^4x \sqrt{\text{det}\thinspace \bar{g}}\thinspace \tilde{\xi}_m\Lambda^{\tilde{\xi}_m \tilde{\xi}_n}\tilde{\xi}_n,\\
		\tilde{\xi}_m\Lambda^{\tilde{\xi}_m \tilde{\xi}_n}\tilde{\xi}_n &= I^{h_{\mu\nu} h_{\alpha\beta}} h_{\mu\nu} D_\rho D^\rho h_{\alpha\beta}+ I^{a_\alpha a_\beta} a_\alpha D_\rho D^\rho a_\beta + h_{\mu\nu} P^{h_{\mu\nu} h_{\alpha\beta}} h_{\alpha\beta}\\
		&\quad + a_\alpha P^{a_\alpha a_\beta} a_\beta+ h_{\mu\nu}P^{h_{\mu\nu}a_\alpha} a_\alpha + a_\alpha P^{a_\alpha h_{\mu\nu}} h_{\mu\nu}\\
		& \quad + h_{\mu\nu}(2\omega^\rho)^{h_{\mu\nu}a_\alpha} (D_\rho a_\alpha) + a_\alpha (2\omega^\rho)^{a_\alpha h_{\mu\nu}} (D_\rho h_{\mu\nu}),
	\end{split}
\end{align}
where $I$, $P$ and $\omega_\rho (=\frac{1}{2}N_\rho)$ hold the forms, 
{
	\allowdisplaybreaks
	\begin{subequations}
		\begin{align}
			I^{h_{\mu\nu} h_{\alpha\beta}} &= \frac{1}{2}(\bar{g}^{\mu\alpha} \bar{g}^{\nu\beta} + \bar{g}^{\mu\beta} \bar{g}^{\nu\alpha}-\bar{g}^{\mu\nu}\bar{g}^{\alpha\beta}), \\
			I^{a_\alpha a_\beta} &= \bar{g}^{\alpha\beta}, \\
			P^{h_{\mu\nu} h_{\alpha\beta}} &=R^{\mu\alpha\nu\beta}+R^{\mu\beta\nu\alpha}-2(\bar{F}^{\mu\alpha}\bar{F}^{\nu\beta}+\bar{F}^{\mu\beta}\bar{F}^{\nu\alpha}) \nonumber\\
			&\quad -\frac{1}{2}\bar{F}_{\rho\sigma}\bar{F}^{\rho\sigma}(\bar{g}^{\mu\alpha} \bar{g}^{\nu\beta} + \bar{g}^{\mu\beta} \bar{g}^{\nu\alpha}-\bar{g}^{\mu\nu}\bar{g}^{\alpha\beta}) \nonumber\\
			&\quad -\frac{1}{2}(\bar{g}^{\mu\alpha}R^{\nu\beta}+ \bar{g}^{\mu\beta}R^{\nu\alpha}+\bar{g}^{\nu\alpha}R^{\mu\beta}+\bar{g}^{\nu\beta}R^{\mu\alpha}),\\
			P^{a_\alpha a_\beta} &=-R^{\alpha\beta},\\
			P^{h_{\mu\nu}a_\alpha} &= P^{a_\alpha h_{\mu\nu}}=\frac{1}{\sqrt{2}}\big( (D^\mu \bar{F}^{\alpha\nu})+ (D^\nu \bar{F}^{\alpha\mu})\big),\\
			(\omega^\rho)^{h_{\mu\nu}a_\alpha} &=-(\omega^\rho)^{a_\alpha h_{\mu\nu}}= \frac{1}{\sqrt{2}}\big(\bar{g}^{\mu\alpha}\bar{F}^{\rho\nu}+\bar{g}^{\nu\alpha}\bar{F}^{\rho\mu}-\bar{g}^{\mu\rho}\bar{F}^{\alpha\nu}\nonumber\\
			&\qquad\qquad\qquad\qquad\quad-\bar{g}^{\nu\rho}\bar{F}^{\alpha\mu}-\bar{g}^{\mu\nu}\bar{F}^{\rho\alpha}\big).\label{pem26}
		\end{align}
	\end{subequations}
}
According to the formulae \eqref{B11}, the above data provide us the following results for $E$ and $\Omega_{\rho\sigma}$
{
	\allowdisplaybreaks
	\begin{subequations}
		\begin{align}
			\tilde{\xi}_m E^{\tilde{\xi}_m \tilde{\xi}_n}\tilde{\xi}_n &=h_{\mu\nu} (R^{\mu\alpha\nu\beta}+ R^{\mu\beta\nu\alpha}-\bar{g}^{\mu\nu}R^{\alpha\beta}-\bar{g}^{\alpha\beta}R^{\mu\nu})h_{\alpha\beta}+\frac{3}{2}\bar{g}^{\alpha\beta} a_\alpha \bar{F}_{\mu\nu}\bar{F}^{\mu\nu}a_\beta \nonumber\\
			&\quad +\frac{1}{\sqrt{2}} h_{\mu\nu}(D^\mu \bar{F}^{\alpha\nu}+ D^\nu \bar{F}^{\alpha\mu})a_\alpha +\frac{1}{\sqrt{2}} a_\alpha(D^\mu \bar{F}^{\alpha\nu}+ D^\nu \bar{F}^{\alpha\mu})h_{\mu\nu},\label{pem27}\\
			\tilde{\xi}_m (\Omega_{\rho\sigma})^{\tilde{\xi}_m \tilde{\xi}_n}\tilde{\xi}_n &= h_{\mu\nu}\Big(\frac{1}{2}\big(\bar{g}^{\mu\alpha} R\indices{^{\nu\beta}_{\rho\sigma}} +\bar{g}^{\mu\beta}R\indices{^{\nu\alpha}_{\rho\sigma}}+\bar{g}^{\nu\alpha}R\indices{^{\mu\beta}_{\rho\sigma}}+\bar{g}^{\nu\beta}R\indices{^{\mu\alpha}_{\rho\sigma}}\big)\nonumber\\
			&\quad +(\omega_\rho)^{h_{\mu\nu} a_{\theta}}(\omega_\sigma)\indices{_{a_{\theta}}^{h_{\alpha\beta}}}-(\omega_\sigma)^{h_{\mu\nu} a_{\theta}}(\omega_\rho)\indices{_{a_{\theta}}^{h_{\alpha\beta}}}\Big) h_{\alpha\beta}\nonumber\\
			&\quad + a_\alpha\Big( R\indices{^{\alpha\beta}_{\rho\sigma}}+(\omega_\rho)^{a_{\alpha} h_{\mu\nu}}(\omega_\sigma)\indices{_{h_{\mu\nu}}^{a_{\beta}}}-(\omega_\sigma)^{a_{\alpha} h_{\mu\nu}}(\omega_\rho)\indices{_{h_{\mu\nu}}^{a_{\beta}}}\Big) a_\beta\nonumber\\
			&\quad +h_{\mu\nu}\Big((D_\rho\omega_\sigma)\indices{^{h_{\mu\nu}a_\alpha}}-(D_\sigma\omega_\rho)\indices{^{h_{\mu\nu}a_\alpha}}\Big) a_\alpha \nonumber\\
			&\quad +a_\alpha\Big((D_\rho\omega_\sigma)\indices{^{a_{\alpha}}^{h_{\mu\nu}}}-(D_\sigma\omega_\rho)\indices{^{a_{\alpha}}^{h_{\mu\nu}}}\Big)h_{\mu\nu},\label{pem28}
		\end{align}
\end{subequations}}
where the appropriate forms of $(\omega^\rho)^{h_{\mu\nu}a_\alpha}$ and $(\omega^\rho)^{a_\alpha h_{\mu\nu}}$ in the result \eqref{pem28} can be arranged from \cref{pem26}. From here, our next challenge is to calculate the crucial trace values $\text{tr}(I)$, $\text{tr}(E)$, $\text{tr}(E^2)$ and $\text{tr}(\Omega_{\rho\sigma}\Omega^{\rho\sigma})$ as urged by the formulae \eqref{B12} for finding  the Seeley-DeWitt coefficients. \Cref{App3} contains a detailed outline of these lengthy trace calculations. In terms of background invariants, the trace results are recorded as
\begin{align}\label{pem29}
	\begin{split}
		\text{tr}(I) &=14,\\
		\text{tr}(E) &= 6\bar{F}_{\mu\nu}\bar{F}^{\mu\nu},\\
		\text{tr}(E^2) &= 3R_{\mu\nu\rho\sigma}R^{\mu\nu\rho\sigma}-7R_{\mu\nu}R^{\mu\nu}+9(\bar{F}_{\mu\nu}\bar{F}^{\mu\nu})^2+3R_{\mu\nu\rho\sigma}\bar{F}^{\mu\nu}\bar{F}^{\rho\sigma},\\
		\text{tr}(\Omega_{\rho\sigma}\Omega^{\rho\sigma})  &= -7R_{\mu\nu\rho\sigma}R^{\mu\nu\rho\sigma}+56R_{\mu\nu}R^{\mu\nu}-54(\bar{F}_{\mu\nu}\bar{F}^{\mu\nu})^2-18R_{\mu\nu\rho\sigma}\bar{F}^{\mu\nu}\bar{F}^{\rho\sigma},
	\end{split}
\end{align}
providing the following Seeley-DeWitt results (without the ghost contribution)
\begin{align}\label{pem30}
	\begin{split}
		{a_0}^{\text{EM,no-ghost}}(x) &= \frac{7}{8\pi^2},\\
		{a_2}^{\text{EM,no-ghost}}(x) &= \frac{3}{8\pi^2}\bar{F}_{\mu\nu}\bar{F}^{\mu\nu},\\
		a_4^{\text{EM,no-ghost}}(x) &= \frac{1}{16\pi^2 \times 180}(179 R_{\mu\nu\rho\sigma}R^{\mu\nu\rho\sigma}+ 196 R_{\mu\nu}R^{\mu\nu}).
	\end{split}
\end{align}
In addition, we must need to include appropriate ghost fields for countering the effect of gauge-fixing terms \eqref{pem14} and \eqref{pem22}. All these ghost fields can be described via a combined action \cite{Banerjee:2011oo},
\begin{align}\label{pemghost}
	\begin{split}
		\delta^2\mathcal{S}_{\text{EM,ghost}} &=\frac{1}{2}\int \mathrm{d}^4x \sqrt{\text{det}\thinspace \bar{g}}\thinspace\Big( \bar{g}^{\mu\nu} b_\mu D_\rho D^\rho c_\nu+ \bar{g}^{\mu\nu} c_\mu D_\rho D^\rho b_\nu +b D_\rho D^\rho c\\
		&\quad +c D_\rho D^\rho b +  b_\mu R^{\mu\nu} c_\nu + c_\mu  R^{\mu\nu} b_\nu-2 b \bar{F}^{\rho\nu} (D_\rho c_\nu) -2 c_\mu \bar{F}^{\mu\rho}(D_\rho b)\Big),
	\end{split}
\end{align}
where $b_\mu,c_\mu$ are vector ghosts that arise due to diffeomorphism invariance of metric fluctuation $h_{\mu\nu}$, and $b,c$ are scalar ghosts induced due to U(1) gauge invariances of gauge fluctuation $a_\mu$. The quadratic fluctuated form \eqref{pemghost} is of Laplace-type and hence, one can read off $I$, $P$ and $\omega_\rho (=\frac{1}{2}N_\rho)$ for the ghost fluctuations $\tilde{\xi}_m = \lbrace b_{\mu},c_\mu, b, c\rbrace $ as
\begin{align}\label{pem31}
	\begin{split}
		\tilde{\xi}_m I^{\tilde{\xi}_m \tilde{\xi}_n}\tilde{\xi}_n &= b_\mu \bar{g}^{\mu\nu}c_\nu+ c_\mu \bar{g}^{\mu\nu}b_\nu+bc+cb,\\
		\tilde{\xi}_m P^{\tilde{\xi}_m \tilde{\xi}_n}\tilde{\xi}_n &=b_\mu R^{\mu\nu}c_\nu + c_\mu R^{\mu\nu}b_\nu,\\
		\tilde{\xi}_m (\omega_\rho)^{\tilde{\xi}_m \tilde{\xi}_n}\tilde{\xi}_n &= -b \bar{F}\indices{_\rho^\nu} c_\nu - c_\mu \bar{F}\indices{^\mu_\rho}b,
	\end{split}
\end{align}
which further yields the following results for $E$ and $\Omega_{\rho\sigma}$ 
\begin{align}\label{pem32}
	\begin{split}
		\tilde{\xi}_m E^{\tilde{\xi}_m \tilde{\xi}_n}\tilde{\xi}_n &=b_\mu R^{\mu\nu}c_\nu + c_\mu R^{\mu\nu}b_\nu,\\
		\tilde{\xi}_m (\Omega_{\rho\sigma})^{\tilde{\xi}_m \tilde{\xi}_n}\tilde{\xi}_n &= b_\mu R\indices{^{\mu\nu}_{\rho\sigma}} c_\nu+ c_\mu R\indices{^{\mu\nu}_{\rho\sigma}} b_\nu + b (D^\nu \bar{F}_{\rho\sigma}) c_\nu-c_\mu (D^\mu \bar{F}_{\rho\sigma})b.
	\end{split}
\end{align}
Then the needful trace results are calculated as
\begin{align}\label{pem33}
	\begin{split}
		\text{tr}(I) &= 10,\\
		\text{tr}(E) &=0,\\
		\text{tr}(E^2) &= 2 R_{\mu\nu}R^{\mu\nu},\\
		\text{tr}(\Omega_{\rho\sigma}\Omega^{\rho\sigma}) &= -2 R_{\mu\nu\rho\sigma}R^{\mu\nu\rho\sigma},
	\end{split}
\end{align}
which serves the following Seeley-DeWitt results for the ghost action \eqref{pemghost}
\begin{align}\label{pem34}
	\begin{split}
		{a_0}^{\text{EM,ghost}}(x) &= -\frac{5}{8\pi^2},\\
		{a_2}^{\text{EM,ghost}}(x) &= 0,\\
		a_4^{\text{EM,ghost}}(x) &= \frac{1}{16\pi^2 \times 18}(2 R_{\mu\nu\rho\sigma}R^{\mu\nu\rho\sigma}-17 R_{\mu\nu}R^{\mu\nu}).
	\end{split}
\end{align}
Note that here we have set $\chi=-1$ in the Seeley-DeWitt formulae \eqref{B12} to account for the reverse spin-statistics of the ghosts. Finally, combining ${a_{2n}}^{\text{EM,no-ghost}}$ and ${a_{2n}}^{\text{EM,ghost}}$, we obtain the net result for the first three Seeley-DeWitt coefficients of the Einstein-Maxwell sector,
\begin{align}\label{E2}
	\begin{split}
		{a_0}^{\text{EM}}(x) &= \frac{1}{4\pi^2},\\
		{a_2}^{\text{EM}}(x) &= \frac{3}{8\pi^2}\bar{F}_{\mu\nu}\bar{F}^{\mu\nu},\\
		a_4^{\text{EM}}(x) &= \frac{1}{16\pi^2 \times 180}(199 R_{\mu\nu\rho\sigma}R^{\mu\nu\rho\sigma}+ 26 R_{\mu\nu}R^{\mu\nu}).
	\end{split}
\end{align}
Corresponding coefficients of trace anomalies (introduced in relation \eqref{next1}) are then extracted from the above $a_4(x)$ result as
\begin{equation}
	(c,a)^{\text{EM}} = \frac{137}{60}, \frac{53}{45}.
\end{equation}
Employing the Seeley-DeWitt and trace anomaly data into \say{\hyperref[ST1]{Strategy A}} and \say{\hyperref[ST2]{Strategy B}}, we find the local corrections to the extremal and non-extremal Kerr-Newman black hole entropy due to the Einstein-Maxwell sector. The results  are
\begin{subequations}
	\begin{align}
		\mathcal{C}^{\text{EM,extremal}}_{\text{local}} &= -\frac{1}{360}\Big( 1233\mathcal{B} + (463-3080 {b^\prime}^2-7960 {b^\prime}^4-3184 {b^\prime}^6)\mathcal{B}^\prime\Big),\\
		\mathcal{C}^{\text{EM,non-extremal}}_{\text{local}} &=  \frac{1}{90}\left(398 +\frac{411\beta Q^4 \mathcal{B}^{\prime\prime}}{32 \pi b^5 r_H^4(b^2+r_H^2)}\right).
	\end{align}
\end{subequations}
Also, the zero-mode corrections receive the only contribution from metric fluctuation in the formula \eqref{B9}. $\mathcal{C}_{\text{zm}}$
values are $(-4,-4,-6)$ in extremal limit and $(-1,-1,-3,-3)$ in non-extremal limit of Kerr-Newman, Kerr, Reissner-Nordstr\"om and Schwarzschild black holes.
\begin{table}[t]
	\renewcommand{\arraystretch}{1.5}
	\centering 
	\begin{tabular}{|l| c|p{2.5in}|} 
		\hline 
		\textbf{Black hole type} & \textbf{Limits} &\hspace{0.7in} \textbf{\bm$\Delta S_{\text{BH}}/\text{ln}\thinspace \mathcal{A}_{H}$}
		\\ [0.7ex]
		\hline\hline 
		&extremal &$-\frac{1}{720}\Big( 1233\mathcal{B} + (463-3080{b^\prime}^2-7960{b^\prime}^4-3184{b^\prime}^6)\mathcal{B}^\prime+1440\Big)$  \\
		\raisebox{1.5ex}{Kerr-Newman} &non-extremal
		& \hspace{0.4in}$\frac{1}{180}\left(308 +\frac{411\beta Q^4 \mathcal{B}^{\prime\prime}}{32 \pi b^5 r_H^4(b^2+r_H^2)}\right)$  \\[1ex]
		\hline
		&extremal &\hspace{1.15in} $\frac{19}{90}$  \\
		\raisebox{1.5ex}{Kerr} &non-extremal
		&\hspace{1.2in}$\frac{77}{45}$  \\[1ex]
		\hline
		&extremal &$\hspace{1.05in}-\frac{241}{45}$  \\
		\raisebox{1.5ex}{Reissner-Nordstr\"om} &non-extremal
		&\hspace{0.5in}$\frac{1}{90}\left(64+\frac{411\beta Q^4}{10\pi r_H^5}\right)$  \\[1ex]
		\hline
		Schwarzschild &non-extremal &\hspace{1.2in}$\frac{32}{45}$  \\[1ex]
		\hline 
	\end{tabular}
	\caption{Logarithmic correction contributions ($\Delta S_{\text{BH}}$) of the Einstein-Maxwell sector to the entropy of Kerr-Newman family of black holes in the $d=4$ ``minimally-coupled" EMT.} \label{tab:em}
\end{table}
The combined formula \eqref{B7} finally provides the logarithmic correction contributions of the Einstein-Maxwell sector to the entropy of the Kerr-Newman family of black holes that are recorded in \cref{tab:em}. The $a_4^{\text{EM}}(x)$ coefficient \eqref{E2} and the extremal logarithmic corrections exactly match with the results given in \cite{Bhattacharyya:2012ss}. 

\subsection{Contributions of the minimally-coupled scalar field }\label{scalar}
In the fluctuation of the action \eqref{E5}, the quadratic fluctuated part of the minimally-coupled massless scalar field $\phi$ is
\begin{subequations}
	\begin{align}\label{scalar1}
		\delta^2\mathcal{S}_{\text{scalar}}[\phi] = \int \mathrm{d}^4x \thinspace \delta^2(\sqrt{\text{det}\thinspace {g}} \thinspace D_\rho \phi D^\rho \phi),
	\end{align}
	where we can adjust up to a total derivative,
	\begin{align}
		\delta^2(\sqrt{\text{det}\thinspace {g}} \thinspace D_\rho \phi D^\rho \phi) = -\sqrt{\text{det}\thinspace \bar{g}}\,\phi D_\rho D^\rho \phi,
	\end{align}
\end{subequations}
and reexpress into the following form of the Laplace-type operator $\Lambda$
\begin{align}\label{scalar2}
	\begin{split}
		\delta^2\mathcal{S}_{\text{scalar}}[\phi] &= \int \mathrm{d}^4x \sqrt{\text{det}\thinspace \bar{g}} \thinspace \phi \Lambda \phi,\\
		\Lambda &=  -D_\rho D^\rho.
	\end{split}
\end{align}
Comparing it with the standard schematic \eqref{Bx}, we read off $P=N^\rho=0$ and then write down the results for the matrices $I, E, \omega_\rho$ and $\Omega_{\rho\sigma}$, 
\begin{align}
	{I}=1, \thinspace E =0, \thinspace\omega_\rho=0,\enspace  \Omega_{\rho\sigma} = 0,
\end{align}
along with their trace values, 
\begin{align}\label{scalar3}
	\text{tr}(I) = 1,\thinspace \text{tr}(E) = 0, \thinspace \text{tr}(E^2) = 0,\thinspace \text{tr}(\Omega_{\rho\sigma}\Omega^{\rho\sigma}) = 0.
\end{align}
Inserting all the trace data in the formulae \eqref{B12}, we find the first three Seeley-DeWitt coefficients for the massless scalar field fluctuation as
\begin{align}\label{scalar4}
	\begin{split}
		a_0^{\text{scalar}}(x) &= \frac{1}{16\pi^2},\\
		a_2^{\text{scalar}}(x) &= 0,\\
		a_4^{\text{scalar}}(x) &= \frac{1}{180\times 16\pi^2}(R_{\mu\nu\rho\sigma}R^{\mu\nu\rho\sigma}-R_{\mu\nu}R^{\mu\nu}),
	\end{split}
\end{align}
followed by the coefficients of trace anomalies,
\begin{equation}\label{scalar5}
	(c,a)^{\text{scalar}} = \frac{1}{120}, \frac{1}{360}.
\end{equation}
The $a_4^{\text{scalar}}(x)$ coefficient \eqref{scalar4} and trace anomaly results \eqref{scalar5} encourage us to follow \say{\hyperref[ST1]{Strategy A}} and \say{\hyperref[ST2]{Strategy B}} and calculate the scalar field contribution in the logarithmic correction to the entropy of extremal and non-extremal Kerr-Newman family of black holes. The results are provided in \cref{tab:scalar}, where there is no $\mathcal{C}_{\text{zm}}$ contribution for the particular scalar field.\footnote{Refer to the discussions of \cref{zeromode}.}
\begin{table}[t]
	\renewcommand{\arraystretch}{1.5}
	\centering 
	\begin{tabular}{|l| c| c|} 
		\hline 
		\textbf{Black hole type} & \textbf{Limits} & \textbf{\bm$\Delta S_{\text{BH}}/\text{ln}\thinspace \mathcal{A}_{H}$}
		\\ [0.7ex]
		\hline\hline 
		&extremal &$-\frac{1}{1440}\Big( 9\mathcal{B} - (1+40 {b^\prime}^2+80 {b^\prime}^4+32 {b^\prime}^6)\mathcal{B}^\prime\Big)$  \\
		\raisebox{1.5ex}{Kerr-Newman} &non-extremal
		& $\frac{1}{180}\left(2+\frac{3\beta Q^4 \mathcal{B}^{\prime\prime}}{64\pi b^5 r_H^4(b^2+r_H^2)}\right)$  \\[1ex]
		\hline
		&extremal & $\frac{1}{90}$  \\
		\raisebox{1.5ex}{Kerr} &non-extremal
		&$\frac{1}{90}$  \\[1ex]
		\hline
		&extremal &$-\frac{1}{180}$  \\
		\raisebox{1.5ex}{Reissner-Nordstr\"om} &non-extremal
		&$\frac{1}{180}\left(2+\frac{3\beta Q^4}{10\pi r_H^5}\right)$  \\[1ex]
		\hline
		Schwarzschild &non-extremal &$\frac{1}{90}$  \\[1ex]
		\hline 
	\end{tabular}
	\caption{Logarithmic correction contributions ($\Delta S_{\text{BH}}$) of the massless scalar field to the entropy of Kerr-Newman family of black holes in the $d=4$ ``minimally-coupled" EMT.} \label{tab:scalar}
\end{table}
\subsection{Contributions of the minimally-coupled vector field }\label{vector}
The quadratic fluctuated part of the additionally-coupled vector field $a^{\prime}_\mu$ in the fluctuation of the action \eqref{E5} is presented as
\begin{subequations}\label{vector1}
	\begin{align}
		\delta^2\mathcal{S}_{\text{vector}}[a^{\prime}_\mu]=-\frac{1}{4}\int \mathrm{d}^4x \, \delta^2(\sqrt{\text{det}\thinspace {g}}\thinspace f^{\prime}_{\mu\nu}{f^\prime}^{\mu\nu}),\thinspace {f}^{\prime}_{\mu\nu} = \partial_\mu a^{\prime}_\nu-\partial_\nu a^{\prime}_\mu,
	\end{align}
	with
	\begin{align}
		\delta^2(\sqrt{\text{det}\thinspace {g}}\thinspace f^{\prime}_{\mu\nu}{f^\prime}^{\mu\nu})=2 \sqrt{\text{det}\thinspace \bar{g}}\thinspace \Big(-a^{\prime}_\mu D_\rho D^\rho {a^{\prime}}^\mu+ a^{\prime}_\mu R^{\mu\nu} a^{\prime}_\nu - D_\mu {a^{\prime}}^\mu D_\nu {a^{\prime}}^\nu \Big),
	\end{align}
\end{subequations}
where we have omitted all total derivative terms and also used the commutation relation \eqref{cv} for the vector field fluctuation $a^{\prime}_\mu$. We then gauge fix the action \eqref{vector1} by incorporating the gauge fixing term,
\begin{align}\label{vector3}
	-\frac{1}{2}\int \mathrm{d}^4x \sqrt{\text{det}\thinspace \bar{g}}\thinspace (D_\mu {a^{\prime}}^\mu) (D_\nu {a^{\prime}}^\nu).
\end{align}
The gauge-fixed action (without ghost) provides the desired Laplace-type operator $\Lambda$ as
\begin{align}\label{vector4}
	\begin{split}
		\delta^2\mathcal{S}_{\text{vector}}[a^{\prime}_\mu] &=\frac{1}{2}\int \mathrm{d}^4x \sqrt{\text{det}\thinspace \bar{g}}\thinspace a^{\prime}_\mu \Lambda^{a^{\prime}_\mu a^{\prime}_\nu} a^{\prime}_\nu ,\\
		\Lambda^{a^{\prime}_\mu a^{\prime}_\nu} &= \bar{g}^{\mu\nu}D_\rho D^\rho - R^{\mu\nu},
	\end{split}
\end{align}
yielding ($N^\rho)^{a^{\prime}_\mu a^{\prime}_\nu}=0,\thinspace P^{a^{\prime}_\mu a^{\prime}_\nu}=-R^{\mu\nu}$ in the schematic \eqref{Bx}. With the aid of the formulae \eqref{B11}, one obtains results for the useful matrices,
\begin{align}\label{vector5}
	I^{a^{\prime}_\mu a^{\prime}_\nu} =  \bar{g}^{\mu\nu}, \thinspace (\omega^\rho)^{a^{\prime}_\mu a^{\prime}_\nu} =0, \thinspace E^{a^{\prime}_\mu a^{\prime}_\nu} = - R^{\mu\nu}, \thinspace (\Omega_{\rho\sigma})^{a^{\prime}_\mu a^{\prime}_\nu}= {R^{\mu\nu}}_{\rho\sigma},
\end{align} 
along with their following trace results
\begin{align}\label{vector6}
	\begin{split}
		\text{tr}(I) &= 4, \\ \text{tr}(E) &= 0,\\ \text{tr}(E^2) &= R_{\mu\nu}R^{\mu\nu}, \\ \text{tr}(\Omega_{\rho\sigma}\Omega^{\rho\sigma}) &=-R_{\mu\nu\rho\sigma}R^{\mu\nu\rho\sigma}.
	\end{split}
\end{align}
Putting everything together in the formulae \eqref{B12}, one can come up with the Seeley-DeWitt coefficients for the additionally-coupled vector field (without ghost),
\begin{align}\label{vector7}
	\begin{split}
		a_0^{\text{vector,no-ghost}}(x) &= \frac{1}{4\pi^2},\\
		a_2^{\text{vector,no-ghost}}(x) &= 0,\\
		a_4^{\text{vector,no-ghost}}(x) &= -\frac{1}{180\times 16\pi^2}(11 R_{\mu\nu\rho\sigma}R^{\mu\nu\rho\sigma}-86 R_{\mu\nu}R^{\mu\nu}).
	\end{split}
\end{align}
The particular kind of gauge-fixing term \eqref{vector3} demands to account for the following ghost action \cite{Banerjee:2011oo}
\begin{align}\label{vector8}
	\delta^2\mathcal{S}_{\text{vector,ghost}} =\frac{1}{2}\int \mathrm{d}^4x \sqrt{\text{det}\thinspace \bar{g}}\thinspace \Big(b D_\rho D^\rho c+ c D_\rho D^\rho b\Big),
\end{align}
where $b$ and $c$ are corresponding scalar ghosts. Since $b$ and $c$ are scalars and minimally coupled to the background metric $\bar{g}_{\mu\nu}$, their total Seeley-DeWitt contribution is -2 times the free massless scalar field result \eqref{scalar4}, i.e.,
\begin{align}\label{vector9}
	\begin{split}
		a_{2n}^{\text{vector,ghost}}(x) &= -2a_{2n}^{\text{scalar}}(x),
	\end{split}
\end{align}
where the minus sign is included due to the fact that ghost fields follow the reverse of usual spin-statistics. Then the complete Seeley-DeWitt coefficient results for the vector field, adding the contributions from the gauge-fixed part \eqref{vector7} and the ghost part \eqref{vector9}, are 
\begin{align}\label{vector10}
	\begin{split}
		a_0^{\text{vector}}(x) &= \frac{1}{8\pi^2},\\
		a_2^{\text{vector}}(x) &= 0,\\
		a_4^{\text{vector}}(x) &= -\frac{1}{180\times 16\pi^2}(13 R_{\mu\nu\rho\sigma}R^{\mu\nu\rho\sigma}-88 R_{\mu\nu}R^{\mu\nu}).
	\end{split}
\end{align}
Also, the trace anomaly form \eqref{next1} of $a_4^{\text{vector}}(x)$ specifies the following $(c, a)$ data
\begin{equation}\label{vector11}
	(c,a)^{\text{vector}} = \frac{1}{10}, \frac{31}{180}.
\end{equation}
In a similar fashion, one can use the data \eqref{vector10} and \eqref{vector11} respectively into \say{\hyperref[ST1]{Strategy A}} and \say{\hyperref[ST2]{Strategy B}} and compute the logarithmic correction contribution of the minimally-coupled vector field to the entropy of Kerr-Newman family of black holes. The extremal and non-extremal results are recorded in \cref{tab:vector}.
\begin{table}[t]
	\renewcommand{\arraystretch}{1.5}
	\centering 
	\begin{tabular}{|l| c| c|} 
		\hline 
		\textbf{Black hole type} & \textbf{Limits} & \textbf{\bm$\Delta S_{\text{BH}}/\text{ln}\thinspace \mathcal{A}_{H}$}
		\\ [0.7ex]
		\hline\hline 
		&extremal &$-\frac{1}{360}\Big( 27\mathcal{B} + (97+ 280 {b^\prime}^2+260 {b^\prime}^4+104 {b^\prime}^6)\mathcal{B}^\prime\Big)$  \\
		\raisebox{1.5ex}{Kerr-Newman} &non-extremal
		& $\frac{1}{180}\left(-26+\frac{9\beta Q^4 \mathcal{B}^{\prime\prime}}{16\pi b^5 r_H^4(b^2+r_H^2)}\right)$  \\[1ex]
		\hline
		&extremal & $-\frac{13}{90}$  \\
		\raisebox{1.5ex}{Kerr} &non-extremal
		&$-\frac{13}{90}$  \\[1ex]
		\hline
		&extremal &$-\frac{31}{90}$  \\
		\raisebox{1.5ex}{Reissner-Nordstr\"om} &non-extremal
		&$\frac{1}{90}\left(-13+\frac{9\beta Q^4}{5\pi r_H^5}\right)$  \\[1ex]
		\hline
		Schwarzschild &non-extremal &$-\frac{13}{90}$  \\[1ex]
		\hline 
	\end{tabular}
	\caption{Logarithmic correction contributions ($\Delta S_{\text{BH}}$) of the massless additionally-coupled vector field to the entropy of Kerr-Newman family of black holes in the $d=4$ ``minimally-coupled" EMT.} \label{tab:vector}
\end{table}
As per the discussion of \cref{zeromode}, the logarithmic correction results do not include any zero-mode correction from the vector field.
\subsection{Contributions of the minimally-coupled spin-1/2 Dirac field}\label{spinor}
In the quadratic fluctuation of the action \eqref{E5}, the contribution of the minimally-coupled massless Dirac spinor $\lambda$ is
\begin{align}\label{c1}
	\delta^2\mathcal{S}_{\text{Dirac}}[\lambda]= \int \mathrm{d}^4x \, \delta^2(\sqrt{\text{det}\thinspace {g}} \thinspace i\bar{\lambda}\gamma^\rho D_\rho\lambda),
\end{align}
where 
\begin{align}
	\delta^2(\sqrt{\text{det}\thinspace {g}} \thinspace \bar{\lambda}\gamma^\rho D_\rho\lambda) = \sqrt{\text{det}\thinspace \bar{g}} \thinspace \bar{\lambda}\gamma^\rho D_\rho\lambda.
\end{align}
The Seeley-DeWitt coefficients for the fluctuation of an elementary free spin-1/2 Dirac field around an arbitrary background have already been reported in our earlier work \cite{Karan:2018ac}. We find it worthy of reviewing the work \cite{Karan:2018ac} in order to obtain necessary results from the quadratic fluctuated action \eqref{c1}. Unlike bosons, the quadratic fluctuations of fermions are characterized by first-order operators. After correctly identifying the first order Dirac-type operator $\slashed{D}$ for the spin-1/2 fluctuation \eqref{c1}, we structure the required Laplace-type operator $\Lambda$ as following
\begin{align}\label{c2}
	\begin{split}
		\delta^2\mathcal{S}_{\text{Dirac}}[\lambda] &= \int \mathrm{d}^4x \sqrt{\text{det}\thinspace \bar{g}} \thinspace \bar{\lambda}\slashed{D}\lambda,\\
		\slashed{D} &= i\gamma^\rho D_\rho,\\
		\Lambda &=\slashed{D}^\dagger\slashed{D}= - \mathbb{I}_4 D_\rho D^\rho,
	\end{split}
\end{align}
where we have employed various gamma matrix identities, the spin-1/2 commutation relation \eqref{cd} and the $R=0$ condition for the Einstein-Maxwell background.
The particular form of $\Lambda$ yields $P=N^\rho=0$, which further expresses the forms of matrices defined in \cref{B11} as
\begin{align}\label{c3}
	\begin{split}
		I&=\mathbb{I}_4, \thinspace \omega^\rho = 0,\thinspace E = 0,\\
		\Omega_{\rho\sigma} &= \frac{1}{4}\gamma^\alpha\gamma^\beta R_{\rho\sigma\alpha\beta}.
	\end{split}
\end{align}
We then compute the following trace results
\begin{align}\label{c4}
	\begin{split}
		\text{tr}(I) &= 4,\thinspace \text{tr}(E) = 0,\thinspace \text{tr}(E^2) = 0,\\ \text{tr}(\Omega_{\rho\sigma}\Omega^{\rho\sigma}) &= -\frac{1}{2}R_{\mu\nu\rho\sigma}R^{\mu\nu\rho\sigma},
	\end{split}
\end{align}
and find the first three Seeley-DeWitt coefficients for the minimally-coupled spin-1/2 Dirac field fluctuation,
\begin{align}\label{c5}
	\begin{split}
		a_0^{\text{Dirac}}(x) &= -\frac{1}{4\pi^2},\\
		a_2^{\text{Dirac}}(x) &=0, \\
		a_4^{\text{Dirac}}(x) &= \frac{1}{360\times 16\pi^2}(7 R_{\mu\nu\rho\sigma}R^{\mu\nu\rho\sigma}+ 8 R_{\mu\nu}R^{\mu\nu}).
	\end{split}
\end{align}
Notice that we have used $\chi=-1$ in the formula \eqref{B12} to encounter the Dirac spinor condition. The coefficients of trace anomalies are extracted from the $a_4^{\text{Dirac}}(x)$ result as
\begin{equation}
	(c,a)^{\text{Dirac}} = \frac{1}{20}, \frac{11}{360}.
\end{equation}
Finally, the logarithmic correction contribution of the minimally-coupled spin-1/2 Dirac field to the entropy of extremal and non-extremal Kerr-Newman black family of holes are calculated as in \cref{tab:dirac}.
\begin{table}[t]
	\renewcommand{\arraystretch}{1.5}
	\centering 
	\begin{tabular}{|l| c| c|} 
		\hline 
		\textbf{Black hole type} & \textbf{Limits} & \textbf{\bm$\Delta S_{\text{BH}}/\text{ln}\thinspace \mathcal{A}_{H}$}
		\\ [0.7ex]
		\hline\hline 
		&extremal &$-\frac{1}{720}\Big( 27\mathcal{B} + (17-40 {b^\prime}^2-140 {b^\prime}^4-56 {b^\prime}^6)\mathcal{B}^\prime\Big)$  \\
		\raisebox{1.5ex}{Kerr-Newman} &non-extremal
		& $\frac{1}{180}\left(7+\frac{9\beta Q^4 \mathcal{B}^{\prime\prime}}{32\pi b^5 r_H^4(b^2+r_H^2)}\right)$  \\[1ex]
		\hline
		&extremal & $\frac{7}{180}$  \\
		\raisebox{1.5ex}{Kerr} &non-extremal
		&$\frac{7}{180}$  \\[1ex]
		\hline
		&extremal &$-\frac{11}{180}$  \\
		\raisebox{1.5ex}{Reissner-Nordstr\"om} &non-extremal
		&$\frac{1}{180}\left(7+\frac{9\beta Q^4}{5\pi r_H^5}\right)$  \\[1ex]
		\hline
		Schwarzschild &non-extremal &$\frac{7}{180}$  \\[1ex]
		\hline 
	\end{tabular}
	\caption{Logarithmic correction contributions ($\Delta S_{\text{BH}}$) of the massless spin-1/2 Dirac field to the entropy of Kerr-Newman family of black holes in the $d=4$ ``minimally-coupled" EMT.} \label{tab:dirac}
\end{table}
In these corrections, the minimally-coupled spin-1/2 Dirac spinor contributes nothing to the $\mathcal{C}_{\text{zm}}$ formula \eqref{B9}.
\subsection{Contributions of the minimally-coupled spin-3/2 Rarita-Schwinger field}\label{RS}
The quadratic fluctuated part of the minimally-coupled massless Rarita-Schwinger field $\psi_\mu$ (Majorana type) is expressed as
\begin{subequations}\label{rs1}
	\begin{align}
		\delta^2\mathcal{S}_{\text{RS}}[\psi_{\mu }]= -\int \mathrm{d}^4x \,\delta^2(\sqrt{\text{det}\thinspace {g}} \thinspace \bar{\psi}_{\mu }\gamma^{\mu\rho\nu}D_\rho\psi_{\nu }),
	\end{align}
	where one can adjust,
	\begin{align}
		\delta^2(\sqrt{\text{det}\thinspace {g}} \thinspace \bar{\psi}_{\mu }\gamma^{\mu\rho\nu}D_\rho\psi_{\nu }) = \frac{1}{2} \sqrt{\text{det}\thinspace \bar{g}}\,\bar{\psi}_{\mu }\left(\gamma^\mu\gamma^\rho\gamma^\nu-\gamma^\nu\gamma^\rho\gamma^\mu\right)D_\rho\psi_\nu,
	\end{align}
\end{subequations}
with the following gauge-fixing term for casting the gauge $\gamma^\rho\psi_\rho=0$
\begin{align}\label{rs2}
	\frac{1}{2}\int \mathrm{d}^4x \sqrt{\text{det}\thinspace \bar{g}} \thinspace\bar{\psi}_\mu \gamma^\mu\gamma^\rho\gamma^\nu D_\rho \psi_\nu.
\end{align}
We again review the work \cite{Karan:2018ac} where Seeley-DeWitt coefficients for a free Rarita-Schwinger Dirac spinor are evaluated around an arbitrary background, without considering any ghost contribution for the particular gauge-fixing term chosen. But we aim to calculate the complete (gauge-fixed \eqref{rs3} and ghost \eqref{rs8}) Seeley-DeWitt coefficient results for the quadratic fluctuated minimally-coupled spin-3/2 Majorana spinor \eqref{rs1}. After combining \eqref{rs1} and \eqref{rs2}, the gauge-fixed action extracts a first-order Dirac-type\footnote{The Seeley-DeWitt computation approach (\cref{sdc}) sets the spin-3/2 spinor as Dirac one, which is restored back into the original Majorana form at the final result \eqref{rs7} by taking $\chi=-1/2$.} operator $\slashed{D}$ that further structures the second-order Laplace-type operator $\Lambda$ as following
\begin{align}\label{rs3}
	\begin{split}
		\delta^2\mathcal{S}_{\text{RS}}[\psi_{\mu }]&= \int \mathrm{d}^4x \sqrt{\text{det}\thinspace \bar{g}} \thinspace \bar{\psi}_{\mu }\slashed{D}^{\bar{\psi}_\mu \psi_\nu}\psi_{\nu },\\
		\slashed{D}^{\bar{\psi}_\mu \psi_\nu}&=\frac{i}{2}\gamma^\nu\gamma^\rho\gamma^\mu D_\rho,\\
		\Lambda^{\bar{\psi}_\mu \psi_\nu} &=(\slashed{D}^{\bar{\psi}_\mu\psi_\alpha})^\dagger{\slashed{D}_{\bar{\psi}_\alpha}}^{\psi_\nu}\\
		&=  -\mathbb{I}_4 \bar{g}^{\mu\nu}{D}_\rho {D}^\rho+\mathbb{I}_4 R^{\mu\nu}-\frac{1}{2}\gamma^\alpha\gamma^\beta {R^{\mu\nu}}_{\alpha\beta} +\frac{1}{2}\gamma^\mu\gamma^\alpha {R^\nu}_\alpha-\frac{1}{2}\gamma^\nu\gamma^\alpha {R^\mu}_\alpha,
	\end{split}
\end{align}
where various gamma identities, spin-3/2 commutation relation \eqref{crs}, and $R=0$ condition are employed to obtain the last line equality. As per the schematic \eqref{Bx}, we read off $N^\rho$ and $P$, 
\begin{align}\label{rs4}
	\begin{split}
		(N^\rho)^{\bar{\psi}_\mu \psi_\nu} &= 0,\\
		P^{\bar{\psi}_\mu \psi_\nu} &= -\mathbb{I}_4 R^{\mu\nu}+\frac{1}{2}\gamma^\alpha\gamma^\beta {R^{\mu\nu}}_{\alpha\beta}-\frac{1}{2}\gamma^\mu\gamma^\alpha {R^\nu}_\alpha+\frac{1}{2}\gamma^\nu\gamma^\alpha {R^\mu}_\alpha,
	\end{split}
\end{align}
and then find the matrices $I$, $\omega_\rho$, $E$ and $\Omega_{\rho\sigma}$,
\begin{align}\label{rs5}
	\begin{split}
		&I^{\bar{\psi}_\mu \psi_\nu} = \mathbb{I}_4 \bar{g}^{\mu\nu}, \enspace (\omega^\rho)^{\bar{\psi}_\mu \psi_\nu} = 0,\\
		&E^{\bar{\psi}_\mu \psi_\nu} = -\mathbb{I}_4 R^{\mu\nu}+\frac{1}{2}\gamma^\alpha\gamma^\beta {R^{\mu\nu}}_{\alpha\beta}-\frac{1}{2}\gamma^\mu\gamma^\alpha {R^\nu}_\alpha+\frac{1}{2}\gamma^\nu\gamma^\alpha {R^\mu}_\alpha, \\
		&(\Omega_{\rho\sigma})^{\bar{\psi}_\mu \psi_\nu}= \mathbb{I}_4{R^{\mu\nu}}_{\rho\sigma}+\frac{1}{4}\bar{g}^{\mu\nu}\gamma^\alpha\gamma^\beta R_{\rho\sigma\alpha\beta}.
	\end{split}
\end{align}
The relevant trace values are evaluated as
\begin{align}\label{rs6}
	\begin{split}
		\text{tr}(I) &= 16, \text{tr}(E)= 0 \\
		\text{tr}(E^2) &= 2R_{\mu\nu\rho\sigma}R^{\mu\nu\rho\sigma}, \\
		\text{tr}(\Omega_{\rho\sigma}\Omega^{\rho\sigma}) &=-6R_{\mu\nu\rho\sigma}R^{\mu\nu\rho\sigma}.
	\end{split}
\end{align}
By setting $\chi=-1/2$ and the above trace data in the formulae \eqref{B12}, we present the Seeley-DeWitt results (without ghost contribution) for the spin-3/2 Majorana spinor as
\begin{align}\label{rs7}
	\begin{split}
		a_0^{\text{RS,no-ghost}}(x) &= -\frac{1}{2\pi^2},\\
		a_2^{\text{RS,no-ghost}}(x) &= 0,\\
		a_4^{\text{RS,no-ghost}}(x) &= -\frac{1}{180\times 16\pi^2}(53 R_{\mu\nu\rho\sigma}R^{\mu\nu\rho\sigma}-8 R_{\mu\nu}R^{\mu\nu}).
	\end{split}
\end{align}
It is also customary to include a ghost action that counters the gauge-fixing term \eqref{rs2},
\begin{align}\label{rs8}
	\delta^2\mathcal{S}_{\text{RS,ghost}} =\frac{1}{2}\int \mathrm{d}^4x \sqrt{\text{det}\thinspace \bar{g}}\thinspace (\bar{\tilde{b}}\gamma^\rho D_\rho \tilde{c} + \bar{\tilde{e}}\gamma^\rho D_\rho \tilde{e}),
\end{align}
where $\tilde{b}$, $\tilde{c}$, and $\tilde{e}$ are three minimally-coupled bosonic ghosts (i.e., spin-1/2 Majorana spinors) \cite{Banerjee:2011oo}. The combined Seeley-DeWitt contribution of these ghosts is equivalently obtained as
\begin{align}\label{rs100}
	a^{\text{RS,ghost}}_{2n} = (-1)\times 3\times \frac{1}{2} \times a^{\text{Dirac}}_{2n},
\end{align}
where the minus sign is included for the reverse spin-statistics of ghosts, $3$ serves the ghost multiplicity and $1/2$ factor converts the single spin-1/2 Dirac results $a^{\text{Dirac}}_{2n}$ (recorded in \cref{c5}) into the Majorana kind. After combining $a^{\text{RS,no-ghost}}_{2n}$ and $a^{\text{RS,ghost}}_{2n}$ contributions, the complete Seeley-DeWitt results for the minimally-coupled Rarita-Schwinger field fluctuation are
\begin{align}\label{rs9}
	\begin{split}
		a_0^{\text{RS}}(x) &= -\frac{1}{8\pi^2},\\
		a_2^{\text{RS}}(x) &= 0,\\
		a_4^{\text{RS}}(x) &= -\frac{1}{720\times 16\pi^2}(233 R_{\mu\nu\rho\sigma}R^{\mu\nu\rho\sigma}-8 R_{\mu\nu}R^{\mu\nu}),
	\end{split}
\end{align}
along with the following trace anomaly data
\begin{equation}
	(c,a)^{\text{RS}} = -\frac{77}{120}, -\frac{229}{720}.
\end{equation}
In the end, we utilize the $a_4^{\text{RS}}(x)$ and trace anomaly data and achieve the logarithmic correction contributions of the minimally-coupled Rarita-Schwinger field to the entropy of extremal and non-extremal Kerr-Newman family of black holes. The results are presented in \cref{tab:rs}. These corrections do not receive any zero-mode contribution from the non-supersymmetric Rarita-Schwinger field, as discussed in \cref{zeromode}.
\begin{table}[t!]
	\renewcommand{\arraystretch}{1.5}
	\centering 
	\begin{tabular}{|l| p{.65in}| c|} 
		\hline 
		\textbf{Black hole type} & \textbf{Limits} & \textbf{\bm$\Delta S_{\text{BH}}/\text{ln}\thinspace \mathcal{A}_{H}$}
		\\ [1ex]
		\hline\hline 
		&extremal &$\frac{1}{1440}\Big( 693\mathcal{B} + (223-1880 {b^\prime}^2-4660 {b^\prime}^4-1864 {b^\prime}^6)\mathcal{B}^\prime\Big)$  \\
		\raisebox{1.5ex}{Kerr-Newman} &non-extremal
		& $-\frac{1}{360}\left(233+\frac{231\beta Q^4 \mathcal{B}^{\prime\prime}}{32\pi b^5 r_H^4(b^2+r_H^2)}\right)$  \\[1ex]
		\hline
		&extremal & $-\frac{233}{360}$  \\
		\raisebox{1.5ex}{Kerr} &non-extremal
		&$-\frac{233}{360}$  \\[1ex]
		\hline
		&extremal &$\frac{229}{360}$  \\
		\raisebox{1.5ex}{Reissner-Nordstr\"om} &non-extremal
		&$-\frac{1}{360}\left(233+\frac{231\beta Q^4}{5\pi r_H^5}\right)$  \\[1ex]
		\hline
		Schwarzschild &non-extremal &$-\frac{233}{360}$  \\[1ex]
		\hline 
	\end{tabular}
	\caption{Logarithmic correction contributions ($\Delta S_{\text{BH}}$) of the massless spin-3/2 Majorana field to the entropy of Kerr-Newman family of black holes in the $d=4$ ``minimally-coupled" EMT.} \label{tab:rs}
\end{table}
\section{Discussions}\label{gem}
This section generalizes the $d=4$ \say{minimally-coupled} EMT further by coupling any arbitrary numbers of massless fields, which leads to a set of generalized Seeley-DeWitt coefficient and logarithmic correction formulae for all the extremal and non-extremal Kerr-Newman family of black holes. We then employ the generalized \say{minimally-coupled} data in successful derivation of the logarithmic corrections to the entropy of extremal black holes in $\mathcal{N} \geq 2,d=4$ Einstein-Maxwell supergravity theories. Finally, we conclude by summarizing and discussing the results.
\subsection{Einstein-Maxwell theory minimally-coupled to arbitrary numbers of massless fields}\label{gen}
Consider a generalized EMT with $n_{\text{EM}}$ numbers of Einstein-Maxwell sectors (each containing one metric and one vector field) is minimally coupled to $n_0$ numbers of scalar fields, $n_1$ numbers of additional vector fields, $n_{1/2}$ numbers of spin-1/2 Dirac fields and $n_{3/2}$ numbers of spin-3/2 Rarita-Schwinger fields (Majorana). Then the first three Seeley-DeWitt coefficients for the quadratic fluctuations of the field content of the generalized $d=4$ \say{minimally-coupled} EMT are expressed as
\begin{align}\label{gem1}
	{a_0}^{\text{EM(mc)}}(x) &= \frac{1}{16\pi^2}(4 n_{\text{EM}}+n_0 + 2n_1 -4 n_{1/2}-2n_{3/2} ),\nonumber\\
	{a_2}^{\text{EM(mc)}}(x) &= \frac{3}{8\pi^2}n_{\text{EM}}\bar{F}_{\mu\nu}\bar{F}^{\mu\nu},\\
	a_4^{\text{EM(mc)}}(x) &= \frac{1}{16\pi^2 \times 360}\bigg\lbrace  \Big(398 n_{\text{EM}}+2n_0 -26n_1 +7 n_{1/2}-\frac{233}{2}n_{3/2} \Big)R_{\mu\nu\rho\sigma}R^{\mu\nu\rho\sigma} \nonumber\\
	&\qquad +\Big(52 n_{\text{EM}}-2n_0 +176n_1+8 n_{1/2}+4n_{3/2} \Big)R_{\mu\nu}R^{\mu\nu}\bigg\rbrace,\nonumber
\end{align}
followed by the trace anomaly data $(c,a)$,
\begin{align}\label{gem2}
	\begin{split}
		c^{\text{EM(mc)}} &= \frac{1}{360}\Big(822 n_{\text{EM}}+3n_0+36n_1+18 n_{1/2}-231 n_{3/2}\Big),\\
		a^{\text{EM(mc)}} &= \frac{1}{360} \Big(424 n_{\text{EM}}+n_0+62n_1+11n_{1/2}-\frac{229}{2}n_{3/2}\Big).
	\end{split}
\end{align}
The data \eqref{gem1} and \eqref{gem2} provide the following local corrections to the entropy of extremal and non-extremal Kerr-Newman black holes
{
	\allowdisplaybreaks
	\begin{subequations}
		\begin{align}
			\mathcal{C}^{\thinspace\text{extremal}}_{\text{local}} &= -\frac{1}{360}\bigg\lbrace  \Big(398 n_{\text{EM}}+2n_0-26n_1+7n_{1/2}-\frac{233}{2}n_{3/2} \Big)\nonumber\\
			&\qquad\qquad\times \Big( 3\mathcal{B}-(8{b^\prime}^6+20 {b^\prime}^4+8 {b^\prime}^2-1)\mathcal{B}^\prime\Big) \nonumber\\
			&\qquad +\Big( 3\mathcal{B} +(8{b^\prime}^2+5)\mathcal{B}^\prime\Big)\Big(13 n_{\text{EM}}-\frac{n_0}{2}+44n_1+2n_{1/2}+n_{3/2} \Big) \bigg\rbrace, \label{gem3a}\\
			\mathcal{C}^{\thinspace\text{non-extremal}}_{\text{local}} &= \frac{1}{90}\bigg\lbrace \Big(398 n_{\text{EM}}+2 n_0-26n_1+7 n_{1/2}-\frac{233}{2} n_{3/2}\Big)\nonumber\\
			& +\frac{\beta Q^4 \mathcal{B}^{\prime\prime}}{64\pi b^5 r_H^4(b^2+r_H^2)}\Big(822 n_{\text{EM}}+3n_0 +36n_1+18 n_{1/2}-231 n_{3/2}\Big)\bigg\rbrace \label{gem3b}.
		\end{align}
\end{subequations}}
On the other hand, the zero-mode correction only includes the contribution from metric in the formula \eqref{B9} and should be added to the Einstein-Maxwell sector. For Kerr-Newman black holes, 
\begin{subequations}
	\begin{align}
		\mathcal{C}^{\thinspace\text{extremal}}_{\text{zm}} &= (2-1)\times (-3-1)=-4, \label{gem4a}\\
		\mathcal{C}^{\thinspace\text{non-extremal}}_{\text{zm}} &= (2-1)\times (-1)=-1.\label{gem4b}
	\end{align}
\end{subequations}
The logarithmic corrections to the entropy of Kerr-Newman black holes in the generalized \say{minimally-coupled} EMT are
{
	\allowdisplaybreaks
	\begin{subequations}\label{gem5}
		\begin{align}
			\Delta S^{\thinspace\text{extremal}}_{\text{BH}} &= -\frac{1}{720}\bigg\lbrace  \Big(398 n_{\text{EM}}+2n_0-26n_1+7n_{1/2}-\frac{233}{2}n_{3/2} \Big) \nonumber\\
			&\quad\qquad\times \Big( 3\mathcal{B}-(8{b^\prime}^6+20{b^\prime}^4+8{b^\prime}^2-1)\mathcal{B}^\prime\Big)+\Big( 3\mathcal{B} +(8{b^\prime}^2+5)\mathcal{B}^\prime\Big) \nonumber\\
			&\qquad\times\Big(13 n_{\text{EM}}-\frac{n_0}{2}+44n_1+2n_{1/2}+n_{3/2} \Big)+1440\bigg\rbrace\text{ln}\thinspace \mathcal{A}_{H}, \label{gem5a}\\
			\Delta S^{\thinspace\text{non-extremal}}_{\text{BH}} &= \frac{1}{180}\bigg\lbrace \Big(398 n_{\text{EM}}+2 n_0-26n_1+7 n_{1/2}-\frac{233}{2} n_{3/2}\Big)\nonumber\\
			& +\frac{\beta Q^4 \mathcal{B}^{\prime\prime}}{64\pi b^5 r_H^4(b^2+r_H^2)}\Big(822 n_{\text{EM}}+3n_0+36n_1+18 n_{1/2}-231 n_{3/2}\Big)-90\bigg\rbrace \text{ln}\thinspace \mathcal{A}_{H}. \label{gem5b}
		\end{align}
\end{subequations}}
In particular limits (refer to \cref{Bc}) of the Kerr-Newman results, one can extract the logarithmic entropy corrections for Kerr and Reissner-Nordstr\"om black holes as below:
\begin{itemize}
	\item[\scalebox{0.6}{$\blacksquare$}]
	For the extremal and non-extremal \textit{Kerr} black holes, we need to set $b^\prime\to\infty$ and $Q=0$ in the $\mathcal{C}_{\text{local}}$ results \eqref{gem3a} and \eqref{gem3b}, respectively. This yields exactly the same extremal and non-extremal local corrections for the Kerr black hole,
	\begin{align}\label{gem6}
		\mathcal{C}^{\thinspace\text{extremal}}_{\text{local}}=\mathcal{C}^{\thinspace\text{non-extremal}}_{\text{local}} &= \frac{1}{90}\Big(398 n_{\text{EM}}+2n_0-26n_1+7n_{1/2}-\frac{233}{2}n_{3/2} \Big).
	\end{align}
	The corresponding zero-mode corrections are obtained as
	\begin{subequations}
		\begin{align}\label{gem7}
			\mathcal{C}^{\thinspace\text{extremal}}_{\text{zm}} &= (2-1)\times (-3-1)=-4,\\
			\mathcal{C}^{\thinspace\text{non-extremal}}_{\text{zm}} &= (2-1)\times (-1)=-1.
		\end{align}
	\end{subequations}
	Hence, the logarithmic corrections to the entropy of Kerr black holes in the generalized $d=4$ \say{minimally-coupled} EMT are
	\begin{subequations}\label{gem8}
		\begin{align}
			\Delta S^{\thinspace\text{extremal}}_{\text{BH}} &= \frac{1}{180}\Big(398 n_{\text{EM}}+2n_0-26n_1+7n_{1/2}-\frac{233}{2}n_{3/2} -360\Big)\text{ln}\thinspace \mathcal{A}_{H}, \label{gem8a}\\
			\Delta S^{\thinspace\text{non-extremal}}_{\text{BH}} &= \frac{1}{180}\Big(398 n_{\text{EM}}+2n_0-26n_1+7n_{1/2}-\frac{233}{2}n_{3/2} -90\Big)\text{ln}\thinspace \mathcal{A}_{H}.\label{gem8b}
		\end{align}
	\end{subequations} 
	
	\item[\scalebox{0.6}{$\blacksquare$}]
	For the extremal and non-extremal \textit{Reissner-Nordstr\"om} black holes, we need to set $b^\prime\to 0$ and $b \to 0$ in the $\mathcal{C}_{\text{local}}$ results \eqref{gem3a} and \eqref{gem3b}, respectively. This provides extremal and non-extremal local corrections for the Reissner-Nordstr\"om black holes,
	\begin{subequations}\label{gem9}
		\begin{align}
			\mathcal{C}^{\thinspace\text{extremal}}_{\text{local}} &=- \frac{1}{90}\Big(424 n_{\text{EM}}+n_0+62n_1+11n_{1/2}-\frac{229}{2}n_{3/2} \Big),\\
			\mathcal{C}^{\thinspace\text{non-extremal}}_{\text{local}} &= \frac{1}{90}\bigg\lbrace 398 n_{\text{EM}}+2 n_0-26n_1+7 n_{1/2}-\frac{233}{2} n_{3/2}\nonumber\\
			&\quad +\frac{\beta Q^4}{10\pi r_H^5}\Big(822 n_{\text{EM}}+3n_0+36n_1+18 n_{1/2}-231 n_{3/2}\Big)\bigg\rbrace.
		\end{align}
	\end{subequations}
	The corresponding zero-mode corrections are obtained as
	\begin{subequations}
		\begin{align}
			\mathcal{C}^{\thinspace\text{extremal}}_{\text{zm}} &= (2-1)\times (-3-3)=-6,\label{gem10a}\\
			\mathcal{C}^{\thinspace\text{non-extremal}}_{\text{zm}} &= (2-1)\times (-3)=-3. \label{gem10b}
		\end{align}
	\end{subequations}
	The logarithmic corrections to the entropy of Reissner-Nordstr\"om black holes in the generalized $d=4$ \say{minimally-coupled} EMT are thus
	\begin{subequations}\label{gem11}
		\begin{align}
			\Delta S^{\thinspace\text{extremal}}_{\text{BH}} &= -\frac{1}{180}\Big(424 n_{\text{EM}}+n_0+62n_1+11 n_{1/2}-\frac{229}{2}n_{3/2} +540\Big)\text{ln}\thinspace \mathcal{A}_{H},\label{gem11a}\\
			\Delta S^{\thinspace\text{non-extremal}}_{\text{BH}} &= \frac{1}{180}\bigg\lbrace 398 n_{\text{EM}}+2 n_0-26n_1+7 n_{1/2}-\frac{233}{2} n_{3/2}\nonumber\\
			& +\frac{\beta Q^4}{10\pi r_H^5}\Big(822 n_{\text{EM}}+3n_0+36n_1 +18 n_{1/2}-231 n_{3/2}\Big)-270\bigg\rbrace \text{ln}\thinspace \mathcal{A}_{H}.\label{gem11b}
		\end{align}
	\end{subequations}
	
	\item[\scalebox{0.6}{$\blacksquare$}]
	For the \textit{Schwarzschild} black holes, we need to set $Q=0$ in \cref{gem3b}, which yields the same $\mathcal{C}_{\text{local}}$ as the non-extremal Kerr result \eqref{gem6}. Also, Schwarzschild black hole is non-rotating and hence the zero-mode correction $\mathcal{C}_{\text{zm}}$ is the same as the non-extremal Reissner-Nordstr\"om result \eqref{gem10b}. Hence, the logarithmic correction to the entropy of Schwarzschild black hole in the generalized $d=4$ \say{minimally-coupled} EMT is
	\begin{align}\label{gem12}
		\Delta S^{\thinspace\text{non-extremal}}_{\text{BH}} = \frac{1}{180}\Big(398 n_{\text{EM}}+2n_0-26n_1+7n_{1/2}-\frac{233}{2}n_{3/2} -270\Big)\text{ln}\thinspace \mathcal{A}_{H}.
	\end{align}
\end{itemize}
The above results are our principal focus in this work. 
The corrections for extremal black holes (\cref{gem5a,gem8a,gem11a}) are novel reports to the literature. The non-extremal black hole results (\cref{gem5b,gem8b,gem11b,gem12}) agree with \cite{Sen:2013ns}, where the $a_4(x)$ formula\footnote{One can find a mismatch of the $R_{\mu\nu}R^{\mu\nu}$ coefficient for spin-3/2 fields with \cite{Sen:2013ns}. But this is justified because the coefficient of $R_{\mu\nu}R^{\mu\nu}$ for the spin-3/2 field is not absolute, rather gauge dependent (while the coefficient of $R_{\mu\nu\rho\sigma}R^{\mu\nu\rho\sigma}$ is gauge-independent) \cite{Christensen:1979md}. After choosing the gauge $\gamma^\rho\psi_\rho=0$, the coefficient of $R_{\mu\nu}R^{\mu\nu}$ in $360(4\pi)^2 a_4(x)$ will always be $4$ for a spin-3/2 Majorana spinor, which is also consistent with \cite{Christensen:1980iy}.} is managed from a set of secondary data provided in \cite{Duff:1977ay,Christensen:1979md,Christensen:1980iy,Duff:1980qv,Christensen:1980ee}. Also, the Schwarzschild corrections due to vector and spin-3/2 Rarita-Schwinger fields in \cref{gem12} are entirely consistent with \cite{Majhi:2009uk}, where results are obtained via the tunneling approach. 

\subsection{A local method for logarithmic correction to the extremal black hole entropy in $\mathcal{N} \geq 2$ Einstein-Maxwell supergravity}\label{localmethod}
Logarithmic entropy corrections for the black holes in Einstein-Maxwell embedded supergravity theories are already reported in various works \cite{Banerjee:2011pp,Sen:2012qq,Gupta:2014ns,Karan:2019sk,Karan:2020sk,Banerjee:2020wbr,Keeler:2014nn,Charles:2015nn,Larsen:2015nx,Ferrara:2012bp}. The basic technical approach is common -- analysis of relevant quadratic fluctuated action via the heat kernel method. But supergravity actions are incredibly complicated and unwieldy. The direct calculations are hardly manageable and so far only accomplished for $\mathcal{N}=1,2,4,8$ Einstein-Maxwell supergravity theories (EMSGTs). In \cite{Charles:2015nn,Karan:2020sk,Keeler:2014nn,Ferrara:2012bp}, the results are estimated for all $\mathcal{N} \geq 3,d=4$ EMSGTs by considering the $\mathcal{N} \geq 3 \to \mathcal{N}=2$ decompositions. For the derivations of the minimally-coupled EMSGT sectors, the \say{minimally-coupled} EMT results of the previous section are found to have essential utility  \cite{Karan:2020sk}. But, the quadratic fluctuated supergravity actions mostly include \say{non-minimal} coupling terms \cite{Karan:2020sk,Sen:2012qq,Banerjee:2011pp,Karan:2019sk,Charles:2015nn}. Consequently, one can not directly employ the \say{minimally-coupled} EMT data in the full reproduction of logarithmic entropy corrections in the EMSGTs. 
However, we find an alternative but indirect way of getting a supersymmetrized form of the $a_4(x)$ formula \eqref{gem1} for near-horizon backgrounds, which can be utilized for the full reproduction of logarithmic corrections to the entropy of extremal black holes in $\mathcal{N} \geq 2,d=4$ EMSGTs. The entire approach is depicted as follows.

At first, we need to express the $a_4(x)$ formula \eqref{gem1} in terms of the Weyl tensor square $W_{\mu\nu\rho\sigma}W^{\mu\nu\rho\sigma}$ and Euler density $E_4$ (using the trace anomaly form \eqref{next1}),
\begin{align}\label{loc1}
	360(4\pi)^2 a_4^{\text{EM(mc)}}(x) &= \Big(822 n_{\text{EM}}+3n_0+36n_1+18 n_{1/2}-231 n_{3/2}\Big)W_{\mu\nu\rho\sigma}W^{\mu\nu\rho\sigma}\nonumber\\
	&\quad- \Big(424 n_{\text{EM}}+n_0+62n_1+11n_{1/2}-\frac{229}{2}n_{3/2}\Big)E_4.
\end{align} 
The supersymmetric completion of the above expression entirely depends on the invariants $W_{\mu\nu\rho\sigma}W^{\mu\nu\rho\sigma}$ and $E_4$. The Euler density $E_4$ is a topological invariant and hence self-supersymmetric. On the other hand, it is well reported that the supersymmetrization of $W_{\mu\nu\rho\sigma}W^{\mu\nu\rho\sigma}$, evaluated on near-horizon black hole backgrounds \cite{Behrndt:1998eq,LopesCardoso:1998tkj,LopesCardoso:1999cv,Mohaupt:2000mj,Sahoo:2006rp}, is surprisingly found to be the same as $E_4$ \cite{Sen:2006iz,Maldacena:1997de}. Again, the quantum entropy function formalism \cite{Sen:2008wa,Sen:2009wb,Sen:2009wc} used in \say{\hyperref[ST1]{Strategy A}} requires only the near-horizon details of extremal black holes. All that mentioned suggests a simple and straightforward way to supersymmetrize any theory by analyzing only the Gauss-Bonnet term $E_4$ as well as bypassing the need for Weyl tensor square term $W_{\mu\nu\rho\sigma}W^{\mu\nu\rho\sigma}$ and then achieve logarithmic entropy corrections for the extremal black holes in supergravity theories. As discussed, substituting $W_{\mu\nu\rho\sigma}W^{\mu\nu\rho\sigma}=E_4$ in \eqref{loc1} will lead to the supersymmetrized $a_4(x)$ formula for near-horizon backgrounds in $\mathcal{N} \geq 2,d=4$ EMSGTs,
\begin{align}\label{loc2}
	a_4^{\mathcal{N}\geq 2}(x) &= \frac{1}{16\pi^2 \times 360}\Big(398 n_{\text{EM}}+2n_0 -26 n_1 +7 n_{1/2}-\frac{233}{2}n_{3/2} \Big)E_4.
\end{align}
This necessarily yields a $\mathcal{C}_{\text{local}}$ formula (via the \say{\hyperref[ST1]{Strategy A}}) for the extremal Kerr-Newman black holes, 
\begin{align}\label{loc3}
	\mathcal{C}^{\mathcal{N}\geq 2\thinspace\text{extremal}}_{\text{local}} &= \frac{1}{90}\Big(398 n_{\text{EM}}+2n_0 -26n_1 +7 n_{1/2}-\frac{233}{2}n_{3/2} \Big)\frac{(2{b^\prime}^6+5{b^\prime}^4+4{b^\prime}^2+1)}{({b^\prime}^2+1)^2(2{b^\prime}^2+1)},
\end{align}
where the equality is obtained after using the extremal near-horizon limits \eqref{ext2} in the form of $E_4$. In addition, one obtains the following zero-mode corrections using the formula \eqref{B9}
\begin{equation}\label{loc4}
	\mathcal{C}^{\mathcal{N}\geq 2\thinspace\text{extremal}}_{\text{zm}} =
	\begin{cases}
		(2-1)\times(-3-1)-(3-1)\times 0=-4 &\text{for Kerr-Newman},\\
		(2-1)\times (-3-1)-(3-1)\times 0=-4 &\text{for Kerr},\\
		(2-1)\times (-3-3)-(3-1)\times (-4)=2 &\text{for Reissner-Nordstr\"om}.
	\end{cases}
\end{equation}
Note that extremal Reissner-Nordstr\"om black holes with near-horizon geometry $AdS_2 \times S^2$ are the only possible BPS solutions in the four-dimensional $\mathcal{N} \geq 2$ EMSGTs. Summing all up, we write a combined logarithmic entropy correction formula that reads as follows
\begin{align}\label{loc5}
	\Delta S^{\mathcal{N}\geq 2\thinspace\text{extremal}}_{\text{BH}} &= \frac{1}{180}\Big(398 n_{\text{EM}}+2n_0 -26n_1 +7 n_{1/2}-\frac{233}{2}n_{3/2} \Big) \nonumber\\
	&\qquad \times\frac{(2{b^\prime}^6+5{b^\prime}^4+4{b^\prime}^2+1)}{({b^\prime}^2+1)^2(2{b^\prime}^2+1)}\thinspace \text{ln}\thinspace \mathcal{A}_{H} + \frac{\mathcal{C}^{\mathcal{N}\geq 2\thinspace\text{extremal}}_{\text{zm}}}{2} \ln \mathcal{A}_{H}.
\end{align}
By setting particular multiplicity values in the above formula, one can extract logarithmic corrections to the entropy of extremal Kerr-Newman, Kerr ($b^\prime\to\infty$) and Reissner-Nordstr\"om ($b^\prime\to 0$) black holes in all $\mathcal{N} \geq 2,d=4$ EMSGTs.

For a matter coupled $\mathcal{N} = 2,d=4$ EMSGT, with the supergravity multiplet ($n_{\text{EM}}=1,n_{3/2}=2$) coupled to $n_V$ vector multiplets ($n_0 =2,n_1=1,n_{1/2}=1$) and  $n_H$ hyper multiplets ($n_0=4,n_{1/2}=1$), one obtains
\begin{subequations}\label{loc6}
	\begin{align}
		\Delta S^{\,\mathcal{N}=2\text{,\,Kerr-Newman}}_{\text{BH}} &=\frac{1}{12}\big(11-n_V+n_H\big)\frac{(2{b^\prime}^6+5{b^\prime}^4+4{b^\prime}^2+1)}{({b^\prime}^2+1)^2(2{b^\prime}^2+1)} \ln \mathcal{A}_{H}-2 \ln \mathcal{A}_{H},\\
		\Delta S^{\,\mathcal{N}=2\text{,\,Kerr}}_{\text{BH}} &=\frac{1}{12}\big(-13-n_V+n_H\big)\ln \mathcal{A}_{H},\\
		\Delta S^{\,\mathcal{N}=2\text{,\,Reissner-Nordstr\"om}}_{\text{BH}} &=\frac{1}{12}\big(23-n_V+n_H\big)\ln \mathcal{A}_{H}.
	\end{align}
\end{subequations}
For $\mathcal{N} \geq 3,d=4$ EMSGTs, it is possible to describe the extremal logarithmic entropy corrections via some generalized formulae (only in terms of $\mathcal{N}$). The field contents of the $\mathcal{N} \geq 3$ branch (recorded in \cref{tab:sugracontent}) are such that we can write a relation connecting their multiplicities with the corresponding $\mathcal{N}$,
\begin{align}\label{loc7}
	\frac{1}{180}\Big(398 n_{\text{EM}}+2n_0 -26n_1 +7 n_{1/2}-\frac{233}{2}n_{3/2} \Big) = 3-\mathcal{N}.
\end{align}
Substituting the above relation into the formula \eqref{loc5}, we get
\begin{subequations}\label{loc8}
	\begin{align}
		\Delta S^{\,\mathcal{N}\geq 3\text{,\,Kerr-Newman}}_{\text{BH}} &=(3-\mathcal{N})\frac{(2{b^\prime}^6+5{b^\prime}^4+4{b^\prime}^2+1)}{({b^\prime}^2+1)^2(2{b^\prime}^2+1)} \ln \mathcal{A}_{H}-2 \ln \mathcal{A}_{H},\\
		\Delta S^{\,\mathcal{N}\geq 3\text{,\,Kerr}}_{\text{BH}} &=(1-\mathcal{N})\ln \mathcal{A}_{H},\\
		\Delta S^{\,\mathcal{N}\geq 3\text{,\,Reissner-Nordstr\"om}}_{\text{BH}} &=(4-\mathcal{N})\ln \mathcal{A}_{H}.
	\end{align}
\end{subequations}
All these logarithmic entropy corrections (\cref{loc6,loc8}) exhibit perfect matching with the available direct approach results in \cite{Karan:2020sk,Sen:2012qq,Banerjee:2011pp,Karan:2019sk,Keeler:2014nn,Gupta:2014ns}. The whole process of estimating logarithmic entropy corrections for extremal Kerr-Newman, Kerr and Reissner-Nordstr\"om black holes in $\mathcal{N} \geq 2,d=4$ EMSGTs is an indirect \say{local method}. It is strictly limited to the analysis of extremal black holes via \say{\hyperref[ST1]{Strategy A}}. As compared to the direct approaches, it is much simpler because one needs not required to deal with overly complicated quadratic fluctuated supergravity actions and execute a mountain of complex trace calculations.
\begin{table}[t!]
	\renewcommand{\arraystretch}{1.8}
	\centering 
	\begin{tabular}{|c| c c c c c|} 
		\hline 
		\bm{$\mathcal{N} \geq 3$} \textbf{EMSGTs} & \bm{$n_{\text{EM}}$} & \bm{$n_0$} & \bm{$n_1$} & \bm{$n_{1/2}$} & \bm{$n_{3/2}$}
		\\ [0.7ex]
		\hline\hline 
		$\mathcal{N} = 3,d=4$ EMSGT &1 & 0 & 2 & $\frac{1}{2}$ & 3  \\[1ex]
		$\mathcal{N} = 4,d=4$ EMSGT &1 & 2 & 5 & 2 & 4  \\[1ex]
		
		$\mathcal{N} = 5,d=4$ EMSGT &1 & 10 & 9 & $\frac{11}{2}$ & 5  \\[1ex]
		$\mathcal{N} = 6,d=4$ EMSGT &1 & 30 & 15 & 13 & 6  \\[1ex]
		$\mathcal{N} = 8,d=4$ EMSGT &1 & 70 & 27 & 28 & 8  \\[1ex]
		\hline 
	\end{tabular}
	\caption{Multiplicities of the content in $\mathcal{N} \geq 3,d=4$ EMSGTs. The fraction values in the fifth column arise while extracting out the Dirac spinors from the odd multiplicities of Majorana or Weyl content by diving a 1/2 factor. Note that $n_{\text{EM}}$ already includes one vector field along with the metric.} \label{tab:sugracontent}
\end{table}

\subsection{Summary and conclusions}
To sum up, we have provided a consolidated manual for investigating logarithmic correction to the entropy of Kerr-Newman family of black holes for both extremal and non-extremal limits. The whole framework is divided into two separate strategies (\say{\hyperref[ST1]{Strategy A}} and \say{\hyperref[ST2]{Strategy B}}) based on the Euclidean gravity approaches \cite{Sen:2008wa,Sen:2009wb,Sen:2009wc,Sen:2013ns}. Seeley-DeWitt coefficients are found to be the crucial and common ingredients in these strategies, where a standard method \cite{Vassilevich:2003ll} computes them in various background invariants. Following this manual, we have calculated the first three Seeley-DeWitt coefficients for the fluctuations of the generalized $d=4$ \say{minimally-coupled} EMT and employ them in obtaining logarithmic correction to the entropy of extremal and non-extremal black holes in the theory. The investigation of a global platform for simultaneously calculating logarithmic corrections to the entropy of all the Kerr-Newman family (Kerr-Newman, Kerr, Reissner-Nordstr\"om and Schwarzschild) of black holes in both extremal and non-extremal limits is novel. The extremal black hole results are mostly new reports. The non-supersymmetric \say{minimally-coupled} data, generalized for arbitrary numbers of minimally-coupled fields, have essential utility in alternative derivations of the results for different $\mathcal{N} \geq 2,d=4$ EMSGTs via both direct (e.g., \cite{Karan:2020sk}) and indirect local methods (see \cref{localmethod}). 

This work reported the presence of both the geometric dependent as well as fully universal logarithmic entropy corrections $\Delta S_{\text{BH}}$ for the black hole solutions in the $d=4$ \say{minimally-coupled} EMT. Our analysis showed that the quantum corrections for the charged (Kerr-Newman and Reissner-Nordstr\"om) black holes are geometric via the parameters $Q, b, b^\prime, \beta, r_H$, while $\Delta S_{\text{BH}}$ for the uncharged (Kerr and Schwarzschild) black holes have no dependence on their parameters, i.e., universal in both extremal and non-extremal limits. But, we found an exception for the extremal Reissner-Nordstr\"om black holes that possess a universal form of logarithmic entropy corrections. Almost a similar outcome is exhibited by the extremal black holes in $\mathcal{N} \geq 2,d=4$ EMSGTs. However, the extremal Kerr-Newman results show a slight deviation: universal only in $\mathcal{N}=3$ EMSGT but have geometric dependence in all other $\mathcal{N}\geq 2$ EMSGTs. Any progress in searching the reason behind the typical patterns of $\Delta S_{\text{BH}}$ would be welcome. We have not witnessed any vanishing $\Delta S_{\text{BH}}$ for black holes in the $d=4$ \say{minimally-coupled} EMT, as reported for extremal Reissner-Nordstr\"om black holes in the $\mathcal{N}=4,d=4$ EMSGT. One should not worry about the logarithmic correction contributions that are found to be negative. The total quantum corrected black hole entropy $\big(\frac{\mathcal{A}_{H}}{4G_N}+\Delta S_{\text{BH}}\big)$ is always positive due to the presence of the leading positive Bekenstein-Hawking term $\frac{\mathcal{A}_{H}}{4G_N}$ in the large charge limit. All the calculated $\Delta S_{\text{BH}}$ results as well as their particular characteristics are significant and can be served as `macroscopic experimental data' to understand the microstructure of the general black holes in any quantum theory of gravity in the future.

\acknowledgments

The authors are thankful to Finn Larsen, Ashoke Sen and Antony Charles for providing some invaluable insights that greatly assisted this work. We also acknowledge Gourav Banerjee and R. Nitish for useful discussions. 
\appendix
\section{Local and zero-mode contributions of one-loop effective action}\label{App1}
Zero modes are described by the normalized eigenfunctions $f^0_{i}(x)$ of the kinetic differential operator $\Lambda$ that provide zero eigenvalues ($\lambda_i = 0$),
\begin{align}\label{ap1}
	\Lambda f^0_{i}(x) = 0.
\end{align} 
The total number of zero modes of the operator $\Lambda$ for all the fluctuations $\tilde{\xi}_m$, denoted by $n_{\text{zm}}$, can be introduced by rewriting the heat trace $D(s)$ (defined in \cref{B4}) as
\begin{align}\label{ap2}
	D(s) = \sum_i e^{-s\lambda_i} = \sideset{}{'}\sum_{i\atop (\lambda_i \neq 0)} e^{-s\lambda_i} + n_{\text{zm}},
\end{align}
with
\begin{align}\label{ap3}
	n_{\text{zm}} = \sum_i\int \mathrm{d}^4x \sqrt{\text{det}\thinspace \bar{g}}\thinspace f^0_{i}(x)f^0_{i}(x),
\end{align}
where the prime indicates that the summation is over non-zero modes only.
The quantum corrected Euclidean path integral partition function $\mathcal{Z}_{\text{1-loop}}$ corresponding to the one-loop effective action $\mathcal{W}$ (defined in \cref{B3}) has the following structure \cite{Gibbons:1977ta,Hawking:1978td,Hawking:1977te,Avramidi:1994th,Denardo:1982tb,Peixoto:2001wx} 
\begin{align}\label{ap4}
	\mathcal{Z}_{\text{1-loop}}=e^{-\mathcal{W}} = \int \mathcal{D}[\tilde{\xi}_m]\exp \bigg(- \int \mathrm{d}^4x \sqrt{\text{det}\thinspace \bar{g}}\thinspace \tilde{\xi}_m\Lambda \tilde{\xi}_m\bigg) = (\text{det}\thinspace \Lambda)^{-\chi/2}.
\end{align}
But for the case of zero modes, the functional integral \eqref{ap4} cannot sustain its Gaussian form, and hence one needs to remove the zero modes from $\mathcal{Z}_{\text{1-loop}}$ to carry out the heat kernel treatment of $\mathcal{W}$. However, we can add their contribution back by substituting the zero-mode part of $\mathcal{Z}_{\text{1-loop}}$ with ordinary volume integrals over different asymptotic symmetries that induce the zero modes \cite{Banerjee:2011oo,Banerjee:2011pp,Sen:2012rr,Sen:2012qq}. As a result, the one-loop partition function $\mathcal{Z}_{\text{1-loop}}$ is disintegrated into a product of two separate parts,
\begin{align}\label{ap5}
	\mathcal{Z}_{\text{1-loop}}= e^{-\mathcal{W}} = (\text{det}^\prime\thinspace\Lambda)^{-\chi/2}\cdot \mathcal{Z}_{\mathrm{zero}}(L),
\end{align}
where $\text{det}^\prime\thinspace\Lambda$ is the determinant over non-zero modes of $\Lambda$ and $\mathcal{Z}_{\mathrm{zero}}(L)$ is the zero-mode integral that scales non-trivially with an overall length scale $L$ of the background metric. For this choice of scaling, the non-zero eigenvalues of the Laplace-type operator $\Lambda$ scale as $L^{-2}$, which essentially sets a new rescaled heat kernel parameter $\bar{s} = s/L^2$ of the integration range $\epsilon/L^2 \ll \bar{s} \ll 1$ (or equivalently $\epsilon \ll {s} \ll L^2$). Then, using the Seeley-DeWitt expansion \eqref{B5} in the relation \eqref{B4}, we express the non-zero mode contribution to the one-loop effective action as
\begin{align}\label{ap6}
	\begin{split}
		\mathcal{W}^\prime =\frac{\chi}{2}\thinspace \text{ln}\thinspace \text{det}^\prime\Lambda &= -\frac{\chi}{2}\int_{\epsilon}^{L^2} \frac{\mathrm{d}{s}}{{s}}\Big(D({s})-n_{\text{zm}}\Big)\\
		&= -\frac{1}{2}\bigg(\int \mathrm{d}^4x \sqrt{\text{det}\thinspace \bar{g}}\thinspace a_4(x) - \chi n_{\text{zm}} \bigg)\ln \left(\frac{L^2}{G_N}\right) +\cdots .
	\end{split}
\end{align}
Here \say{$\cdots$} represents all non-logarithmic terms containing the other Seeley-DeWitt coefficients. 
On the other hand, the zero mode contribution to the path integral from all fluctuations $\tilde{\xi}_m$, each having $n^0_{\tilde{\xi}_m}$ zero modes, can be represented as \cite{Banerjee:2011oo,Banerjee:2011pp,Sen:2012rr,Sen:2012qq} 
\begin{align}\label{ap7}
	\mathcal{Z}_{\mathrm{zero}}(L) = L^{ \sum_{\tilde{\xi}_m}\chi \beta_{\tilde{\xi}_m} n^0_{\tilde{\xi}_m}}\mathcal{Z}_0,
\end{align}
where $\mathcal{Z}_0$ does not scale with $L$. $\beta_{\tilde{\xi}_m}$ are numbers that depend on the type of fluctuation and space-time dimensions. In a $d$-dimensional theory, it is found that $\beta_1 =\frac{d-2}{2}$ for vector, $\beta_2 =\frac{d}{2}$ for metric, $\beta_{3/2} = d-1$ for gravitino, etc. Finally, substituting both \eqref{ap6} and \eqref{ap7} contributions in the relation \eqref{ap5} and using  $n_{\text{zm}}=\sum_{\tilde{\xi}_m}n^0_{\tilde{\xi}_m}$, we write the following restructured one-loop effective action form
\begin{align}\label{ap8}
	\mathcal{W} = -\frac{1}{2}\bigg( \int \mathrm{d}^4x \sqrt{\text{det}\thinspace \bar{g}}\thinspace a_4(x)+ \sum_{\tilde{\xi}_m}\chi(\beta_{\tilde{\xi}_m}- 1)n^0_{\tilde{\xi}_m}\bigg) \ln \left(\frac{L^2}{G_N}\right) + \cdots,
\end{align}
where the first part containing $a_4(x)$ coefficient is termed as the \say{local} contribution and the second part controlled by the zero-mode parameters $\beta_{\tilde{\xi}_m},n^0_{\tilde{\xi}_m}$ is recognized as the \say{zero-mode} contribution. We refer the readers to \cite{Bhattacharyya:2012ye,Gupta:2014ns} for more details regarding the above analysis.

If $\mathcal{W}$ is identified as the one-loop quantum effective action to the partition function describing the macroscopic horizon degeneracy of a black hole with horizon area $\mathcal{A}_{H}$, then the form \eqref{ap8} provides the particular logarithmic correction formula \eqref{B7t}. One needs to consider only the terms proportional to $\text{ln}\thinspace \mathcal{A}_{H}$ in the relation $\Delta S_{\text{BH}}= \ln \mathcal{Z}_{\text{1-loop}}= -\mathcal{W}$ for the large-charge limit $\mathcal{A}_{H} \sim L^2$.

\subsection*{Note on the strategy for non-extremal black holes:}
In the Euclidean gravity approach \cite{Sen:2013ns}, the non-extremal black holes are in equilibrium with thermal gas present in the theory. In order to identify only the particular piece of black hole partition function, one must subtract the thermal gas contributions. The special treatment is to consider two black hole solutions (same type but different scaling) in the same theory --  
\begin{itemize}
	\item[] \textbf{System 1:} black hole with the length parameter $a$ confined in a box of size $\zeta$,
	\item[] \textbf{System 2:} black hole with the length parameter $a^\prime$ confined in a box of size $\zeta a^\prime/a$.
\end{itemize}
It is argued in \cite{Sen:2013ns} that the thermal gas contribution remains invariant in both systems and hence, the difference between corresponding one-loop effective actions becomes
\begin{align}
	\Delta \mathcal{W} = \mathcal{W}_1 (\mathrm{BH1} + \mathrm{gas})- \mathcal{W}_2 (\mathrm{BH2} + \mathrm{gas}) = \mathcal{W}_1 (\mathrm{BH1})- \mathcal{W}_2 (\mathrm{BH2}).
\end{align}
Now the eigenvalues of system 2 are scaled in terms of those in system 1 as
\begin{align}
	\lambda^\prime_i = \lambda_i a^2/{a^\prime}^2,
\end{align}
which appropriately fits the $\Delta \mathcal{W}$ into the form \eqref{ap6} for an upper integration limit $\epsilon^\prime =\epsilon/ L^2$ with $L= {a^\prime}/a$. This finally leads us to the same effective action form \eqref{ap8} and logarithmic correction formula \eqref{B7t} for the non-extremal black hole after extracting only the terms proportional to $\ln a^2$. The above note is mostly based on \cite{Sen:2013ns,Charles:2018yey}.\footnote{Refer to section 2.3, appendix B of \cite{Sen:2013ns} and section 2.2.3 of \cite{Charles:2018yey} for the detailed analysis.}

\section{The four-dimensional Einstein-Maxwell background: EOMs and identities}\label{App2}
It is necessary to achieve the Einstein equation for background solutions of the particular four-dimensional Einstein-Maxwell theory \eqref{E1}. At first, we need to reshape the action \eqref{E1} as
\begin{align}
	\begin{split}
		\mathcal{S}_{\text{EM}}= \int \mathrm{d}^4x \sqrt{\text{det}\thinspace {g}} \thinspace \mathcal{R} + \mathcal{S}_{M},
	\end{split}
\end{align}
where $\mathcal{S}_{M}$ is the action for the matter term,
\begin{equation}
	\mathcal{S}_{M}= -\int \mathrm{d}^4x \sqrt{\text{det}\thinspace {g}} \thinspace F_{\rho\sigma}F^{\rho\sigma}.
\end{equation}
The Einstein equation satisfying the classical solution ($\bar{g}_{\mu\nu},\bar{A}_\mu$) is defined as
\begin{align}
	R_{\mu\nu}-\frac{1}{2}\bar{g}_{\mu\nu}R= \frac{1}{2}T_{\mu\nu},
\end{align}
where the energy-momentum tensor $T_{\mu\nu}$ for the matter term is
\begin{align}
	T_{\mu\nu} = -\frac{2}{\sqrt{\text{det}\thinspace {g}}}\frac{\delta \mathcal{S}_{M}}{\delta g^{\mu\nu}}\Bigr\rvert_{(\bar{g}_{\mu\nu},\bar{A}_\mu)}.
\end{align}
Avoiding all the boundary terms, one can obtain
\begin{align}
	\delta \mathcal{S}_{M} &= -\int \mathrm{d}^4x \Big(\delta(\sqrt{\text{det}\thinspace {g}})F_{\rho\sigma}F^{\rho\sigma}+ \sqrt{\text{det}\thinspace {g}}\thinspace \delta (g^{\mu\rho}g^{\nu\sigma}F_{\rho\sigma}F_{\mu\nu})\Big) \nonumber\\
	&=\int \mathrm{d}^4x \sqrt{\text{det}\thinspace {g}}\thinspace \Big(\frac{1}{2}g_{\mu\nu}F_{\rho\sigma}F^{\rho\sigma}- 2 F_{\mu\rho}{F_\nu}^\rho \Big)\delta g^{\mu\nu},
\end{align}
and finally express the particular Einstein equation, 
\begin{equation}\label{trace}
	R_{\mu\nu}-\frac{1}{2}\bar{g}_{\mu\nu}R= 2\bar{F}_{\mu\rho}{\bar{F_\nu}}^\rho-\frac{1}{2}\bar{g}_{\mu\nu}\bar{F}_{\rho\sigma}\bar{F}^{\rho\sigma},
\end{equation}
where $\bar{F}_{\mu\nu}= \partial_\mu \bar{A}_\nu-\partial_\nu \bar{A}_\mu$. Trace over \eqref{trace} allows us to consider the constraint $R=0$ for the Einstein-Maxwell theory. Therefore all the terms proportional to the background Ricci scalar $R$ vanish, and this argument is used throughout this paper. For the four-dimensional Einstein-Maxwell background, the relevant Riemannian identities, Bianchi identities and equations of motion are listed as
{
	\allowdisplaybreaks
	\begin{align}\label{process2}
		\bar{F}_{\mu\rho}{\bar{F_\nu}}^\rho &= \frac{1}{2}R_{\mu\nu}+\frac{1}{4}\bar{g}_{\mu\nu}\bar{F}_{\rho\sigma}\bar{F}^{\rho\sigma}, \nonumber\\
		R&=0, \enspace R_{\mu[\nu\rho\sigma]}=0, \nonumber\\
		D^\mu \bar{F}_{\mu\nu} &=0,\enspace D_{[\mu}\bar{F}_{\nu\rho]}=0,\\
		(D_\rho \bar{F}_{\mu\nu})(D^\rho \bar{F}^{\mu\nu}) &= R_{\mu\nu\rho\sigma}\bar{F}^{\mu\nu}\bar{F}^{\rho\sigma}-R_{\mu\nu}R^{\mu\nu},\nonumber\\
		(D_\mu \bar{{F}_\rho}^\nu)(D_\nu \bar{F}^{\rho\mu}) &= \frac{1}{2}R_{\mu\nu\rho\sigma}\bar{F}^{\mu\nu}\bar{F}^{\rho\sigma}-\frac{1}{2}R_{\mu\nu}R^{\mu\nu}.\nonumber
\end{align}}

\section{Trace calculations for the pure Einstein-Maxwell sector}\label{App3}
The trace calculations for obtaining the Seeley-DeWitt contribution of the Einstein-Maxwell sector \eqref{pem25} are extremely tedious and complex compared to the minimally-coupled fields' contributions. For the pure Einstein-Maxwell fluctuations $\tilde{\xi}_m = \lbrace h_{\mu\nu}, a_\mu \rbrace $, we set up the following traces
{
	\allowdisplaybreaks
	\begin{subequations}\label{a11}
		\begin{align}
			\text{tr}(E) &= \text{tr}\Big(E\indices{^{h_{\mu\nu}}_{h_{\alpha\beta}}} + E\indices{^{a_\alpha}_{a_\beta}} + E\indices{^{h_{\mu\nu}}_{a_{\alpha}}} + E\indices{^{a_{\alpha}}_{h_{\mu\nu}}} \Big),\\
			\text{tr}(E^2) &= \text{tr}\Big(E\indices{^{h_{\mu\nu}}_{h_{\theta\phi}}}E\indices{^{h_{\theta\phi}}_{h_{\alpha\beta}}}+ E\indices{^{a_\alpha}_{a_\gamma}}E\indices{^{a_\gamma}_{a_\beta}}\nonumber\\
			&\qquad\quad + E\indices{^{h_{\mu\nu}}_{a_{\theta}}}E\indices{^{a_{\theta}}_{h_{\alpha\beta}}}+ E\indices{^{a_{\alpha}}_{h_{\mu\nu}}}E\indices{^{h_{\mu\nu}}_{a_{\beta}}} \Big),\\
			\text{tr}(\Omega_{\rho\sigma}\Omega^{\rho\sigma}) &= \text{tr}\Big((\Omega_{\rho\sigma})\indices{^{h_{\mu\nu}}_{h_{\theta\phi}}}(\Omega^{\rho\sigma})\indices{^{h_{\theta\phi}}_{h_{\alpha\beta}}}+ (\Omega_{\rho\sigma})\indices{^{a_\alpha}_{a_\gamma}}(\Omega^{\rho\sigma})\indices{^{a_\gamma}_{a_\beta}}\nonumber\\
			&\qquad\quad + (\Omega_{\rho\sigma})\indices{^{h_{\mu\nu}}_{a_{\theta}}}(\Omega^{\rho\sigma})\indices{^{a_{\theta}}_{h_{\alpha\beta}}}+ (\Omega_{\rho\sigma})\indices{^{a_{\alpha}}_{h_{\mu\nu}}}(\Omega^{\rho\sigma})\indices{^{h_{\mu\nu}}_{a_{\beta}}}\Big).
		\end{align}
	\end{subequations}
}
Simplifying \cref{pem27} further, the following components of $E$ are extracted as
{
	\allowdisplaybreaks
	\begin{subequations}
		\begin{align}
			E\indices{^{h_{\mu\nu}}_{h_{\alpha\beta}}} &= \underbrace{R\indices{^\mu_\alpha^\nu_\beta} + R\indices{^\mu_\beta^\nu_\alpha} -\bar{g}^{\mu\nu}R_{\alpha\beta}}_{X_1}, \\
			E\indices{^{a_\alpha}_{a_\beta}} &= \underbrace{\frac{3}{2}\bar{g}^\alpha_\beta \bar{F}_{\mu\nu}\bar{F}^{\mu\nu}}_{X_2},\\
			E\indices{^{h_{\mu\nu}}_{a_{\alpha}}} &= \underbrace{\frac{1}{\sqrt{2}}D^\mu \bar{F}\indices{_\alpha^\nu}+\frac{1}{\sqrt{2}} D^\nu \bar{F}\indices{_\alpha^\mu}}_{X_3},\\
			E\indices{^{a_{\alpha}}_{h_{\mu\nu}}} &= \underbrace{\frac{1}{\sqrt{2}} D_\mu \bar{F}\indices{^\alpha_\nu}+ \frac{1}{\sqrt{2}}D_\nu \bar{F}\indices{^\alpha_\mu}}_{X_4}. 
		\end{align}
	\end{subequations}
}
Also, all the components of $\Omega_{\rho\sigma}$ are extracted from \cref{pem28} as
{
	\allowdisplaybreaks
	\begin{subequations}
		\begin{align}
			(\Omega_{\rho\sigma})\indices{^{h_{\mu\nu}}_{h_{\alpha\beta}}} &= \underbrace{\frac{1}{2}(\bar{g}^\mu_\alpha R\indices{^\nu_\beta_\rho_\sigma}+ \bar{g}^\mu_\beta R\indices{^\nu_\alpha_\rho_\sigma}+ \bar{g}^\nu_\alpha R\indices{^\mu_\beta_\rho_\sigma} + \bar{g}^\nu_\beta R\indices{^\mu_\alpha_\rho_\sigma})}_{Y_1} \nonumber\\
			&\quad-\frac{1}{2}(\bar{g}^{\mu\theta}\bar{F}\indices{_\rho^\nu}+\bar{g}^{\nu\theta}\bar{F}\indices{_\rho^\mu}-\bar{g}^\mu_\rho\bar{F}^{\theta\nu}-\bar{g}^\nu_\rho\bar{F}^{\theta\mu}- \bar{g}^{\mu\nu}\bar{F}\indices{_\rho^\theta})\nonumber \\
			&\qquad \underbrace{\times (\bar{g}_{\alpha\theta}\bar{F}_{\sigma\beta} + \bar{g}_{\beta\theta}\bar{F}_{\sigma\alpha}-\bar{g}_{\alpha\sigma}\bar{F}_{\theta\beta}-\bar{g}_{\beta\sigma}\bar{F}_{\theta\alpha}-\bar{g}_{\alpha\beta}\bar{F}_{\sigma\theta}) }_{Y_2}\nonumber \\
			&\quad+\frac{1}{2}(\bar{g}^{\mu\theta}\bar{F}\indices{_\sigma^\nu}+\bar{g}^{\nu\theta}\bar{F}\indices{_\sigma^\mu}-\bar{g}^\mu_\sigma\bar{F}^{\theta\nu}-\bar{g}^\nu_\sigma\bar{F}^{\theta\mu}- \bar{g}^{\mu\nu}\bar{F}\indices{_\sigma^\theta})\nonumber \\
			&\qquad \underbrace{\times (\bar{g}_{\alpha\theta}\bar{F}_{\rho\beta} + \bar{g}_{\beta\theta}\bar{F}_{\rho\alpha}-\bar{g}_{\alpha\rho}\bar{F}_{\theta\beta}-\bar{g}_{\beta\rho}\bar{F}_{\theta\alpha}-\bar{g}_{\alpha\beta}\bar{F}_{\rho\theta}) }_{Y_3}, \\
			(\Omega_{\rho\sigma})\indices{^{a_\alpha}_{a_\beta}} &= \underbrace{R\indices{^\alpha_{\beta\rho\sigma}}}_{Y_4}\nonumber\\
			&\quad-\frac{1}{2}(\bar{g}^{\mu\alpha}\bar{F}\indices{_\rho^\nu}+\bar{g}^{\nu\alpha}\bar{F}\indices{_\rho^\mu}-\bar{g}^{\mu}_{\rho}\bar{F}\indices{^\alpha^\nu}-\bar{g}^{\nu}_{\rho}\bar{F}^{\alpha\mu}-\bar{g}^{\mu\nu}\bar{F}\indices{_\rho^\alpha}) \nonumber\\
			&\qquad \underbrace{\times(\bar{g}_{\mu\beta}\bar{F}_{\sigma\nu}+\bar{g}_{\nu\beta}\bar{F}_{\sigma\mu}-\bar{g}_{\mu\sigma}\bar{F}_{\beta\nu}-\bar{g}_{\nu\sigma}\bar{F}_{\beta\mu}-\bar{g}_{\mu\nu}\bar{F}_{\sigma\beta})}_{Y_5} \nonumber\\
			&\quad+\frac{1}{2}(\bar{g}^{\mu\alpha}\bar{F}\indices{_\sigma^\nu}+\bar{g}^{\nu\alpha}\bar{F}\indices{_\sigma^\mu}-\bar{g}^{\mu}_{\sigma}\bar{F}\indices{^\alpha^\nu}-\bar{g}^{\nu}_{\sigma}\bar{F}^{\alpha\mu}-\bar{g}^{\mu\nu}\bar{F}\indices{_\sigma^\alpha}) \nonumber\\
			&\qquad \underbrace{\times(\bar{g}_{\mu\beta}\bar{F}_{\rho\nu}+\bar{g}_{\nu\beta}\bar{F}_{\rho\mu}-\bar{g}_{\mu\rho}\bar{F}_{\beta\nu}-\bar{g}_{\nu\rho}\bar{F}_{\beta\mu}-\bar{g}_{\mu\nu}\bar{F}_{\rho\beta})}_{Y_6}, \\
			(\Omega_{\rho\sigma})\indices{^{h_{\mu\nu}}_{a_{\alpha}}} &= \underbrace{\frac{1}{\sqrt{2}}(\bar{g}^{\mu}_{\alpha}D_\rho \bar{F}\indices{_\sigma^\nu} + \bar{g}^{\nu}_{\alpha}D_\rho \bar{F}\indices{_\sigma^\mu}-\bar{g}^{\mu}_{\sigma}D_\rho\bar{F}\indices{_\alpha^\nu}-\bar{g}^{\nu}_{\sigma}D_\rho\bar{F}\indices{_\alpha^\mu}-\bar{g}^{\mu\nu}D_\rho\bar{F}_{\sigma\alpha})}_{Y_7}\nonumber\\
			&\quad  \underbrace{-\frac{1}{\sqrt{2}}(\bar{g}^{\mu}_{\alpha}D_\sigma \bar{F}\indices{_\rho^\nu} + \bar{g}^{\nu}_{\alpha}D_\sigma \bar{F}\indices{_\rho^\mu}-\bar{g}^{\mu}_{\rho}D_\sigma\bar{F}\indices{_\alpha^\nu}-\bar{g}^{\nu}_{\rho}D_\sigma\bar{F}\indices{_\alpha^\mu}-\bar{g}^{\mu\nu}D_\sigma\bar{F}_{\rho\alpha})}_{Y_8},\\
			(\Omega_{\rho\sigma})\indices{^{a_{\alpha}}_{h_{\mu\nu}}} &= \underbrace{-\frac{1}{\sqrt{2}}(\bar{g}^{\alpha}_{\mu}D_\rho \bar{F}_{\sigma\nu}+ \bar{g}^{\alpha}_{\nu}D_\rho \bar{F}_{\sigma\mu}-\bar{g}_{\mu\sigma}D_\rho \bar{F}\indices{^\alpha_\nu}-\bar{g}_{\nu\sigma}D_\rho\bar{F}\indices{^\alpha_\mu}-\bar{g}_{\mu\nu}D_\rho\bar{F}\indices{_\sigma^\alpha})}_{Y_9}\nonumber\\
			&\enspace \underbrace{+\frac{1}{\sqrt{2}}(\bar{g}^{\alpha}_{\mu}D_\sigma \bar{F}_{\rho\nu}+ \bar{g}^{\alpha}_{\nu}D_\sigma \bar{F}_{\rho\mu}-\bar{g}_{\mu\rho}D_\sigma \bar{F}\indices{^\alpha_\nu}-\bar{g}_{\nu\rho}D_\sigma\bar{F}\indices{^\alpha_\mu}-\bar{g}_{\mu\nu}D_\sigma\bar{F}\indices{_\rho^\alpha})}_{Y_{10}}. 
		\end{align}
	\end{subequations}
}
To calculate different needful traces, we employ the following definitions
\begin{align}\label{process1}
	\begin{split}
		A\indices{^{\tilde{\xi}_m}_{\tilde{\xi}_n}} &= A^{\tilde{\xi}_m \tilde{\xi}_p}I_{\tilde{\xi}_p \tilde{\xi}_n},\\
		\mathrm{tr}(A) &= A\indices{^{\tilde{\xi}_m}_{\tilde{\xi}_m}} = A^{\tilde{\xi}_m \tilde{\xi}_p}I_{\tilde{\xi}_p \tilde{\xi}_m},\\
		\mathrm{tr}(A^2) &= A\indices{^{\tilde{\xi}_m}_{\tilde{\xi}_n}}A\indices{^{\tilde{\xi}_n}_{\tilde{\xi}_m}}= \left(A^{\tilde{\xi}_m \tilde{\xi}_p}I_{\tilde{\xi}_p \tilde{\xi}_n}\right)\left(A^{\tilde{\xi}_n \tilde{\xi}_q}I_{\tilde{\xi}_q \tilde{\xi}_m}\right),\\
		\mathrm{tr}(AB) &= A\indices{^{\tilde{\xi}_m}_{\tilde{\xi}_n}}B\indices{^{\tilde{\xi}_n}_{\tilde{\xi}_m}}= \left(A^{\tilde{\xi}_m \tilde{\xi}_p}I_{\tilde{\xi}_p \tilde{\xi}_n}\right)\left(B^{\tilde{\xi}_n \tilde{\xi}_q}I_{\tilde{\xi}_q \tilde{\xi}_m}\right)=\mathrm{tr}(BA),
	\end{split}
\end{align}
where $I^{\tilde{\xi}_m \tilde{\xi}_n}$ acts as an effective metric corresponding to any two arbitrary matrices $A^{\tilde{\xi}_m \tilde{\xi}_n}$ and $B^{\tilde{\xi}_m \tilde{\xi}_n}$. The traces can be simplified and reduced to different background invariants with the help of identities \eqref{process2}. Following the systematic process \eqref{process1} and the relations \eqref{process2}, we then calculate,
{
	\allowdisplaybreaks
	\begin{equation}
		\begin{split}
			\text{tr}(X_1) &=  0,\\
			\text{tr}(X_2) &= 6\bar{F}_{\mu\nu}\bar{F}^{\mu\nu},\\
			\text{tr}(X_3) &= 0,\\
			\text{tr}(X_4) &= 0,\\
			\text{tr}(X_1X_1) &= 3R_{\mu\nu\rho\sigma}R^{\mu\nu\rho\sigma}-4R_{\mu\nu}R^{\mu\nu},\\
			\text{tr}(X_2X_2) &=  9(\bar{F}_{\mu\nu}\bar{F}^{\mu\nu})^2,\\
			\text{tr}(X_3X_4) &=  \frac{3}{2}R_{\mu\nu\rho\sigma}\bar{F}^{\mu\nu}\bar{F}^{\rho\sigma}-\frac{3}{2}R_{\mu\nu}R^{\mu\nu},\\
			\text{tr}(Y_1Y_1) &=  -6R_{\mu\nu\rho\sigma}R^{\mu\nu\rho\sigma},\\
			\text{tr}(Y_2Y_2) &=  7R_{\mu\nu}R^{\mu\nu}-3(\bar{F}_{\mu\nu}\bar{F}^{\mu\nu})^2,\\
			\text{tr}(Y_3Y_3) &=  7R_{\mu\nu}R^{\mu\nu}-3(\bar{F}_{\mu\nu}\bar{F}^{\mu\nu})^2,\\
			\text{tr}(Y_4Y_4) &=  -R_{\mu\nu\rho\sigma}R^{\mu\nu\rho\sigma},\\
			\text{tr}(Y_5Y_5) &=  7R_{\mu\nu}R^{\mu\nu}-3(\bar{F}_{\mu\nu}\bar{F}^{\mu\nu})^2,
		\end{split}
		\hspace{0.10in}
		\begin{split}
			\text{tr}(Y_6Y_6) &=  7R_{\mu\nu}R^{\mu\nu}-3(\bar{F}_{\mu\nu}\bar{F}^{\mu\nu})^2,\\
			\text{tr}(Y_1Y_2) &=  0,\\
			\text{tr}(Y_1Y_3) &=  0,\\
			\text{tr}(Y_2Y_3) &=  -R_{\mu\nu}R^{\mu\nu}-9(\bar{F}_{\mu\nu}\bar{F}^{\mu\nu})^2,\\
			\text{tr}(Y_4Y_5) &=  R_{\mu\nu}R^{\mu\nu},\\
			\text{tr}(Y_4Y_6) &=  R_{\mu\nu}R^{\mu\nu},\\
			\text{tr}(Y_5Y_6) &=  4R_{\mu\nu}R^{\mu\nu}-12(\bar{F}_{\mu\nu}\bar{F}^{\mu\nu})^2,\\
			\text{tr}(Y_7Y_9) &=  6R_{\mu\nu}R^{\mu\nu}-6R_{\mu\nu\rho\sigma}\bar{F}^{\mu\nu}\bar{F}^{\rho\sigma},\\
			\text{tr}(Y_7Y_{10}) &= \frac{3}{2}R_{\mu\nu\rho\sigma}\bar{F}^{\mu\nu}\bar{F}^{\rho\sigma}-\frac{3}{2}R_{\mu\nu}R^{\mu\nu},\\
			\text{tr}(Y_8Y_9) &= \frac{3}{2}R_{\mu\nu\rho\sigma}\bar{F}^{\mu\nu}\bar{F}^{\rho\sigma}-\frac{3}{2}R_{\mu\nu}R^{\mu\nu},\\
			\text{tr}(Y_8Y_{10}) &= 6R_{\mu\nu}R^{\mu\nu}-6R_{\mu\nu\rho\sigma}\bar{F}^{\mu\nu}\bar{F}^{\rho\sigma},
		\end{split}
	\end{equation}
}
and find the following results
{
	\allowdisplaybreaks
	\begin{align}
		\text{tr}\Big(E\indices{^{h_{\mu\nu}}_{h_{\alpha\beta}}}\Big) &=\text{tr}(X_1)=0,\enspace \text{tr}\Big(E\indices{^{a_\alpha}_{a_\beta}}\Big)=\text{tr}(X_2) =6\bar{F}_{\mu\nu}\bar{F}^{\mu\nu}, \nonumber\\
		\text{tr}\Big(E\indices{^{h_{\mu\nu}}_{a_{\alpha}}}\Big)&=\text{tr}(X_3)=0, \enspace \text{tr}\Big( E\indices{^{a_{\alpha}}_{h_{\mu\nu}}} \Big)= \text{tr}(X_4)=0,\nonumber\\
		\text{tr}\Big(E\indices{^{h_{\mu\nu}}_{h_{\theta\phi}}}E\indices{^{h_{\theta\phi}}_{h_{\alpha\beta}}}\Big) &= \text{tr}({X_1}^2)= 3R_{\mu\nu\rho\sigma}R^{\mu\nu\rho\sigma}-4 R_{\mu\nu}R^{\mu\nu},\nonumber\\
		\text{tr}\Big( E\indices{^{a_\alpha}_{a_\gamma}}E\indices{^{a_\gamma}_{a_\beta}}\Big) &= \text{tr}({X_2}^2)= 9(\bar{F}_{\mu\nu}\bar{F}^{\mu\nu})^2,\nonumber\\
		\text{tr}\Big( E\indices{^{h_{\mu\nu}}_{a_{\theta}}}E\indices{^{a_{\theta}}_{h_{\alpha\beta}}}\Big) &= \text{tr}\Big( E\indices{^{a_{\alpha}}_{h_{\mu\nu}}}E\indices{^{h_{\mu\nu}}_{a_{\beta}}} \Big) \nonumber \\
		& = \text{tr}(X_3X_4)= \frac{3}{2}R_{\mu\nu\rho\sigma}\bar{F}^{\mu\nu}\bar{F}^{\rho\sigma}-\frac{3}{2}R_{\mu\nu}R^{\mu\nu},\nonumber\\
		\text{tr}\Big((\Omega_{\rho\sigma})\indices{^{h_{\mu\nu}}_{h_{\theta\phi}}}(\Omega^{\rho\sigma})\indices{^{h_{\theta\phi}}_{h_{\alpha\beta}}}\Big) &= \sum_{i,j=1}^3 \text{tr}(Y_i Y_j)\\
		& = -6R_{\mu\nu\rho\sigma}R^{\mu\nu\rho\sigma}+12 R_{\mu\nu}R^{\mu\nu}-24(\bar{F}_{\mu\nu}\bar{F}^{\mu\nu})^2,\nonumber\\
		\text{tr}\Big((\Omega_{\rho\sigma})\indices{^{a_\alpha}_{a_\gamma}}(\Omega^{\rho\sigma})\indices{^{a_\gamma}_{a_\beta}}\Big) &=\sum_{i,j=4}^6 \text{tr}(Y_i Y_j)\nonumber\\
		& =-R_{\mu\nu\rho\sigma}R^{\mu\nu\rho\sigma}+26 R_{\mu\nu}R^{\mu\nu}-30(\bar{F}_{\mu\nu}\bar{F}^{\mu\nu})^2 ,\nonumber\\
		\text{tr}\Big((\Omega_{\rho\sigma})\indices{^{h_{\mu\nu}}_{a_{\theta}}}(\Omega^{\rho\sigma})\indices{^{a_{\theta}}_{h_{\alpha\beta}}}\Big) &= \text{tr}\Big((\Omega_{\rho\sigma})\indices{^{a_{\alpha}}_{h_{\mu\nu}}}(\Omega^{\rho\sigma})\indices{^{h_{\mu\nu}}_{a_{\beta}}}\Big)\nonumber\\
		&= \sum_{i=7}^8\sum_{j=9}^{10} \text{tr}(Y_i Y_j)= 9R_{\mu\nu}R^{\mu\nu}-9R_{\mu\nu\rho\sigma}\bar{F}^{\mu\nu}\bar{F}^{\rho\sigma}.\nonumber
\end{align}}

{The above results produce the exact list} \eqref{pem29} via the definitions \eqref{a11}.



\begin{thebibliography}{99}



\bibitem{Solodukhin:1995na} S.N. Solodukhin, \emph{The conical singularity and quantum corrections to entropy of black hole}, \href{https://journals.aps.org/prd/abstract/10.1103/PhysRevD.51.609}{\emph{Phys. Rev.} \textbf{D51} (1995) 609} [\href{https://arxiv.org/abs/hep-th/9407001}{arXiv:hep-th/9407001}] [\href{https://inspirehep.net/search?p=find+eprint+hep-th/9407001}{\scshape{in}SPIRE}].

\bibitem{Solodukhin:1995nb} S.N. Solodukhin, \emph{\enquote{Nongeometric} contribution to the entropy of a black hole due to quantum corrections}, \href{https://journals.aps.org/prd/abstract/10.1103/PhysRevD.51.618}{\emph{Phys. Rev.} \textbf{D51} (1995) 618} [\href{https://arxiv.org/abs/hep-th/9408068}{arXiv:hep-th/9408068}][\href{https://inspirehep.net/search?p=find+eprint+hep-th/9408068}{\scshape{in}SPIRE}].

\bibitem{Fursaev:1995df} D.V. Fursaev, \emph{Temperature and entropy of a quantum black hole and conformal anomaly}, \href{https://journals.aps.org/prd/abstract/10.1103/PhysRevD.51.R5352}{\emph{Phys. Rev.} \textbf{D51} (1995) R5352} [\href{https://arxiv.org/abs/hep-th/9412161}{arXiv:hep-th/9412161}] [\href{https://inspirehep.net/search?p=find+eprint+hep-th/9412161}{\scshape{in}SPIRE}].

\bibitem{Mavromatos:1996kc} N.E. Mavromatos and E. Winstanley, \emph{Aspects of hairy black holes in spontaneously broken Einstein-Yang-Mills systems: Stability analysis and entropy considerations}, \href{https://journals.aps.org/prd/abstract/10.1103/PhysRevD.53.3190}{\emph{Phys.Rev.} \textbf{D53} (1996) 3190} [\href{https://inspirehep.net/record/400168}{arXiv:hep-th/9510007}] [\href{https://inspirehep.net/record/400168}{\scshape{in}SPIRE}].

\bibitem{Mann:1996bi} R.B. Mann and S.N. Solodukhin, \emph{Conical geometry and quantum entropy of a charged Kerr black hole}, \href{https://journals.aps.org/prd/abstract/10.1103/PhysRevD.54.3932}{\emph{Phys. Rev.} \textbf{D54} (1996) 3932} [\href{https://arxiv.org/abs/hep-th/9604118}{arXiv:hep-th/9604118}] [\href{https://inspirehep.net/search?p=find+eprint+hep-th/9604118}{\scshape{in}SPIRE}].

\bibitem{Mann:1998hm} R.B. Mann and S.N. Solodukhin, \emph{Universality of quantum entropy for extreme black holes}, \href{https://www.sciencedirect.com/science/article/pii/S0550321398000947?via%3Dihub}{\emph{Nucl. Phys.} \textbf{B523} (1998) 293} [\href{https://arxiv.org/abs/hep-th/9709064}{arXiv:hep-th/9709064}] [\href{https://inspirehep.net/search?p=find+eprint+hep-th/9709064}{\scshape{in}SPIRE}].
	
\bibitem{Kaul:2000rk} R.K. Kaul and P. Majumdar, \emph{Logarithmic correction to the Bekenstein-Hawking entropy}, \href{https://journals.aps.org/prl/abstract/10.1103/PhysRevLett.84.5255}{\emph{Phys. Rev. Lett.} \textbf{84} (2000) 5255-5257} [\href{https://arxiv.org/abs/gr-qc/0002040}{arXiv:gr-qc/0002040}] [\href{https://inspirehep.net/search?p=find+eprint+gr-qc/0002040}{\scshape{in}SPIRE}].	
	
\bibitem{Carlip:2000nv} S. Carlip, \emph{Logarithmic corrections to black hole entropy, from the Cardy formula}, \href{https://iopscience.iop.org/article/10.1088/0264-9381/17/20/302}{\emph{Class. Quant. Grav.} \textbf{17} (2000) 4175} [\href{https://arxiv.org/abs/gr-qc/0005017}{arXiv:gr-qc/0005017}] [\href{https://inspirehep.net/search?p=find+eprint+gr-qc/0005017}{\scshape{in}SPIRE}].
	
\bibitem{Govindarajan:2001ee} T.R. Govindarajan, R.K. Kaul and V. Suneeta, \emph{Logarithmic correction to the Bekenstein-Hawking entropy of the BTZ black hole}, \href{https://iopscience.iop.org/article/10.1088/0264-9381/18/15/303}{\emph{Class. Quant. Grav.} \textbf{18} (2001) 2877} [\href{https://arxiv.org/abs/gr-qc/0104010}{arXiv:gr-qc/0104010}] [\href{https://inspirehep.net/search?p=find+eprint+gr-qc/0104010}{\scshape{in}SPIRE}].
	
\bibitem{Gupta:2002bg} K.S. Gupta and S. Sen, \emph{Further evidence for the conformal structure of a Schwarzschild black hole in an algebraic approach}, \href{https://www.sciencedirect.com/science/article/pii/S0370269301015015?via%3Dihub}{\emph{Phys. Lett.} \textbf{B526} (2002) 121} [\href{https://arxiv.org/abs/hep-th/0112041}{arXiv:hep-th/0112041}] [\href{https://inspirehep.net/search?p=find+eprint+hep-th/0112041}{\scshape{in}SPIRE}].
	
\bibitem{Mukherji:2002de} S. Mukherji and S.S. Pal, \emph{Logarithmic corrections to black hole entropy and AdS/CFT correspondence}, \href{https://iopscience.iop.org/article/10.1088/1126-6708/2002/05/026}{\emph{JHEP} \textbf{05} (2002) 026} [\href{https://arxiv.org/abs/hep-th/0205164}{arXiv:hep-th/0205164}] [\href{https://inspirehep.net/literature?sort=mostrecent&size=25&page=1&q=find%20eprint%20hep-th%2F0205164}{\scshape{in}SPIRE}].	
		
\bibitem{Medved:2004eh} A.J.M. Medved, \emph{A comment on black hole entropy or does Nature abhor a logarithm?}, \href{https://iopscience.iop.org/article/10.1088/0264-9381/22/1/009}{\emph{Class. Quant. Grav.} \textbf{22} (2005) 133} [\href{https://arxiv.org/abs/gr-qc/0406044}{arXiv:gr-qc/0406044}] [\href{https://inspirehep.net/search?p=find+eprint+gr-qc/0406044}{\scshape{in}SPIRE}].
		
\bibitem{Page:2005xp} D.N. Page, \emph{Hawking radiation and black hole thermodynamics}, \href{https://iopscience.iop.org/article/10.1088/1367-2630/7/1/203}{\emph{New J. Phys.} \textbf{7} (2005) 203} [\href{https://arxiv.org/abs/hep-th/0409024}{arXiv:hep-th/0409024}] [\href{https://inspirehep.net/search?p=find+eprint+hep-th/0409024}{\scshape{in}SPIRE}].
		
\bibitem{Banerjee:2008cf} R. Banerjee and B.R. Majhi, \emph{Quantum tunneling beyond semiclassical approximation}, \href{https://iopscience.iop.org/article/10.1088/1126-6708/2008/06/095}{\emph{JHEP} \textbf{06} (2008) 095} [\href{https://arxiv.org/abs/0805.2220}{arXiv:0805.2220}] [\href{https://inspirehep.net/search?p=find+eprint+0805.2220}{\scshape{in}SPIRE}].	
		
\bibitem{Banerjee:2009fz} R. Banerjee and B.R. Majhi, \emph{Quantum tunneling and trace anomaly}, \href{https://www.sciencedirect.com/science/article/pii/S0370269309002834?via%3Dihub}{\emph{Phys. Lett.} \textbf{B674} (2009) 218} [\href{https://arxiv.org/abs/0808.3688}{arXiv:0808.3688}] [\href{https://inspirehep.net/search?p=find+eprint+0808.3688}{\scshape{in}SPIRE}].
			
\bibitem{Majhi:2009gi} B.R. Majhi, \emph{Fermion tunneling beyond semiclassical approximation}, \href{https://journals.aps.org/prd/abstract/10.1103/PhysRevD.79.044005}{\emph{Phys. Rev.} \textbf{D79} (2009) 044005} [\href{https://arxiv.org/abs/0809.1508}{arXiv:0809.1508}] [\href{https://inspirehep.net/search?p=find+eprint+0809.1508}{\scshape{in}SPIRE}].
			
\bibitem{Cai:2010ua} R.G. Cai, L.M. Cao and N. Ohta, \emph{Black holes in gravity with conformal anomaly and logarithmic term in black hole entropy}, \href{https://link.springer.com/article/10.1007%2FJHEP04%282010%29082}{\emph{JHEP} \textbf{04} (2010) 082} [\href{https://arxiv.org/abs/0911.4379}{arXiv:0911.4379}] [\href{https://inspirehep.net/search?p=find+eprint+0911.4379}{\scshape{in}SPIRE}].	
				
\bibitem{Aros:2010jb} R. Aros, D.E. Diaz and A. Montecinos, \emph{Logarithmic correction to BH entropy as Noether charge}, \href{https://link.springer.com/article/10.1007%2FJHEP07%282010%29012}{\emph{JHEP} \textbf{07} (2010) 012} [\href{https://arxiv.org/abs/1003.1083}{arXiv:1003.1083}] [\href{https://inspirehep.net/search?p=find+eprint+1003.1083}{\scshape{in}SPIRE}].	
					
\bibitem{Solodukhin:2010pk} S.N. Solodukhin, \emph{Entanglement entropy of round spheres}, \href{https://www.sciencedirect.com/science/article/pii/S0370269310010932?via%3Dihub}{\emph{Phys. Lett.} \textbf{B693} (2010) 605} [\href{https://arxiv.org/abs/1008.4314}{arXiv:1008.4314}] [\href{https://inspirehep.net/search?p=find+eprint+1008.4314}{\scshape{in}SPIRE}].
						
\bibitem{Banerjee:2011oo} S. Banerjee, R.K. Gupta and A. Sen, \emph{Logarithmic corrections to extremal black hole entropy from quantum entropy function},  \href{https://link.springer.com/article/10.1007%2FJHEP03%282011%29147}
{\emph{JHEP} \textbf{03} (2011) 147} [\href{https://arxiv.org/abs/1005.3044}{arXiv:1005.3044}] [\href{https://inspirehep.net/search?p=find+EPRINT+arXiv:1005.3044}{\scshape{in}SPIRE}].
							
\bibitem{Banerjee:2011pp} S. Banerjee, R. K. Gupta, I. Mandal and A. Sen, \emph{Logarithmic corrections to N=4 and N=8 black hole entropy: a one loop test of quantum gravity}, \href{https://link.springer.com/article/10.1007%2FJHEP11%282011%29143}
{\emph{JHEP} \textbf{11} (2011) 143} [\href{https://arxiv.org/abs/1106.0080}{arXiv:1106.0080}] [\href{https://inspirehep.net/search?p=find+EPRINT+arXiv:1106.0080}{\scshape{in}SPIRE}].
								
\bibitem{Sen:2012qq} A. Sen, \emph{Logarithmic corrections to N=2 black hole entropy: an infrared window into the microstates}, \href{https://link.springer.com/article/10.1007%2Fs10714-012-1336-5}
{\emph{Gen. Rel. Grav.} \textbf{44} (2012) 1207} [\href{https://arxiv.org/abs/1108.3842}{arXiv:1108.3842}] [\href{https://inspirehep.net/search?p=find+EPRINT+arXiv:1108.3842}{\scshape{in}SPIRE}].
									
									
\bibitem{Gupta:2014ns} R.K. Gupta, S. Lal and S. Thakur, \emph{Logarithmic corrections to extremal black hole entropy in N = 2, 4 and 8 supergravity}, \href{https://link.springer.com/article/10.1007%2FJHEP11%282014%29072}
{\emph{JHEP} \textbf{11} (2014) 072} [\href{https://arxiv.org/abs/1402.2441}{arXiv:1402.2441}] [\href{https://inspirehep.net/search?p=find+EPRINT+arXiv:1402.2441}{\scshape{in}SPIRE}].
										
\bibitem{Keeler:2014nn} C. Keeler, F. Larsen and P. Lisbao, \emph{Logarithmic corrections to $\mathcal{N}\geq 2$ black hole entropy}, \href{https://journals.aps.org/prd/abstract/10.1103/PhysRevD.90.043011}{\emph{Phys. Rev.} \textbf{D90} (2014) 043011} [\href{https://arxiv.org/abs/1404.1379}{arXiv:1404.1379}] [\href{https://inspirehep.net/search?p=find+EPRINT+arXiv:1404.1379}{\scshape{in}SPIRE}].
										
\bibitem{Larsen:2015nx} F. Larsen and P. Lisbao, \emph{Quantum corrections to supergravity on $AdS_2 \times S^2$}, \href{https://journals.aps.org/prd/abstract/10.1103/PhysRevD.91.084056}{\emph{Phys. Rev.} \textbf{D91} (2015) 084056} [\href{https://arxiv.org/abs/1411.7423}{arXiv:1411.7423}] [\href{https://inspirehep.net/search?p=find+EPRINT+arXiv:1411.7423}{\scshape{in}SPIRE}].							
										
\bibitem{Karan:2019sk} S. Karan, G. Banerjee, and B. Panda, \emph{Seeley-DeWitt coefficients in $\mathcal{N} = 2$ Einstein-Maxwell supergravity theory and logarithmic corrections to $\mathcal{N} = 2$ extremal black hole entropy},  \href{https://link.springer.com/article/10.1007%2FJHEP08%282019%29056}{\emph{JHEP} \textbf{08}  (2019) 056} [\href{https://arxiv.org/abs/1905.13058}{arXiv:1905.13058}] [\href{https://inspirehep.net/record/1737533}{\scshape{in}SPIRE}].
											
\bibitem{Banerjee:2020wbr} G. Banerjee, S. Karan and B. Panda, \emph{Logarithmic correction to the entropy of extremal black holes in $\mathcal{N}=1$ Einstein-Maxwell supergravity},  \href{https://link.springer.com/article/10.1007/JHEP01(2021)090}
{\emph{JHEP} \textbf{01} (2021) 090} [\href{https://arxiv.org/abs/2007.11497}{arXiv:2007.11497}] [\href{https://inspirehep.net/literature?sort=mostrecent&size=25&page=1&q=find%20eprint%202007.11497}{\scshape{in}SPIRE}].	
												
\bibitem{Karan:2020sk} S. Karan and B. Panda, \emph{Logarithmic correction to the entropy of Kerr-Newman family of black holes in the matter coupled $\mathcal{N} \geq 1$ Einstein-Maxwell supergravity theories},  [\href{https://arxiv.org/abs/2012.12227}{arXiv:2012.12227}] [\href{https://inspirehep.net/literature?sort=mostrecent&size=25&page=1&q=find%20eprint%202012.12227}{\scshape{in}SPIRE}].
													
\bibitem{Sen:2012rr} A. Sen, \emph{Logarithmic corrections to rotating extremal black hole entropy in four and five dimensions}, \href{https://link.springer.com/article/10.1007%2Fs10714-012-1373-0}
{\emph{Gen. Rel. Grav.} \textbf{44} (2012) 1947} [\href{https://arxiv.org/abs/1109.3706}{arXiv:1109.3706}] [\href{https://inspirehep.net/search?p=find+EPRINT+arXiv:1109.3706}{\scshape{in}SPIRE}].
														
\bibitem{Bhattacharyya:2012ss} S. Bhattacharyya, B. Panda and A. Sen, \emph{Heat kernel expansion and extremal Kerr-Newmann black hole entropy in Einstein-Maxwell theory},  \href{https://link.springer.com/article/10.1007%2FJHEP08%282012%29084}
{\emph{JHEP} \textbf{08}  (2012) 084} [\href{https://arxiv.org/abs/1204.4061}{arXiv:1204.4061}] [\href{https://inspirehep.net/search?p=find+EPRINT+arXiv:1204.4061}{\scshape{in}SPIRE}].	
															
															
\bibitem{Chowdhury:2014np} A. Chowdhury, R.K. Gupta, S. Lal, M. Shyani and S. Thakur, \emph{Logarithmic corrections to twisted indices from the quantum entropy function}, \href{https://link.springer.com/article/10.1007%2FJHEP11%282014%29002}
{\emph{JHEP} \textbf{11} (2014) 002} [\href{https://arxiv.org/abs/1404.6363}{arXiv:1404.6363}] [\href{https://inspirehep.net/search?p=find+EPRINT+arXiv:1404.6363}{\scshape{in}SPIRE}].	
																
\bibitem{Sen:2013ns} A. Sen, \emph{Logarithmic corrections to Schwarzschild and other non-extremal black hole entropy in different dimensions}, \href{https://link.springer.com/article/10.1007%2FJHEP04%282013%29156}{\emph{JHEP} \textbf{04} (2013) 156} [\href{https://arxiv.org/abs/1205.0971v2}{arXiv:1205.0971}] [\href{https://inspirehep.net/record/1113590}{\scshape{in}SPIRE}].			
																	
\bibitem{Charles:2015nn} A.M. Charles and F. Larsen, \emph{Universal corrections to non-extremal black hole entropy in $\mathcal{N}\geq 2$ supergravity}, \href{https://link.springer.com/article/10.1007%2FJHEP06%282015%29200}
{\emph{JHEP} \textbf{06} (2015) 200} [\href{https://arxiv.org/abs/1505.01156}{arXiv:1505.01156}] [\href{https://inspirehep.net/search?p=find+EPRINT+arXiv:1505.01156}{\scshape{in}SPIRE}].
																		
																		
\bibitem{Castro:2018tg} A. Castro, V. Godet, F. Larsen and Y. Zeng, \emph{Logarithmic corrections to black hole entropy: the non-BPS branch}, \href{https://link.springer.com/article/10.1007%2FJHEP05%282018%29079}
{\emph{JHEP} \textbf{05} (2018) 079} [\href{https://arxiv.org/abs/1801.01926}{arXiv:1801.01926}] [\href{https://inspirehep.net/search?p=find+EPRINT+arXiv:1801.01926}{\scshape{in}SPIRE}].
																			
\bibitem{Sen:2008wa} A. Sen, \emph{Entropy function and AdS(2) / CFT(1) correspondence}, \href{https://iopscience.iop.org/article/10.1088/1126-6708/2008/11/075}{\emph{JHEP} \textbf{11} (2008) 075} [\href{https://arxiv.org/abs/0805.0095}{arXiv:0805.0095}] [\href{https://inspirehep.net/search?p=find+EPRINT+arXiv:0805.0095}{\scshape{in}SPIRE}].
																			
\bibitem{Sen:2009wb} A. Sen, \emph{Quantum entropy function from AdS(2)/CFT(1) correspondence}, \href{https://www.worldscientific.com/doi/abs/10.1142/S0217751X09045893}{\emph{Int. J. Mod. Phys.} \textbf{A24} (2009) 4225} [\href{https://arxiv.org/abs/0809.3304}{arXiv:0809.3304}] [\href{https://inspirehep.net/search?p=find+EPRINT+arXiv:0809.3304}{\scshape{in}SPIRE}].
																			
\bibitem{Sen:2009wc} A. Sen, \emph{Arithmetic of quantum entropy function}, \href{https://iopscience.iop.org/article/10.1088/1126-6708/2009/08/068}{\emph{JHEP} \textbf{08} (2009) 068} [\href{https://arxiv.org/abs/0903.1477}{arXiv:0903.1477}] [\href{https://inspirehep.net/search?p=find+EPRINT+arXiv:0903.1477}{\scshape{in}SPIRE}].
\bibitem{Majhi:2015pra} B.R. Majhi, \emph{Entropy function from the gravitational surface action for an extremal near horizon black hole}, \href{https://doi.org/10.1140/epjc/s10052-015-3744-7}{\emph{Eur. Phys. J. C} \textbf{75} (2015) 521} [\href{https://arxiv.org/abs/1503.08973}{arXiv:1503.08973}] [\href{https://inspirehep.net/literature?sort=mostrecent&size=25&page=1&q=find%20eprint%201503.08973}{\scshape{in}SPIRE}].


\bibitem{Schwinger:1951sp} J.S. Schwinger, \emph{On gauge invariance and vacuum polarization}, \href{https://journals.aps.org/pr/abstract/10.1103/PhysRev.82.664}{\emph{Phys. Rev.} \textbf{82} (1951) 664} [\href{https://inspirehep.net/search?p=find+J+%22Phys.Rev.,82,664%22}{\scshape{in}SPIRE}].
																				
\bibitem{DeWitt:1975ps}
B.S. DeWitt, \emph{Quantum field theory in curved space-time}, \href{https://www.sciencedirect.com/science/article/abs/pii/0370157375900514?via%3Dihub}{\emph{Phys. Rept.} \textbf{19} (1975) 295} [\href{https://inspirehep.net/search?p=find+J+%22Phys.Rept.,19,295%22}{\scshape{in}SPIRE}].

\bibitem{DeWitt:1965ff} B.S. DeWitt, \emph{Dynamical theory of groups and fields}, Gordon and Breach, New York, NY, U.S.A. (1965).
																					
\bibitem{DeWitt:1967gg} B.S. DeWitt, \emph{Quantum theory of gravity. 1. The canonical theory}, \href{https://journals.aps.org/pr/abstract/10.1103/PhysRev.160.1113}{\emph{Phys. Rev.} \textbf{160} (1967) 1113} [\href{https://inspirehep.net/search?p=find+J+%22Phys.Rev.,160,1113%22}{\scshape{in}SPIRE}].
																						
\bibitem{DeWitt:1967hh} B.S. DeWitt, \emph{Quantum theory of gravity. 2. The manifestly covariant theory}, \href{https://journals.aps.org/pr/abstract/10.1103/PhysRev.162.1195}{\emph{Phys. Rev.} \textbf{162} (1967) 1195} [\href{https://inspirehep.net/search?p=find+J+%22Phys.Rev.,162,1195%22}{\scshape{in}SPIRE}].
																							
\bibitem{DeWitt:1967ii} B.S. DeWitt, \emph{Quantum theory of gravity. 3. Applications of the covariant theory}, \href{https://journals.aps.org/pr/abstract/10.1103/PhysRev.162.1239}{\emph{Phys. Rev.} \textbf{162} (1967) 1239} [\href{https://inspirehep.net/search?p=find+J+%22Phys.Rev.,162,1239%22}{\scshape{in}SPIRE}].
																								
\bibitem{Seeley:1966tt} R.T. Seeley, \emph{Singular integrals and boundary value problems}, \href{https://www.jstor.org/stable/2373078?origin=crossref&seq=1#page_scan_tab_contents}{\emph{Amer. J. Math.} \textbf{88} (1966) 781}.
																								
\bibitem{Seeley:1969uu} R. Seeley, \emph{The resolvent of an elliptic boundary value problem}, \href{https://www.jstor.org/stable/2373309?origin=crossref&seq=1#page_scan_tab_contents}{\emph{Amer. J. Math.} \textbf{91} (1969) 889}.
																								
\bibitem{Vassilevich:2003ll} D.V. Vassilevich, \emph{Heat kernel expansion: user’s manual}, \href{https://www.sciencedirect.com/science/article/pii/S0370157303003545?via%3Dihub}{\emph{Phys. Rept.} \textbf{388} (2003) 279} [\href{https://arxiv.org/abs/hep-th/0306138}{arXiv:hep-th/0306138}] [\href{https://inspirehep.net/search?p=find+EPRINT+hep-th/0306138}{\scshape{in}SPIRE}].
																									
\bibitem{Adamo:2014lk} T. Adamo, and E.T. Newman, \emph{The Kerr-Newman metric: A review}, \href{http://www.scholarpedia.org/article/Kerr-Newman_metric}{\emph{Scholarpedia} \textbf{9} (2014) 31791} [\href{https://arxiv.org/abs/1410.6626}{arXiv:1410.6626 }] [\href{https://inspirehep.net/record/1323609}{\scshape{in}SPIRE}].

\bibitem{Grana:2006mg} M. Grana, \emph{Flux compactifications in string theory: a comprehensive review}, \emph{Phys. Rept.} \textbf{423} (2006) 91 [\href{https://arxiv.org/abs/hep-th/0509003}{arXiv:hep-th/0509003}] [\href{https://inspirehep.net/literature?sort=mostrecent&size=25&page=1&q=find%20eprint%20hep-th%2F0509003}{\scshape{in}SPIRE}].
	
\bibitem{Freedman:2012xp} D. Z. Freedman and A. Van Proeyen, \emph{Supergravity}. Cambridge University Press, Cambridge, UK, 2012.
																									
\bibitem{Astefanesei:2019pk} D. Astefanesei, C. Herdeiro, A. Pombo and E. Radu, \emph{Einstein-Maxwell-scalar black holes: classes of solutions, dyons and extremality}, \href{https://link.springer.com/article/10.1007/JHEP10(2019)078}{\emph{JHEP} \textbf{10} (2019) 078} [\href{https://arxiv.org/abs/1905.08304v1}{arXiv:1905.08304}] [\href{https://inspirehep.net/record/1735793}{\scshape{in}SPIRE}].
																									
																								
																									
																									
\bibitem{Duff:1977ay} M.J. Duff, \emph{Observations on conformal anomalies}, \href{https://www.sciencedirect.com/science/article/pii/0550321377904102?via%3Dihub}{\emph{Nucl. Phys.} \textbf{B125} (1977) 334} [\href{https://inspirehep.net/literature/119306}{\scshape{in}SPIRE}].
																										
\bibitem{Christensen:1979md} S.M. Christensen and  M.J. Duff, \emph{New gravitational index theorems and supertheorems}, \href{https://www.sciencedirect.com/science/article/pii/0550321379905169?via%3Dihub}{\emph{Nucl. Phys.} \textbf{B154} (1979) 301} [\href{https://inspirehep.net/literature/133020}{\scshape{in}SPIRE}].
																											
\bibitem{Christensen:1980iy} S.M. Christensen and  M.J. Duff, \emph{Quantizing gravity with a cosmological constant}, \href{https://www.sciencedirect.com/science/article/pii/055032138090423X?via%3Dihub}{\emph{Nucl. Phys.} \textbf{B170} (1980) 480} [\href{https://inspirehep.net/literature/142745}{\scshape{in}SPIRE}].
																												
\bibitem{Duff:1980qv} M.J. Duff and P. van Nieuwenhuizen, \emph{Quantum inequivalence of different field representations}, \href{https://www.sciencedirect.com/science/article/abs/pii/0370269380908527?via%3Dihub}{\emph{Phys. Lett.} \textbf{B94} (1980) 179} [\href{https://inspirehep.net/literature/153368}{\scshape{in}SPIRE}].
																													
\bibitem{Christensen:1980ee} S.M. Christensen, M.J. Duff, G.W. Gibbons and M. Rocek, \emph{Vanishing one-loop $\beta$ function in gauged $N > 4$ supergravity}, \href{https://journals.aps.org/prl/abstract/10.1103/PhysRevLett.45.161}{\emph{Phys. Rev. Lett.} \textbf{45} (1980) 161} [\href{https://inspirehep.net/literature/152582}{\scshape{in}SPIRE}].

\bibitem{Majhi:2009uk} B.R. Majhi and S. Samanta, \emph{Hawking radiation due to photon and gravitino tunneling}, \href{https://doi.org/10.1016/j.aop.2010.06.010}{\emph{Annals Phys.} \textbf{325} (2010) 2410} [\href{https://arxiv.org/abs/0901.2258}{	arXiv:0901.2258}] [\href{https://inspirehep.net/literature?sort=mostrecent&size=25&page=1&q=find%20eprint%200901.2258}{\scshape{in}SPIRE}].


\bibitem{Henry:2000wd} R.C. Henry, \emph{Kretschmann scalar for a kerr-newman black hole}, \href{https://iopscience.iop.org/article/10.1086/308819}{\emph{Astrophys. J.} \textbf{535} (2000) 350} [\href{https://arxiv.org/abs/astro-ph/9912320}{arXiv:astro-ph/9912320}] [\href{https://inspirehep.net/search?p=find+EPRINT+astro-ph/9912320}{\scshape{in}SPIRE}].

\bibitem{Cherubini:2002we} C. Cherubini, D. Bini, S. Capozziello and R. Ruffini, \emph{Second order scalar invariants of the Riemann tensor: Applications to black hole space-times}, \href{https://www.worldscientific.com/doi/abs/10.1142/S0218271802002037}{\emph{Int. J. Mod. Phys.} \textbf{D11} (2002) 827} [\href{https://arxiv.org/abs/gr-qc/0302095}{arXiv:gr-qc/0302095}] [\href{https://inspirehep.net/search?p=find+EPRINT+gr-qc/0302095}{\scshape{in}SPIRE}].

\bibitem{Karan:2018ac} S. Karan, S. Kumar and B. Panda, \emph{General heat kernel coefficients for massless free spin-3/2 Rarita-Schwinger field}, \href{https://www.worldscientific.com/doi/abs/10.1142/S0217751X1850063X}{\emph{Int. J. Mod. Phys.} \textbf{A33} (2018) 1850063} [\href{https://arxiv.org/abs/1709.08063}{arXiv:1709.08063}] [\href{https://inspirehep.net/search?p=find+EPRINT+arXiv:1709.08063}{\scshape{in}SPIRE}].
\bibitem{Ferrara:2012bp} S. Ferrara and A. Marrani, \emph{Generalized mirror symmetry and quantum black hole entropy}, \href{https://www.sciencedirect.com/science/article/pii/S0370269311014481?via%3Dihub}{\emph{Phys. Lett.} \textbf{B707} (2012) 173-177} [\href{https://arxiv.org/abs/1109.0444}{arXiv:1109.0444}] [\href{https://inspirehep.net/search?p=find+eprint+1109.0444}{\scshape{in}SPIRE}].			
																													
	
\bibitem{Behrndt:1998eq} K. Behrndt, G. Lopes Cardoso, B. de Wit, D. L\"ust, T. Mohaupt and W. A. Sabra, \emph{Higher order black hole solutions in N=2 supergravity and Calabi-Yau string backgrounds}, \href{https://www.sciencedirect.com/science/article/abs/pii/S0370269398004134?via%3Dihub}
{\emph{Phys. Lett. B} \textbf{429} (1998) 289} [\href{https://arxiv.org/abs/hep-th/9801081}{arXiv:hep-th/9801081}] [\href{https://inspirehep.net/literature?sort=mostrecent&size=25&page=1&q=find%20eprint%20hep-th%2F9801081}{\scshape{in}SPIRE}].
	
\bibitem{LopesCardoso:1998tkj} G. Lopes Cardoso, B. de Wit and T. Mohaupt, \emph{Corrections to macroscopic supersymmetric black hole entropy}, \href{https://www.sciencedirect.com/science/article/abs/pii/S0370269399002270?via%3Dihub}{\emph{Phys. Lett. B} \textbf{451} (1999) 309} [\href{https://arxiv.org/abs/hep-th/9812082}{arXiv:hep-th/9812082}] [\href{https://inspirehep.net/literature?sort=mostrecent&size=25&page=1&q=find%20eprint%20hep-th%2F9812082}{\scshape{in}SPIRE}].
	
\bibitem{LopesCardoso:1999cv} G. Lopes Cardoso, B. de Wit and T. Mohaupt, \emph{Deviations from the area law for supersymmetric black holes}, {\emph{Fortsch. Phys.} \textbf{48} (2000) 49} [\href{https://arxiv.org/abs/hep-th/9904005}{arXiv:hep-th/9904005}] [\href{https://inspirehep.net/literature?sort=mostrecent&size=25&page=1&q=find%20eprint%20hep-th%2F9904005}{\scshape{in}SPIRE}].	
	
\bibitem{Mohaupt:2000mj} T. Mohaupt, \emph{Black hole entropy, special geometry and strings}, {\emph{Fortsch. Phys.} \textbf{49} (2001) 3} [\href{https://arxiv.org/abs/hep-th/0007195}{arXiv:hep-th/0007195}] [\href{https://inspirehep.net/literature?sort=mostrecent&size=25&page=1&q=find%20eprint%20hep-th%2F0007195}{\scshape{in}SPIRE}].
	
\bibitem{Sahoo:2006rp} B. Sahoo and A. Sen, \emph{Higher derivative corrections to non-supersymmetric extremal black holes in N=2 supergravity}, \href{https://iopscience.iop.org/article/10.1088/1126-6708/2006/09/029}
{\emph{JHEP} \textbf{09} (2006) 029} [\href{https://arxiv.org/abs/hep-th/0603149}{arXiv:hep-th/0603149}] [\href{https://inspirehep.net/literature?sort=mostrecent&size=25&page=1&q=find%20eprint%20hep-th%2F0603149}{\scshape{in}SPIRE}].						

\bibitem{Sen:2006iz} A. Sen, \emph{Entropy function for heterotic black holes}, \href{https://iopscience.iop.org/article/10.1088/1126-6708/2006/03/008}
{\emph{JHEP} \textbf{03} (2006) 008} [\href{https://arxiv.org/abs/hep-th/0508042v4}{arXiv:hep-th/0508042}] [\href{https://inspirehep.net/literature?sort=mostrecent&size=25&page=1&q=find%20eprint%20hep-th%2F0508042}{\scshape{in}SPIRE}].
	
\bibitem{Maldacena:1997de} J. Maldacena, A. Strominger and E. Witten, \emph{Black hole entropy in M-theory}, \href{https://iopscience.iop.org/article/10.1088/1126-6708/1997/12/002}
{\emph{JHEP} \textbf{12} (1997) 002} [\href{https://arxiv.org/abs/hep-th/9711053}{arXiv:hep-th/9711053}] [\href{https://inspirehep.net/literature?sort=mostrecent&size=25&page=1&q=find%20eprint%20hep-th%2F9711053}{\scshape{in}SPIRE}].	
	
\bibitem{Gibbons:1977ta} G.W. Gibbons and S.W. Hawking, \emph{Action integrals and partition functions in quantum gravity}, \href{https://journals.aps.org/prd/abstract/10.1103/PhysRevD.15.2752}{\emph{Phys. Rev.} \textbf{D15} (1977) 2752} [\href{https://inspirehep.net/search?p=find+J+%22Phys.Rev.,D15,2752%22}{\scshape{in}SPIRE}].
																																
\bibitem{Hawking:1978td} S.W. Hawking, \emph{Quantum gravity and path integrals}, \href{https://journals.aps.org/prd/abstract/10.1103/PhysRevD.18.1747}{\emph{Phys. Rev.} \textbf{D18 } (1978) 1747} [\href{https://inspirehep.net/search?p=find+J+%22Phys.Rev.,D18,1747%22}{\scshape{in}SPIRE}].
																																	
\bibitem{Hawking:1977te} S.W. Hawking, \emph{Zeta function regularization of path integrals in curved space-time}, \href{https://link.springer.com/article/10.1007%2FBF01626516}{\emph{Commun. Math. Phys.} \textbf{55} (1977) 133} [\href{https://inspirehep.net/search?p=find+J+%22Comm.Math.Phys.,55,133%22}{\scshape{in}SPIRE}].
																																		
\bibitem{Denardo:1982tb} G. Denardo and E. Spallucci, \emph{Induced quantum gravity from heat kernel expansion}, \href{https://link.springer.com/article/10.1007%2FBF02902652}{\emph{Nuovo Cim.} \textbf{A69} (1982) 151} [\href{https://inspirehep.net/search?p=find+J+%22NuovoCim.,A69,151%22}{\scshape{in}SPIRE}].
																																			
\bibitem{Avramidi:1994th} I.G. Avramidi, \emph{The heat kernel approach for calculating the effective action in quantum field theory and quantum gravity} [\href{https://arxiv.org/abs/hep-th/9509077}{arXiv:hep-th/9509077}] [\href{https://inspirehep.net/search?p=find+EPRINT+hep-th/9509077}{\scshape{in}SPIRE}].	
																																			
\bibitem{Peixoto:2001wx} G. De Berredo-Peixoto, \emph{A note on the heat kernel method applied to fermions}, \href{https://www.worldscientific.com/doi/abs/10.1142/S0217732301005965}{\emph{Mod. Phys. Lett.} \textbf{A16} (2001) 2463} [\href{https://arxiv.org/abs/hep-th/0108223}{arXiv:hep-th/0108223}] [\href{https://inspirehep.net/search?p=find+EPRINT+hep-th/0108223}{\scshape{in}SPIRE}].		
																																			
\bibitem{Bhattacharyya:2012ye} S. Bhattacharyya, A. Grassi, M. Marino and A. Sen, \emph{A one-loop test of quantum supergravity}, \href{https://iopscience.iop.org/article/10.1088/0264-9381/31/1/015012}{\emph{Class. Quant. Grav.} \textbf{31} (2014) 015012} [\href{https://arxiv.org/abs/1210.6057v3}{arXiv:1210.6057}] [\href{https://inspirehep.net/literature?sort=mostrecent&size=25&page=1&q=find%20eprint%201210.6057}{\scshape{in}SPIRE}].
	
\bibitem{Charles:2018yey}  A.M. Charles, \emph{Explorations of non-supersymmetric black holes in supergravity}, \href{https://deepblue.lib.umich.edu/handle/2027.42/144036}{PhD thesis}, Michigan U., 2018  [\href{https://inspirehep.net/literature/1707891}{\scshape{in}SPIRE}].

	

\end{thebibliography}
\end{document}